\documentclass[conf]{new-aiaa}
\usepackage[utf8]{inputenc}

\usepackage{graphicx}
\usepackage{amsmath}
\usepackage[version=4]{mhchem}
\usepackage{siunitx}
\usepackage{longtable,tabularx}
\usepackage{float}
\usepackage{subfig}
\usepackage{placeins} 
\usepackage{xcolor}
\usepackage{multirow}

\title{A Comparison of LES Inlet Boundary Conditions for Supersonic Jet Flows}

\author{Diego F. Abreu\footnote{Ph.D. Candidate, Graduate Program in Space Sciences 
and Technologies, Departamento de Ciência e Tecnologia Aeroespacial, DCTA/ITA; 
E-mail: mecabreu@yahoo.com.br.}}
\affil{Instituto Tecnológico de Aeronáutica, 12228--900, São José dos Campos, SP, Brazil}
\author{Carlos Junqueira-Junior\footnote{Research Engineer, Arts et Métiers Institute of 
Technology, DynFluid laboratory; E-mail: junior.junqueira@ensam.eu.}}
\affil{Arts et Métiers Institute of Technology, DynFluid, CNAM, 
151 Boulevard de l'Hôpital, 75013, Paris, France}
\author{Eron T. V. Dauricio\footnote{Ph.D. Candidate, Graduate Program in Space Sciences and 
Technologies, Departamento de Ciência e Tecnologia Aeroespacial, DCTA/ITA; 
E-mail: eron.tiago90@gmail.com.}}
\affil{Instituto Tecnológico de Aeronáutica, 12228--900, São José dos Campos, SP, Brazil}
\author{João Luiz F. Azevedo\footnote{Senior Research Engineer, Aerodynamics Division, 
Departamento de Ciência e Tecnologia Aeroespacial, DCTA/IAE/ALA; 
E-mail: joaoluiz.azevedo@gmail.com. Fellow AIAA.}}
\affil{Instituto de Aeronáutica e Espaço, 12228--904, São José dos Campos, SP, Brazil}

\begin{document}

\maketitle

\begin{abstract}
The present work evaluates the effects 
of three inflow boundary conditions on large-eddy simulations of 
supersonic jet flows. The three inlet flow configurations considered are an inviscid profile, a stationary turbulent
profile extracted from a RANS calculation, and a time-dependent tripped
turbulent profile. The study applies the nodal discontinuous Galerkin spatial
discretization to perform the supersonic jet flow simulations. Initially, a
mesh and polynomial resolution study is performed to identify the requisites
for simulating the supersonic jet flow. The velocity profiles resulting from the three
numerical simulations with the different inflow conditions are compared with
experimental data. The velocity profiles extracted from the calculation results
indicate a major role of the steady turbulent inlet profile on the jet flow
behavior, even for the tripped inlet condition. The imposition of a steady 
turbulent inlet profile could approximate the velocity distributions in the
region close to the jet inlet section with the experimental reference. 
Negligible changes were observed to the velocity profiles far from the inlet
section, where a good agreement with the experimental reference was already
obtained from the simulations even with the inviscid profile.
\end{abstract}


\section{Introduction}

The computer power currently available in scientific computing centers 
enables the application of large-eddy simulations (LES) for realistic flow 
conditions. Such an approach can provide valuable intel on aerospace 
configurations such as shear layer and detached flows due to their capacity to
generate unsteady data for flow and temperature fields with high-frequency
fluctuations are necessary for aerodynamics, acoustics, loads, and heat 
transfer analyses. The current paper studies the effects of inlet boundary 
conditions on the LES of supersonic jet flows.

The influence of sub-grid scale modeling on supersonic jet flows was 
investigated in Ref. \cite{Junior2018} using a highly scalable finite
difference solver \cite{Junior2019,Junior2020}. Recent work highlights the
effects of polynomial and mesh resolution on the large-eddy simulation of a
perfectly expanded supersonic jet flow at 1.4 Mach number when using an
inviscid inlet boundary condition (BC) \cite{Abreuetal2022, Abreu2022-ecommas,
Abreu2021}. The results indicate a good agreement with experimental data, but
in the region near the nozzle exit where numerical and experimental velocity
fluctuation profiles differ. Hence, the present effort addresses the effects 
of different inlet BC profiles in the supersonic jet flow simulation to
evaluate its influence on the velocity fluctuation at the nozzle exit vicinity.

The research group performs LES calculations using unstructured nodal 
discontinuous-Galerkin (DG) \cite{Kopriva2010, Hindenlang2012} spatial 
discretizations with three different inlet conditions: the flat-hat inviscid
profile, the steady profile from a RANS calculation, and a time-dependent 
profile generated with a tripping technique \cite{Bogeyetal2011} superimposed
with the steady profile from RANS calculation. In the present study, the
resolution study \cite{Abreuetal2022} is extended with one other simulation
that resolved the most refined mesh with a third-order accurate spatial
discretization. The identification of the mesh and polynomial {\it{}hp}
requisites for the simulation of supersonic jet flows is employed for the
simulations with the different inflow conditions. Velocity fields from
numerical simulations are compared with experimental data to verify the
differences from the velocity data produced by each of the numerical
simulations.



\section{Theoretical and Numerical Formulation}

The work has an interest in the solution of the filtered Navier-Stokes 
equations. The formulation is based on a spatial filtering process 
separating the flow into resolved and non-resolved parts. The filtering
process is implicitly performed using the mesh size as the spatial filter.
The filtered Navier-Stokes equations are utilized in their conservative form
and employ the ``system I'' of Vreman \cite{Vreman1995} to filter the energy
equation. The ideal gas equation of state is utilized, and Sutherland's law is used to
obtain the dynamic viscosity coefficient. The modeling of the subgrid-scale contribution is
performed by an additional stress tensor where the subgrid-scale viscosity
coefficient is employed. The Boussinesq hypothesis is the modeling reference.
The static Smagorinsky model \cite{Smagorinsky1963} is employed to calculate
the subgrid-scale viscosity coefficient.

The nodal discontinuous Galerkin method used in the present work is based on the 
formulation discussed in Refs.\ \cite{Kopriva2010, Hindenlang2012}. In the
methodology presented in these references, the domain is divided into non-overlapping hexahedral
elements. This choice of elements permits the interpolating polynomial to be 
defined as a tensor-product basis with degree $N$ in each space direction. The
implementation is simpler and improves the computational efficiency of the
code. For the nodal form of the discontinuous Galerkin formulation, the 
solution and fluxes in each element are approximated by a polynomial
interpolation, which is constructed from 1-D interpolating polynomials and that is extended to 3-D using a tensor product, as indicated. The nodal discontinuous Galerkin method 
is implemented in the FLEXI numerical framework \cite{Krais2021}. 

The numerical scheme used in the simulations additionally implements the split formulation 
presented in Ref.\ \cite{Pirozzoli2011}, with the discrete form discussed in the work 
presented in Ref.\ \cite{Gassner2016}. Such formulation is useful in order to enhance 
the stability of the simulation. The split formulation is employed for Euler fluxes only. 
The solution and the fluxes are interpolated and integrated at the nodes of a 
Gauss-Lobatto-Legende quadrature, which implements the summation-by-parts property 
necessary to employ the split formulation. The Riemann solver used in the simulations 
is the Roe scheme with entropy fix \cite{Harten1983}. The lifting scheme presented 
in Ref.\ \cite{BassiRebay1997} is used, which is also known as the BR2 scheme. The 
time marching method chosen is an explicit Runge-Kutta scheme with three steps and 
third-order accuracy \cite{Kopriva2009}. The shock waves in the simulation are 
stabilized using the finite-volume sub-cell shock capturing method presented in 
Ref.\ \cite{Sonntag2017}.



\section{Supersonic Jet Flows}

\subsection{Simulation Model Description}

The current study simulates a perfectly expanded supersonic free jet flow with 
a Mach number of 1.4 and a Reynolds number based on a jet inlet diameter,
equal to $1.58 \times 10^6$. This operating condition reproduces one of the
configurations analyzed in the work presented in Ref.~\cite{BridgesWernet2008},
whose data is used as a reference.

\begin{figure}[htb!]
\centering
	\includegraphics[width=0.65\linewidth]{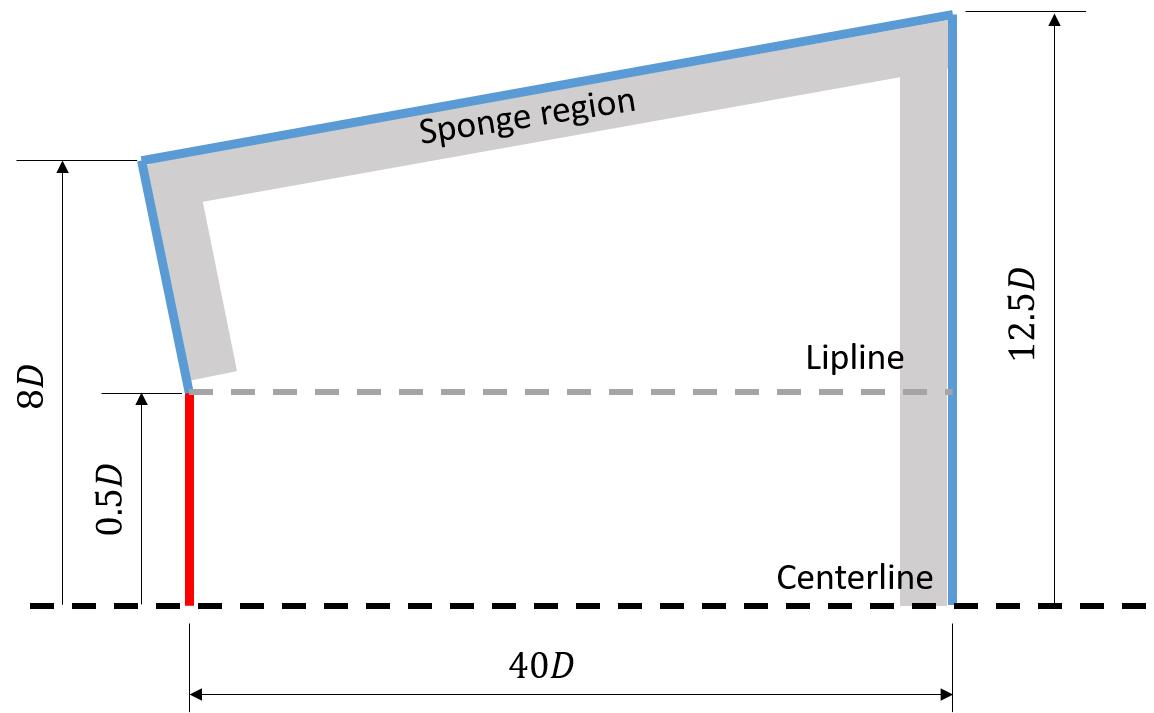}
    \caption{2-D schematic representation of the computational domain used on 
    the jet flow simulations.}
    \label{fig.geo}
\end{figure}

The geometry used for the calculations presents a divergent shape and axis 
length of $40D$, where $D$ is the jet inlet diameter and has external diameters
of $16D$ and $25D$. Figure~\ref{fig.geo} illustrates a 2-D representation of
the computational domain, indicating the inlet surface in red, the far-field
region in blue, the lipline in grey, and the centerline in black. 

\begin{figure}[htb!]
\centering
\subfloat[M-1 mesh.]{
	\includegraphics[width=0.48\linewidth]{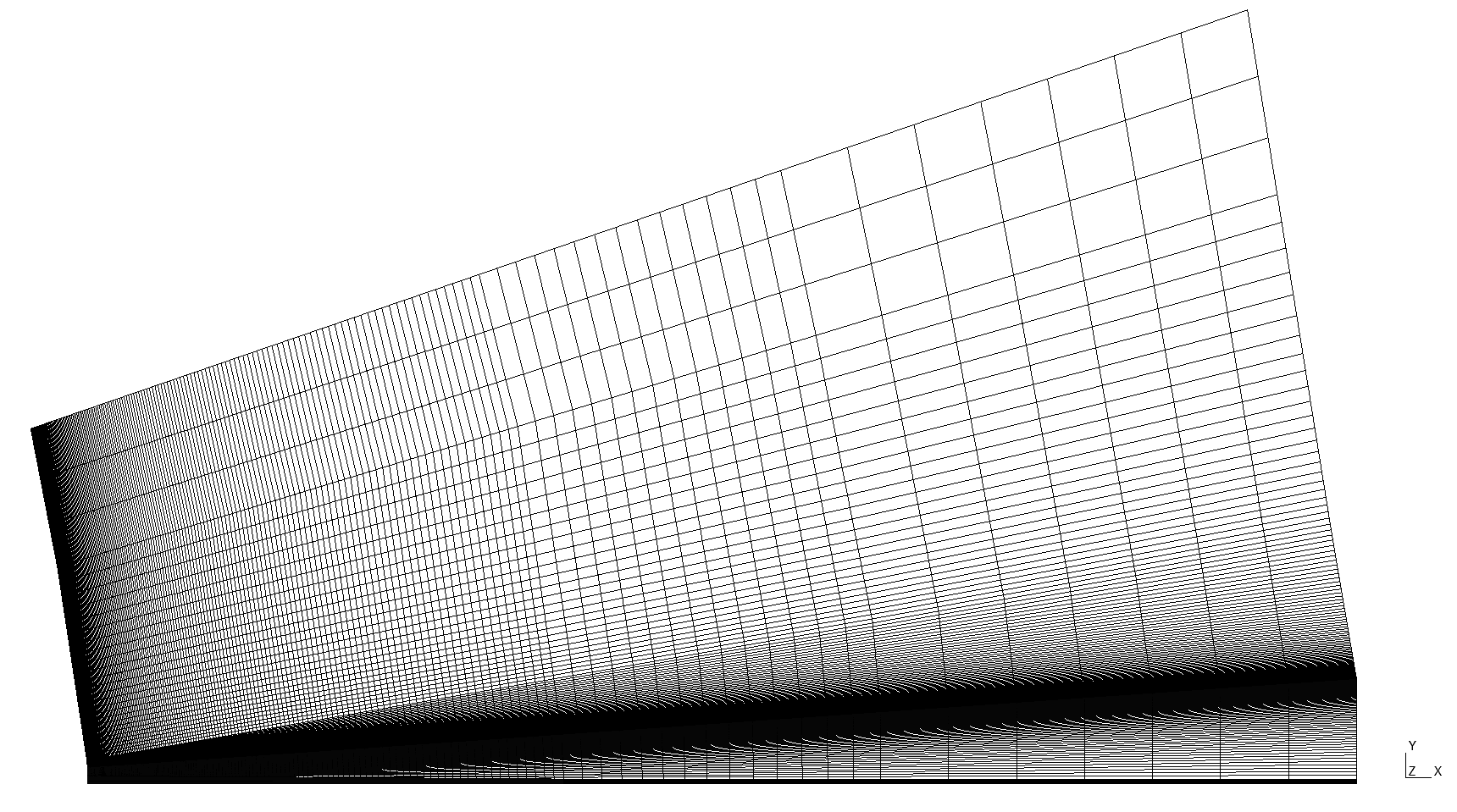}
	\label{fig.mesh1}
	}
\subfloat[M-2 mesh.]{
	\includegraphics[width=0.48\linewidth]{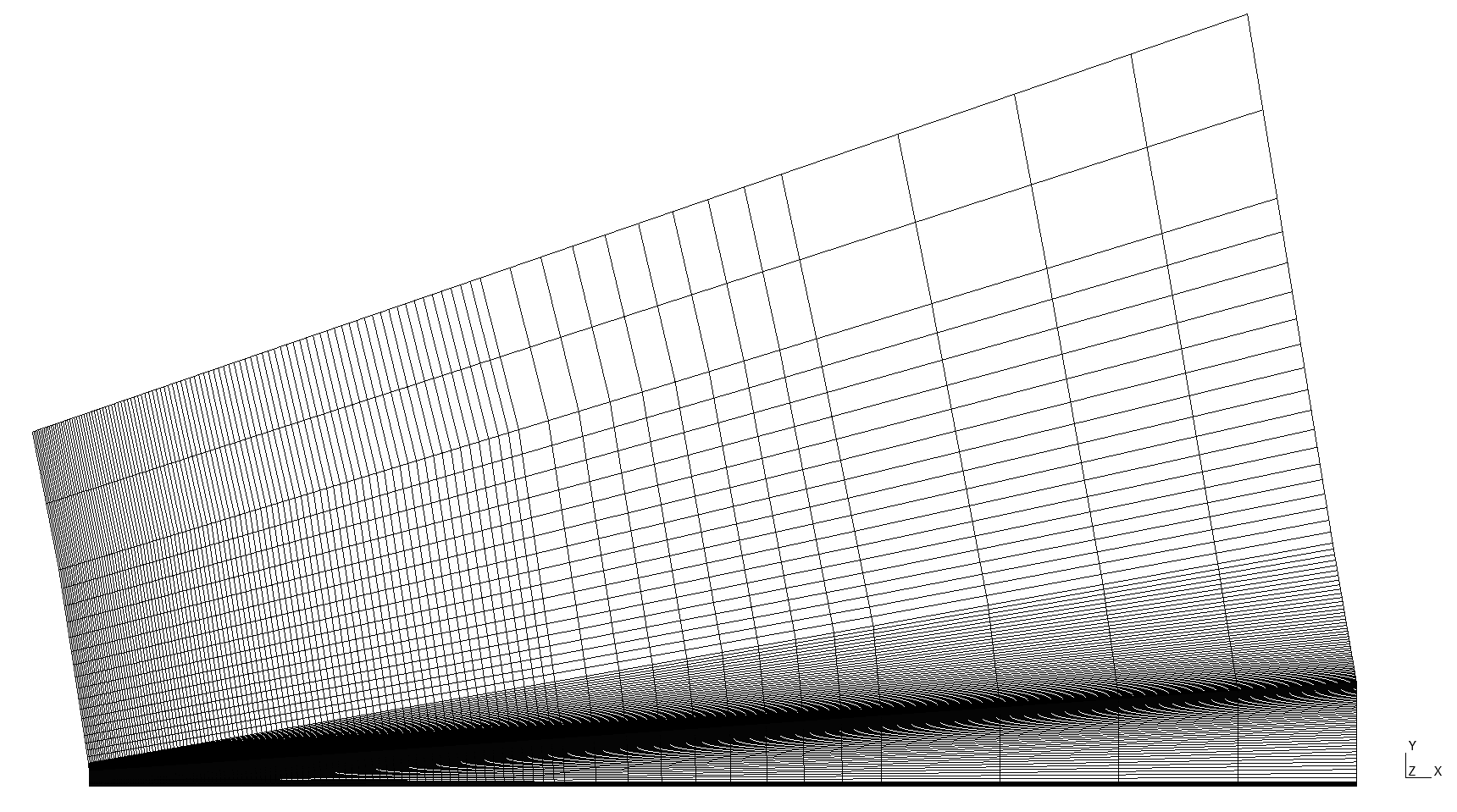}
	\label{fig.mesh2}
	}
\\
\subfloat[M-3 mesh.]{
	\includegraphics[width=0.48\linewidth]{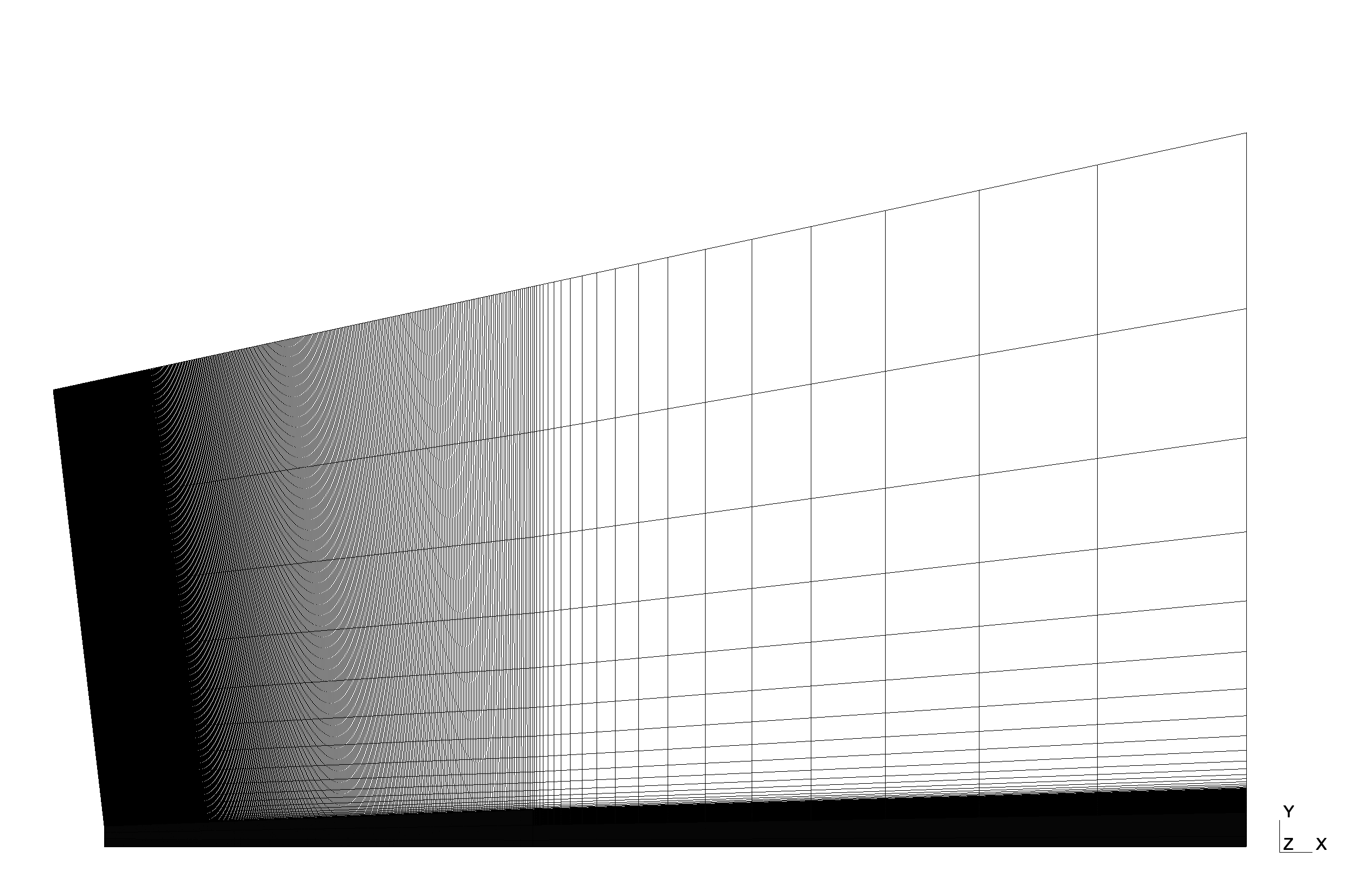}
	\label{fig.mesh3}
	}	
\caption{Visualization of the longitudinal cut planes for the meshes used in
         the present work.}
\label{fig.mesh}
\end{figure}

The study performs simulations using three computational meshes: M-1, M-2, and 
M-3. The first two meshes present the same topology, with the M-1 mesh being
more refined than the M-2 mesh. The latter has a different topology and is the 
most refined mesh employed in this study. Additional information about the mesh
generation is presented in Ref.~\cite{Abreuetal2022}. The grids are created
using the GMSH mesh generator \cite{Geuzaine2009} applying a multiblock
strategy to handle hexahedral elements. Figure~\ref{fig.mesh} exhibits a cut
plane of the meshes described here. The M-1 and M-2 grids have $6.2 \times
10^6$ and $1.8 \times 10^6$ elements, while the M-3 mesh has $15.4 \times 10^6$
elements.

\subsection{Boundary Conditions}

The jet inflow, $(\cdot)_{jet}$, and far-field, $(\cdot)_{ff}$, surfaces are
indicated in Fig.~\ref{fig.geo} in red and blue, respectively. A weakly
enforced solution to a Riemann problem with a Dirichlet condition is enforced
at the boundaries. The flow is characterized as perfectly expanded and
isothermal, {\it i.e.} $p_{jet}/p_{ff}=T_{jet}/T_{ff}=1$, where $p$ stands for
pressure and $T$ for temperature. The Mach number of the jet at the inlet is 
$M_{jet}=1.4$, and the Reynolds number based on the diameter of the nozzle is 
$Re_{jet} = 1.58 \times 10^6$. A small velocity component with $M_{ff}=0.01$ 
in the streamwise direction is imposed at the far-field surfaces. A sponge zone
\cite{Flad2014} is employed around the far-field boundaries, the gray area
illustrated in Fig.~\ref{fig.geo}, to dampen any oscillations that could be
reflected in the jet flow. The simulations apply three different inlet boundary
conditions to study their effect on the calculations: an inviscid profile, a
turbulent profile generated from a steady RANS calculation, and a turbulent 
profile that contains time-dependent fluctuation.

\subsubsection*{Inviscid Profile}

The inviscid profile is the same as imposed in previous studies 
\cite{Abreuetal2022} where properties of the flow are constant all over the
boundary condition with values that exactly match the specified Mach number 
and Reynolds number. The inviscid profile is the simplest and cheapest way 
to implement the boundary condition, however, it is the condition that lacks
physical representation due to the absence of a boundary layer and
unsteadiness. 

\subsubsection*{Mean RANS Profile}

The turbulent steady velocity profile introduces more physical information 
to the jet inlet condition. An {\it{}a priori} RANS simulation is performed 
to provide mean profiles for all properties in the jet inlet condition. At 
this point, the mean profiles are imposed with no fluctuation. An interpolation
is necessary due to differences between the LES and RANS grids applied in each
calculation.

The RANS simulation is solved by a finite volume formulation implemented in 
the commercial software CFD++ \cite{Chakravarthy1999}. The RANS equations are
solved with a second-order accurate discretization using an HLLE Riemman solver
\cite{Einfeldt1988} and minmod limiter \cite{VanLeer1973}. The equation of
state is solved for a perfect gas. The relaxation procedure utilizes an
implicit backward Euler, modified for implementation within a Message Passing 
Interface (MPI). The Spalart-Allmaras turbulence model calculates the turbulent
variables \cite{SpalartAllmaras94}.

\begin{figure}[htb!]
\centering
	\subfloat[2-D representation of the computational domain of the nozzle.] {
	\includegraphics[width=0.47\linewidth]{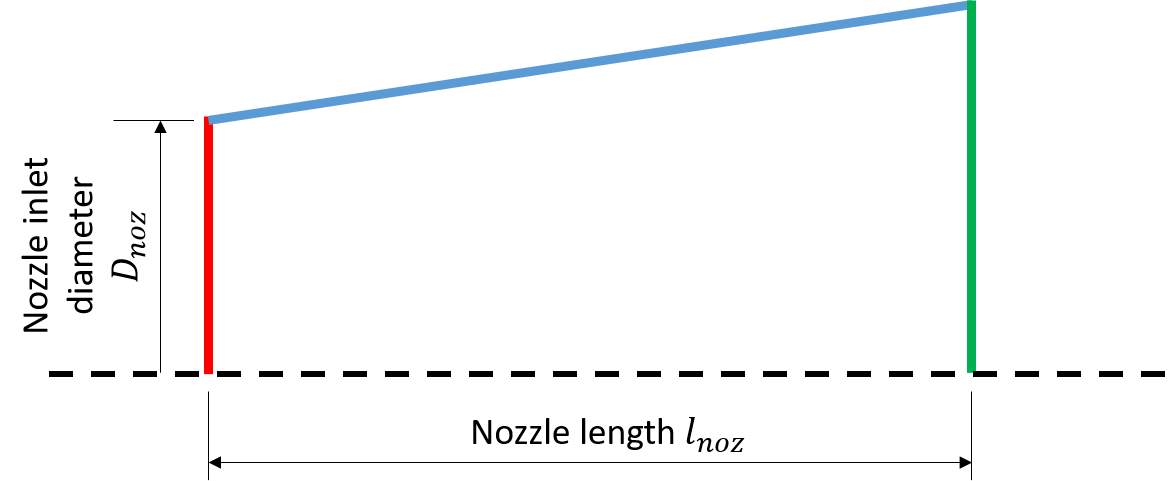}
    \label{fig.geo_sketch}
	}
	\subfloat[Cut plane of the nozzle mesh utilized in the simulations.]{
	\includegraphics[width=0.47\linewidth]{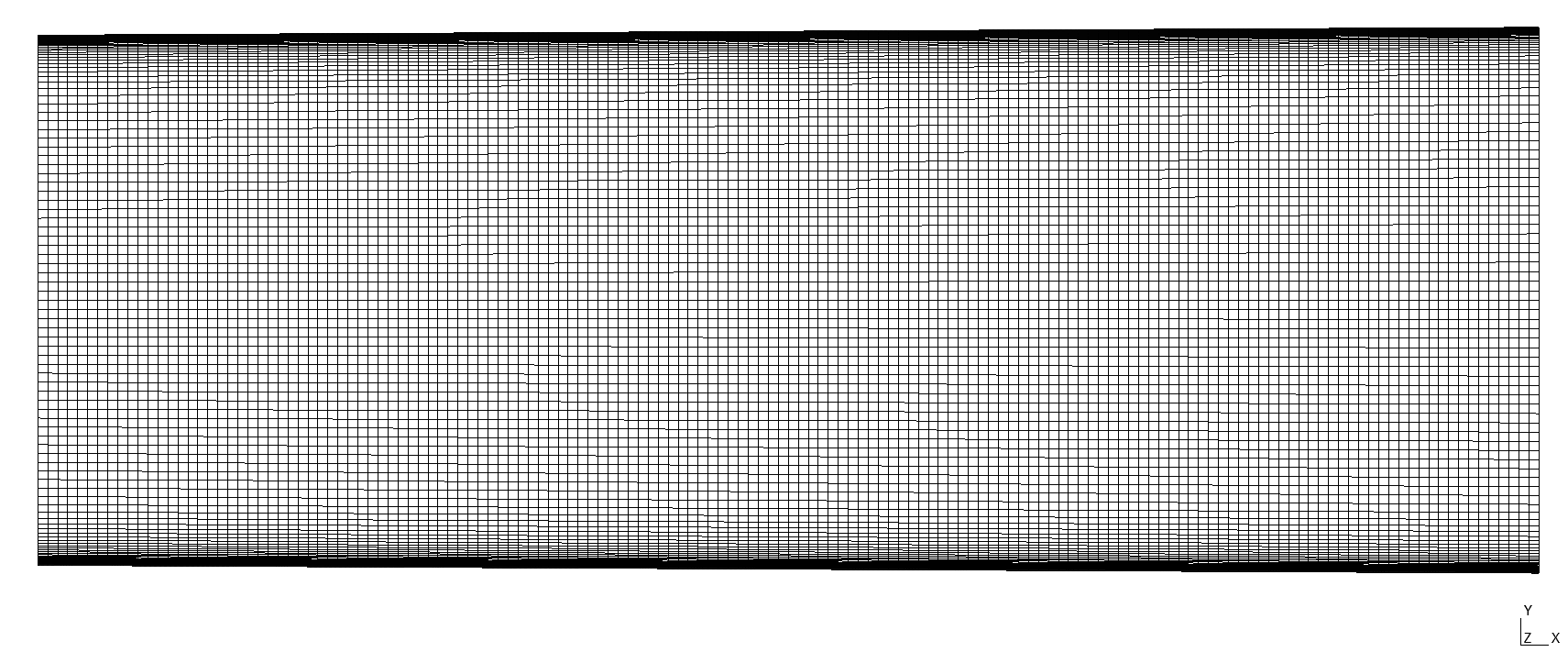}
    \label{fig.mesh_rans}
	}
    \caption{2-D schematic representation and cut plane of the grid used 
    for the RANS calculation of the nozzle flow.}
    \label{fig.rans_sketch}
\end{figure}

The study applies a nozzle geometry considering only the divergent section to
enable control over the boundary layer growth by varying the nozzle length. 
The nozzle inlet diameter is $D_{noz}=0.9686D_j$ while the nozzle length is
$L_{noz}=2.75D_j$. A sketch of the RANS domain is presented in 
Fig.~\ref{fig.geo_sketch}. 

The mesh generated presents a topology based on hexahedral elements. 
The mesh utilized for the RANS simulations is generated with a $Y^+$ close to 
unity to allow the near-wall solution. The discretization in the azimuthal
direction presents one element every $2^o$, meaning that $180$ elements are
equally distributed. The nozzle axis has approximately $50$ elements per nozzle
exit diameter. The mesh size has a total of $1.6 \times 10^6$ elements. The cut
plane of the RANS mesh is illustrated in Fig.~\ref{fig.mesh_rans}.

The boundary condition utilized in the supersonic inlet is a Dirichlet-type 
condition with all properties prescribed. The properties of the nozzle inlet
are adjusted to provide the same properties of the inviscid profile condition
outside of the boundary layer. The outlet condition is a supersonic outflow
with no properties prescribed. The simulation settings apply a CFL ramp
reaching a maximum of $100$, the total number of iterations is $2000$, and one
can observe good convergence with maximum residuals of the order of $10^{-6}$.

\begin{figure}[htb!]
\centering
	\subfloat[2-D Cut plane of the nozzle colored by Mach number contours.]{
	\includegraphics[width=0.5\linewidth]{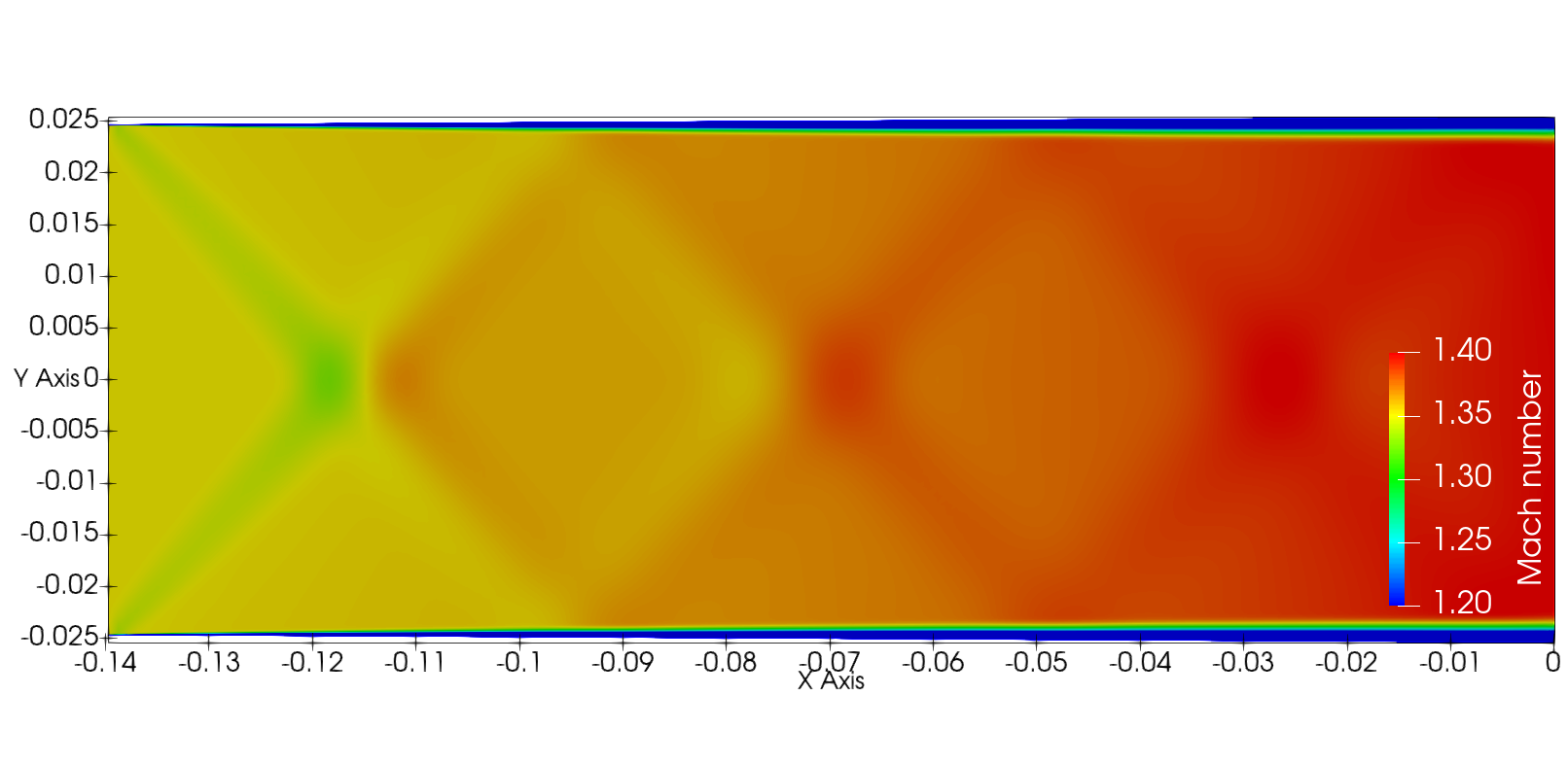}
    \label{fig:noz_contours}
	}
    \subfloat[Longitudinal velocity profile at the output of the nozzle.]{
	\includegraphics[width=0.38\linewidth]{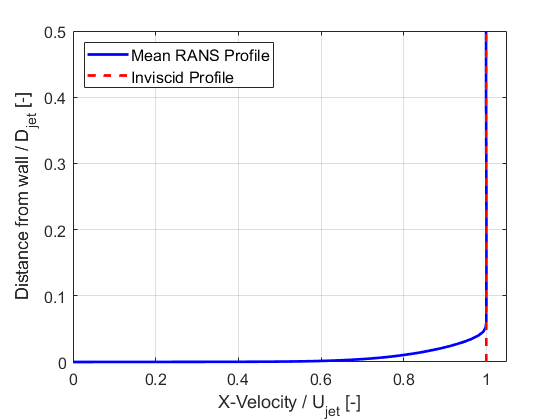}
	\label{fig:noz_velx}
	}
\caption{Contours and profiles of longitudinal velocity obtained from the RANS
         simulation of the nozzle flow.}
\end{figure}

Figure~\ref{fig:noz_contours} presents the contours of the Mach number from the
RANS calculations. The shock waves are weak and do not generate high
distortion in the exit plane. The nozzle design allows the turbulent boundary 
layer thickness growth to values of $\delta_{BL}=0.05D_j$. Flow properties in
the exit plane are extracted as a function of nozzle radius, providing only 
one profile for each variable. The longitudinal velocity profile is presented
in Fig.~\ref{fig:noz_velx}.

\subsubsection*{Tripped Mean RANS Profile}

The mean profile from the RANS simulation, presented in the previous section,
is utilized with the noise-tripping procedure presented in 
Ref.~\cite{Bogeyetal2011}. Such a technique was originally employed in a region
inside the pipe nozzle. In this work, the formulation is employed in the jet
inlet condition. The noise-tripping procedure introduces random disturbance in
the three velocity components of the flow that are not divergence-free. In the
work presented in Ref.~\cite{Bogeyetal2011}, the noise-tripping procedure is
compared to a ring-tripping procedure, which is divergence-free, and the latter
is quieter for the same results. Due to its simplicity, the first method was 
chosen for this work.

The velocity components with the noise-tripping procedure are calculated by
\begin{equation}
\left[ \begin{array}{c} 
             \tilde{u}_x \\ \tilde{u}_r \\ \tilde{u}_{\theta}
       \end{array} \right] = 
       \left[ \begin{array}{c} 
              \tilde{u}_{x,RANS} \\ \tilde{u}_{r,RANS} \\ 
              \tilde{u}_{\theta,RANS}
       \end{array} \right]
       + \alpha u_j
       \left[ \begin{array}{c} 
              3\epsilon_x(x,r,\theta,t) \\ \epsilon_r(x,r,\theta,t) \\ 
              2\epsilon_{\theta}(x,r,\theta,t)
       \end{array} \right],                 
\end{equation}
where the velocity components with the subscript $RANS$ are the mean velocity 
components from the RANS calculations, $\alpha$ is a constant to adjust the 
intensity of the velocity fluctuations, $u_j$ is the jet velocity and 
$\epsilon_x(x,r,\theta,t)$, $\epsilon_r(x,r,\theta,t)$, $\epsilon_{\theta}
(x,r,\theta,t)$ are random numbers between $-1$ and $1$. The noise-tripping
procedure is introduced only inside the boundary layer of the RANS profile.

\subsection{Simulation Settings}

Six numerical studies are presented in the current work, named S-1 to S-6. 
Previous studies \cite{Abreu2021,Abreuetal2022} compared the effects of 
\textit{hp} resolution on the interested jet flow for S-1 to S-4, while two
additional simulations, S-5 and S-6, were performed to evaluate the effects 
of the inlet boundary condition on statistical flow properties. 
Table~\ref{tab.simu} presents the settings used in the six numerical
calculations conducted in this work. Computation S-1 uses the M-1 mesh with 
a second-order accurate spatial discretization, while simulation S-2 employs
the M-2 mesh with a third-order accurate spatial discretization. Calculations 
S-3 and S-4 utilize the M-3 mesh with second- and third-order accurate
discretizations, respectively. The first four computations (S-1 to S-4) use 
an inviscid profile for the inlet condition. Simulation S-5 adopts the M-3 
mesh with third-order accurate discretization and imposes a mean profile from
RANS simulations as the jet inlet condition. Calculation S-6 utilizes the M-3
mesh with third-order accurate discretization and a mean profile from RANS
simulations, incorporating a noise-tripping at the jet inlet condition. The
investigations span from $50$ to $410$ million degrees of freedom (DOFs).
\begin{table}[htb!]
\centering
\caption{Summary of simulations settings.}
\begin{tabular}{ c c c c c c c } \hline\hline
Simulation & Meshes & Order of  & Jet Inlet & Elements  & Total \# 
& Interval data \\ 
 & & Accuracy & Condition & ($10^{6}$) &  DOF ($10^{6}$) & acquisition (FTT)
 \\ \hline
S-1 & M-1 & 2nd order & Inviscid Profile & $6.2$ & $\approx 50$ 
& $\approx 114$ \\
S-2 & M-2 & 3rd order & Inviscid Profile & $1.8$ & $\approx 50$ 
& $\approx 114$ \\
S-3 & M-3 & 2nd order & Inviscid Profile & $15.4$ & $\approx 120$ 
& $\approx 114$ \\ 
S-4 & M-3 & 3rd order & Inviscid Profile & $15.4$ & $\approx 410$ 
& $\approx 114$ \\
S-5 & M-3 & 3rd order & Mean RANS Profile & $15.4$ & $\approx 410$ 
& $\approx 31$ \\
\multirow{ 2}{*}{S-6} & \multirow{ 2}{*}{M-3} & \multirow{ 2}{*}{3rd order} 
& Mean RANS Profile  & \multirow{ 2}{*}{$15.4$} 
& \multirow{ 2}{*}{$\approx 410$} & \multirow{ 2}{*}{$\approx 62$} \\
 & & & + noise-tripping & & & \\
\hline\hline
\end{tabular}
\label{tab.simu}
\end{table}

\subsection{Calculation of Statistical Properties}

\label{chap.stats}

The data acquisition procedure involves two steps. The first is to clean
off the domain since the computation starts with a steady flow initial
condition and develops a statistically steady flow. The non-dimensional 
reference for the simulation time is the flow-through time (FTT) based on the
jet inlet velocity and the nozzle exit diameter. It is necessary to about $240$
FFT to reach the statistically steady state when starting with a steady flow or
about $100$ FTT when starting with a previously developed simulation. The start 
from a steady flow is employed for the S-1 to S-3 simulations, and the restart
is employed for the S-4 to S-6 simulations.

An additional $110$ FTT is simulated to acquire data from each simulation.
The S-1 to S-4 simulations have already completed the acquisition data
interval, while the S-5 has $31$ FTT, and the S-6 simulation has $62$ FTT. The
acquisition frequency interval is also presented in Tab.~\ref{tab.simu}. The
acquisition frequency is $200$ kHz. The maximum time sample of $110$ FTT and
the acquisition frequency of $200$ kHz are associated with the maximum
Strouhal number of $St_{max}=10.5$ and a minimum Strouhal number of
$St_{min}=0.01$.

\begin{figure}[htb!]
\centering
\includegraphics[width=0.8\linewidth]{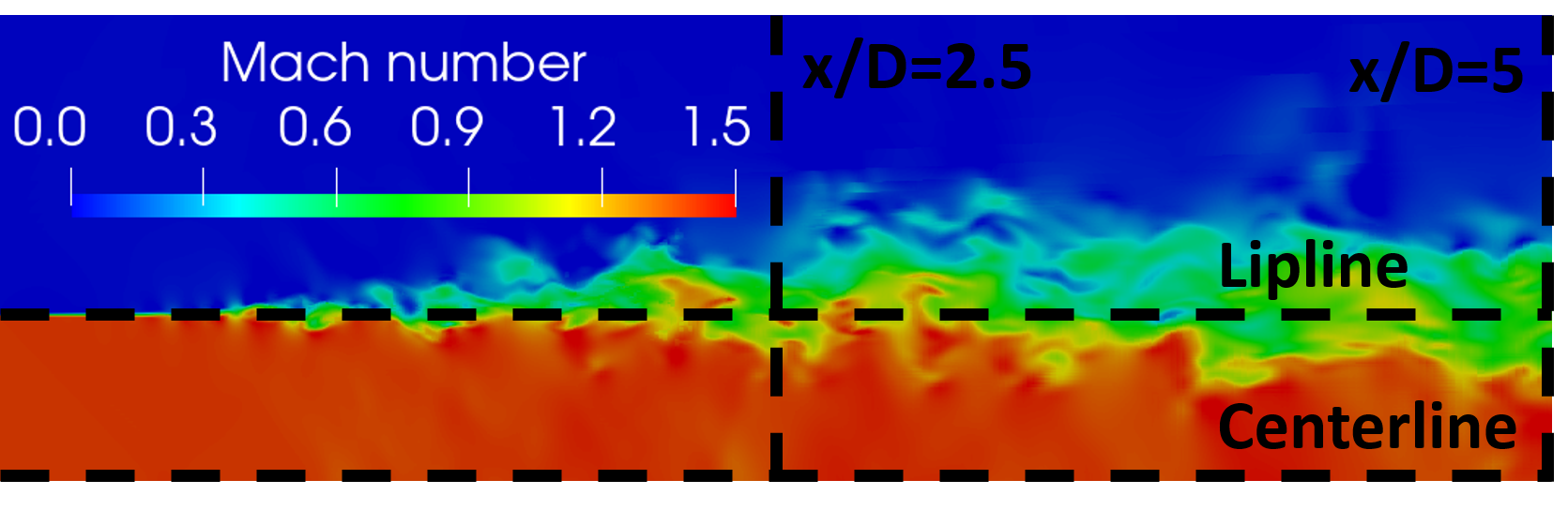}
\caption{Snapshot of the jet simulation with the two longitudinal lines and
         three crossflow lines along which data is extracted. Mach number
         contours are shown.}
\label{fig.jet_data_extract}
\end{figure}

Mean flow properties and root mean square (RMS) fluctuations are calculated
along the centerline, lipline, and different domain surfaces in the streamwise
direction. The centerline is defined as the line in the center of the geometry
$y/D=0$, whereas the lipline is a surface parallel to the centerline and
located at the nozzle diameter, $y/D=0.5$. The data from the lipline are an
azimuthal mean from six equally spaced positions. The four surfaces in the
streamwise positions are $x/D=2.5$, $x/D=5.0$, $x/D=10.0$, and $x/D=15.0$.
Surface properties are averaged using six equally spaced positions in the
azimuthal direction. Figure~\ref{fig.jet_data_extract} illustrates a Mach
number contours snapshot of the jet flow with the lines and surfaces of data
extraction. 

\section{Numerical Results}

This section presents the numerical results from the six simulations 
distributed in two analyses: the resolution study and the inflow condition
study. The S-1 to S-4 simulations are investigated in the resolution study
analysis. In the first analysis, a qualitative investigation of the contours of
the mean longitudinal velocity, RMS longitudinal velocity fluctuation, and mean
pressure is performed. A quantitative analysis of the distribution of mean
longitudinal velocity, RMS of longitudinal and radial velocity fluctuations,
and shear-stress tensor component is also performed, and the numerical results
are compared to experimental data. In the second study, in which the influence
of the different inflow conditions is investigated, the qualitative analysis
from the velocity distributions of the S-4 to S-6 simulations is performed and
compared to the experimental data. 

\subsection{Resolution Study}

\begin{figure}[htb!]
\subfloat[S-1 simulation.]{	
	\includegraphics[width=0.48\linewidth]{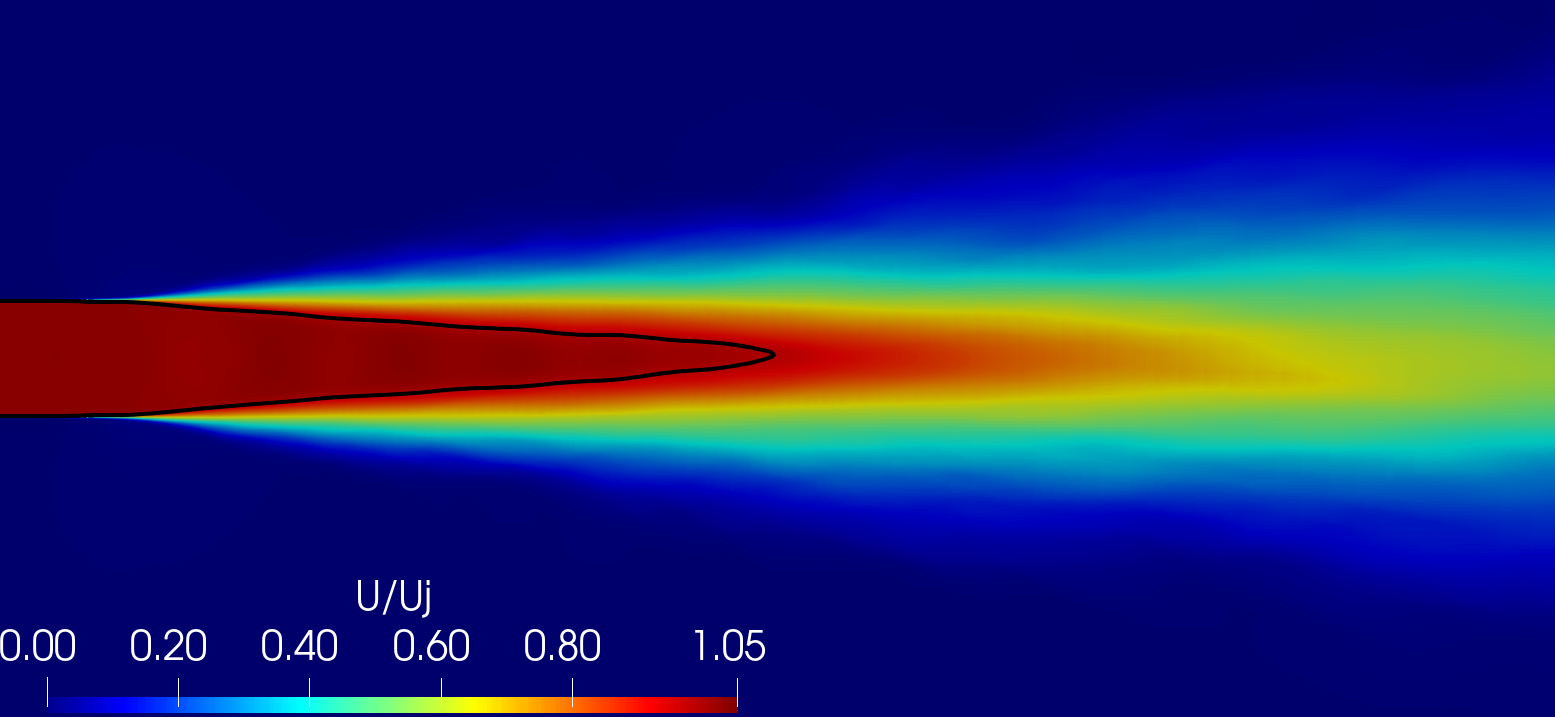}
	\label{fig.res_mvelxa}
	}
\subfloat[S-2 simulation.]{
	\includegraphics[width=0.48\linewidth]{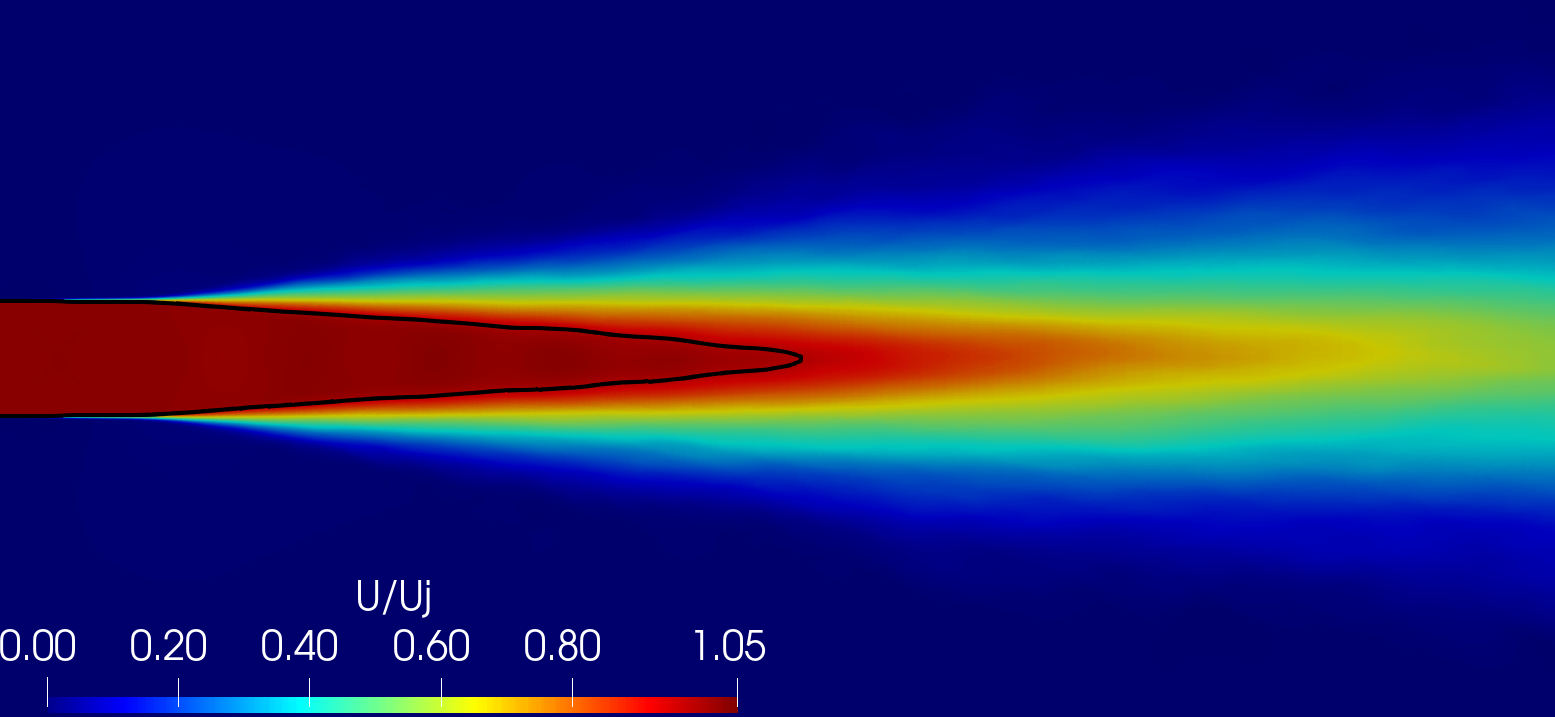}
	\label{fig.res_mvelxb}
	}
\\
\subfloat[S-3 simulation.]{	
	\includegraphics[width=0.48\linewidth]{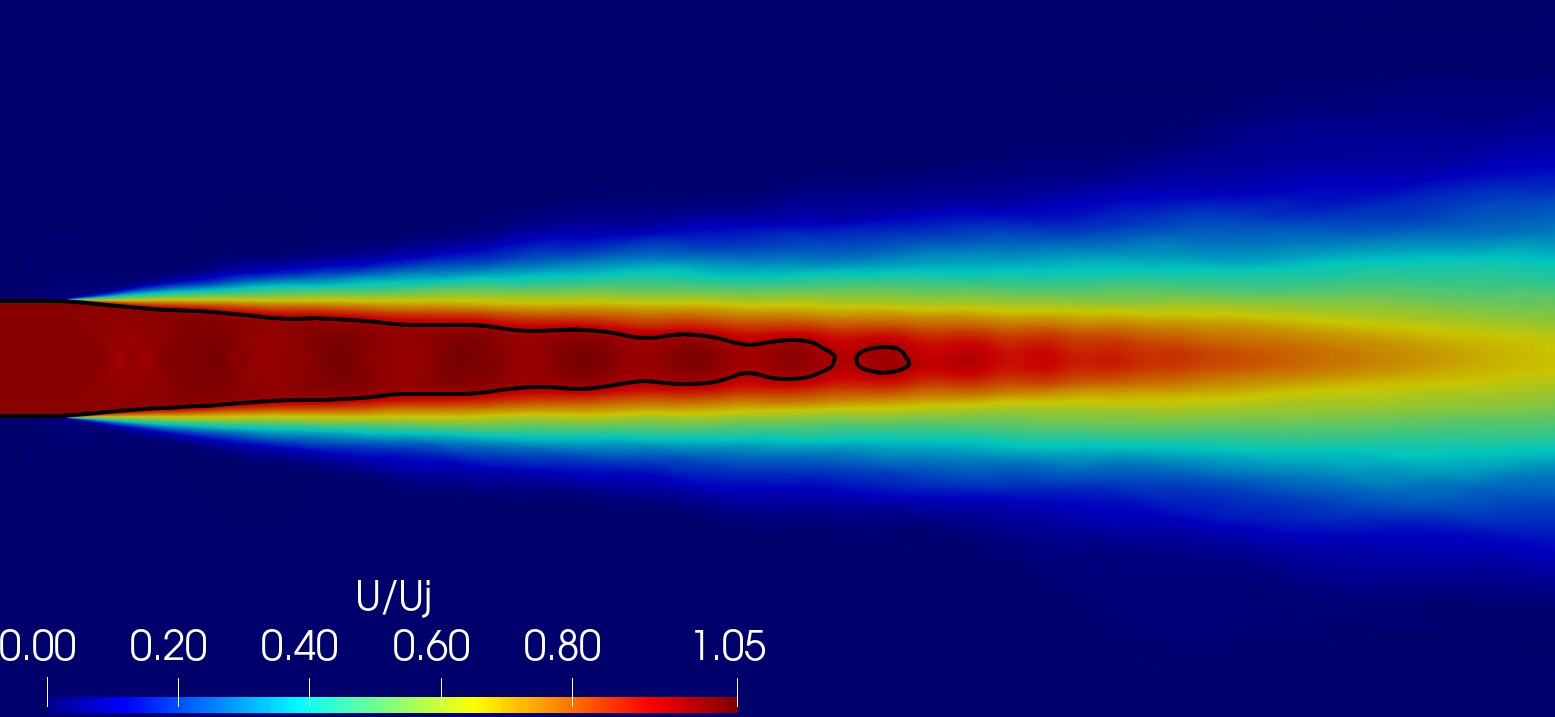}
	\label{fig.res_mvelxc}
	}
\subfloat[S-4 simulation.]{
	\includegraphics[width=0.48\linewidth]{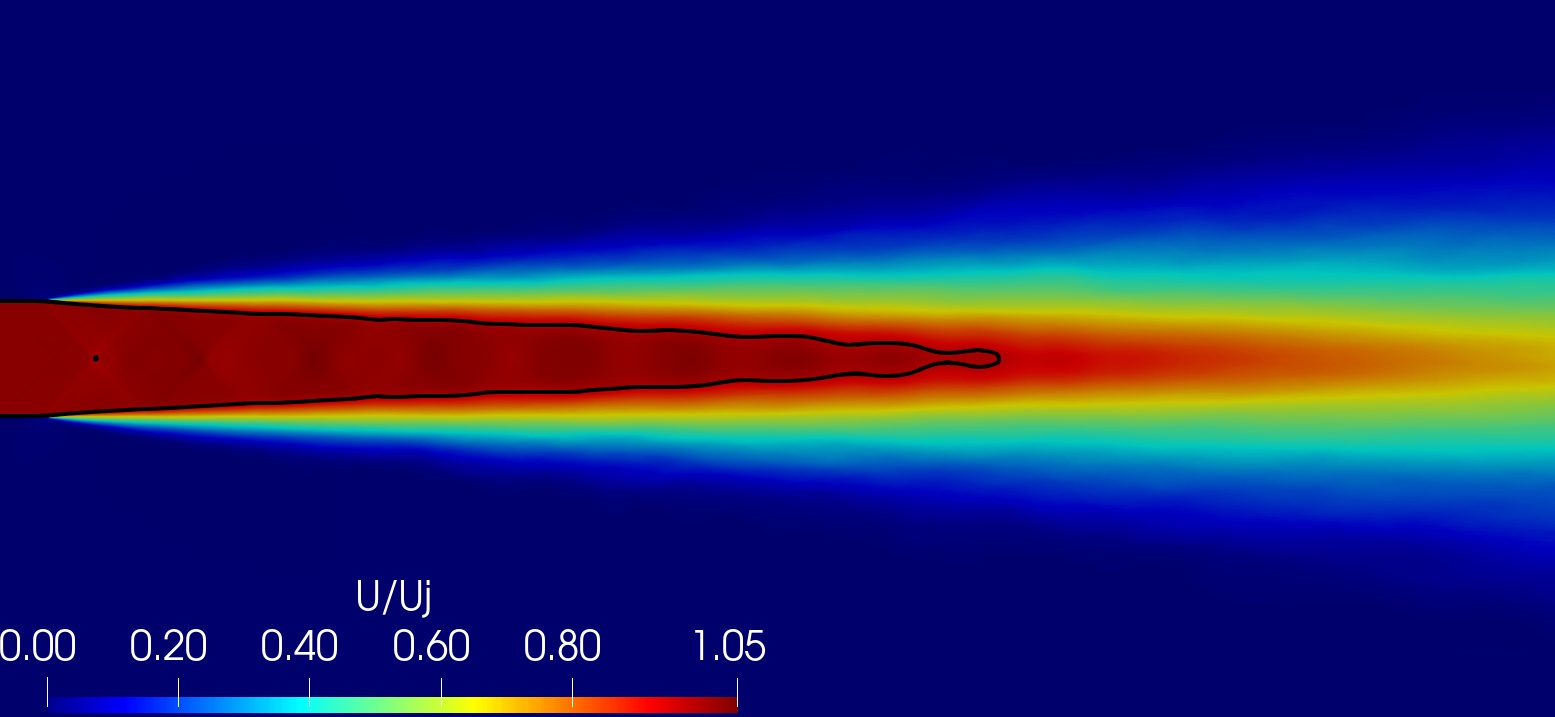}
	\label{fig.res_mvelxd}
	}
\caption{Contours of mean longitudinal velocity component on a cut plane in  
         $z/D=0$. The black line highlights the boundaries of the jet potential
         core.}
\label{fig.res_mvelx}
\end{figure}

Figure~\ref{fig.res_mvelx} presents the mean longitudinal velocity component
contours from the S-1 to S-4 simulations with a highlight of the
boundaries of the jet potential core. The potential core is the region where
the jet velocity is 95\% of the mean inlet velocity, $U=0.95U_j$, indicated 
by a back line in Fig.~\ref{fig.res_mvelx}. The potential core length is the
distance in the centerline from the inlet section to the boundary of the jet
potential core. In Figs.~\ref{fig.res_mvelxa} and \ref{fig.res_mvelxb}, the
mean velocity contours are similar, and the shape of the potential cores has
the same length. The velocity contours in Fig.~\ref{fig.res_mvelxc}, from the
S-3 simulation presents a longer jet potential core when compared to the 
velocity contours from the S-1 and S-2 simulations. The mean velocity contours
from the S-4 simulation are presented in Fig.~\ref{fig.res_mvelxd}. The larger
high-velocity region can be observed when compared to the other three
simulations with the longest potential core. It can also be observed from the
mean velocity contours a larger streamwise velocity spreading from the S-1 
and S-2 simulations than from the S-3 and S-4 simulations.

\begin{figure}[htb!]
\subfloat[S-1 simulation.]{	
	\includegraphics[width=0.48\linewidth]{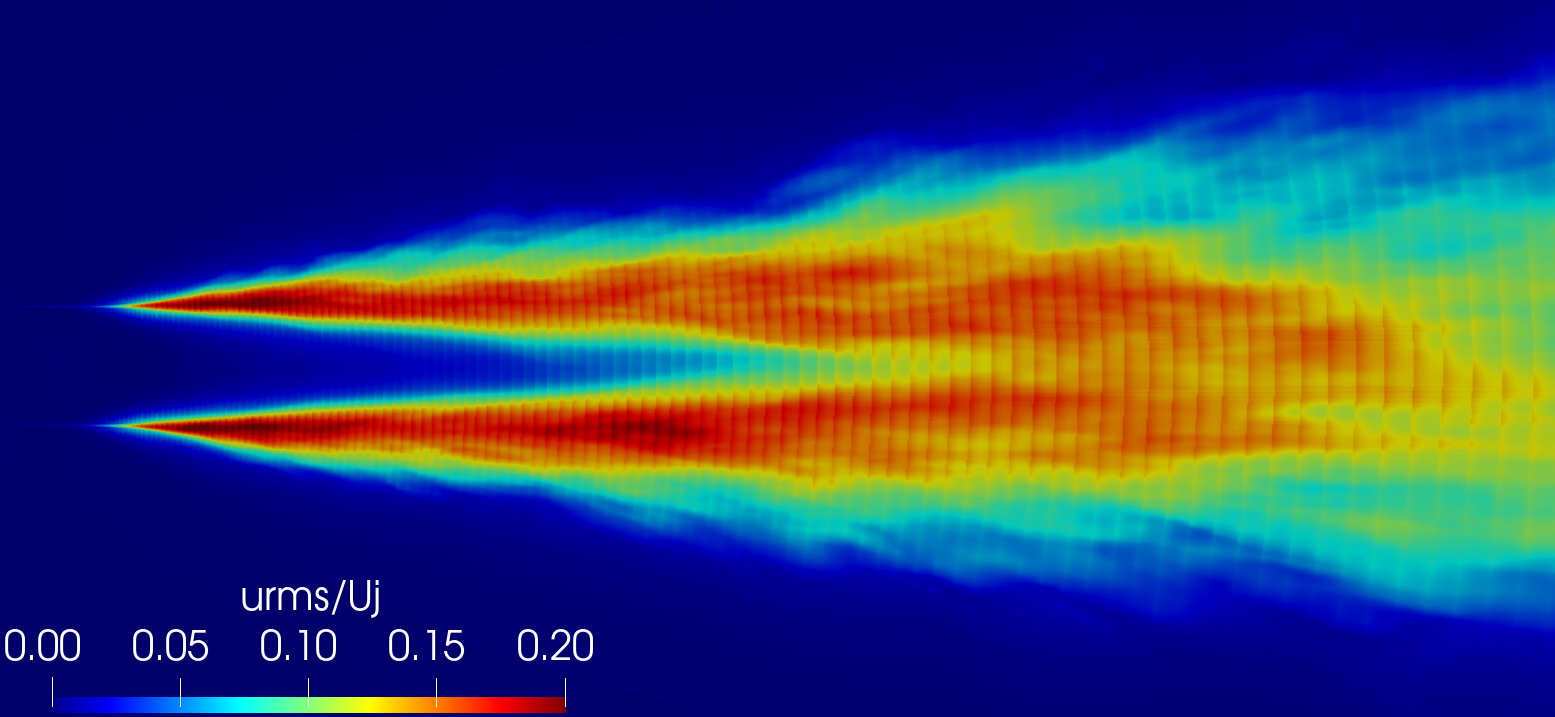}
	\label{fig.res_rvelxa}
	}
\subfloat[S-2 simulation.]{
	\includegraphics[width=0.48\linewidth]{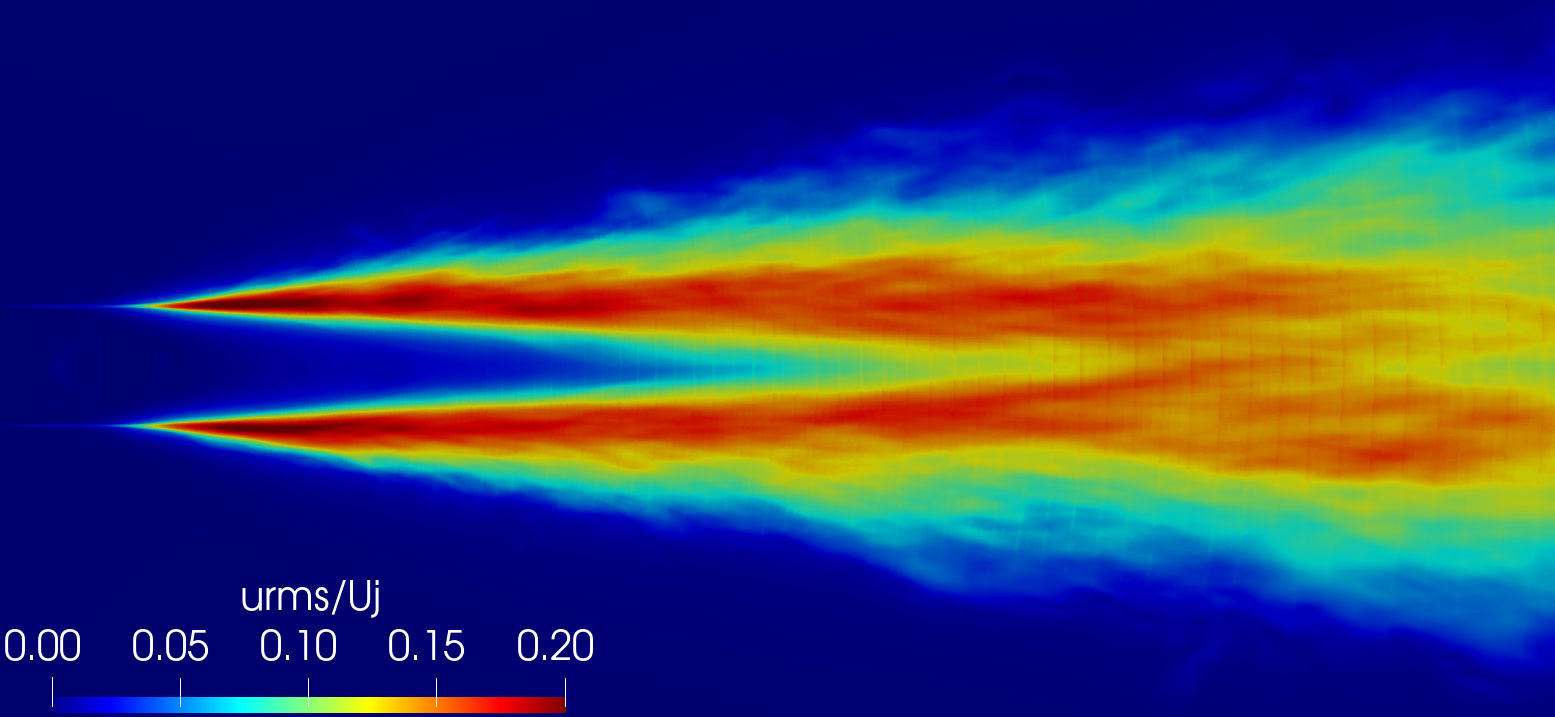}
	\label{fig.res_rvelxb}
	}
\\
\subfloat[S-3 simulation.]{	
	\includegraphics[width=0.48\linewidth]{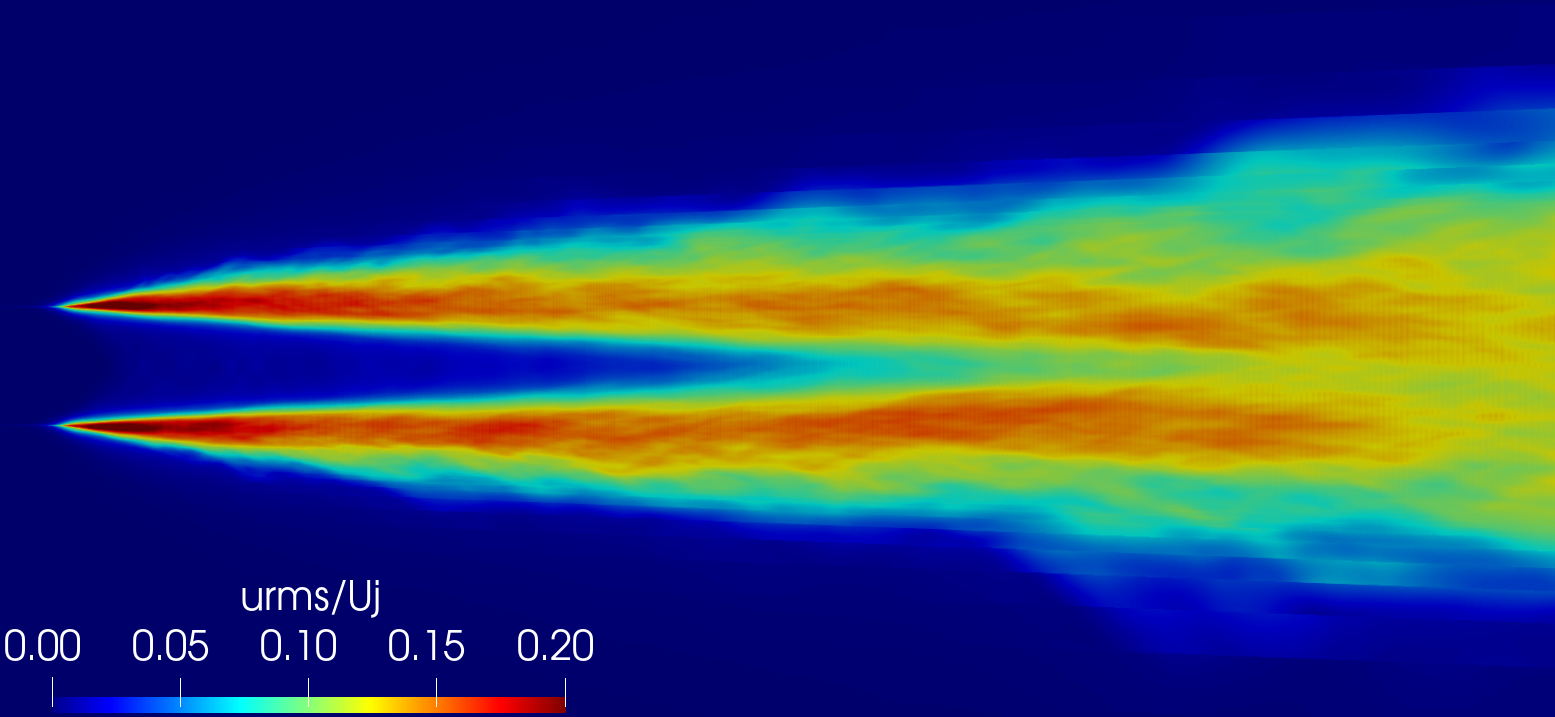}
	\label{fig.res_rvelxc}
	}
\subfloat[S-4 simulation.]{
	\includegraphics[width=0.48\linewidth]{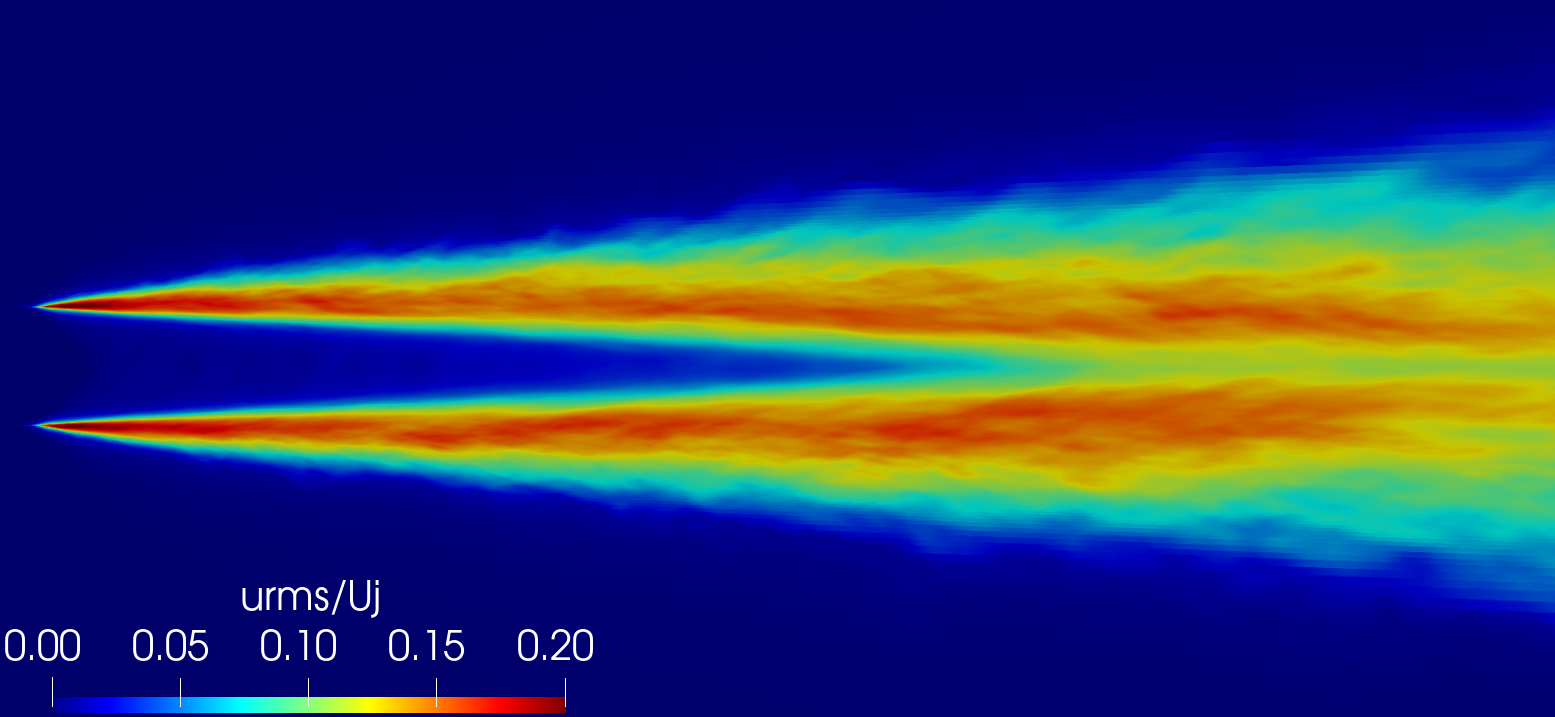}
	\label{fig.res_rvelxd}
	}
\caption{Contours of RMS of longitudinal velocity fluctuation on a cut plane
         in $z/D=0$.}
\label{fig.res_rvelx}
\end{figure}

The contours of RMS of longitudinal velocity fluctuation are presented in
Fig.~\ref{fig.res_rvelx}. One can observe a large region with high values of 
velocity fluctuation for S-1 and S-2 calculations, Figs.~\ref{fig.res_rvelxa}
and \ref{fig.res_rvelxb}. The RMS of the longitudinal velocity fluctuation
contours from the S-3 and S-4 simulations, Figs.~\ref{fig.res_rvelxc} and
\ref{fig.res_rvelxd}, have important differences to the contours from the 
S-1 and S-2 simulations. The region where the velocity fluctuation values 
in the lipline present an increase in the values occurs closer to the
inlet section in the S-3 and S-4 simulations than in the S-1 and S-2 
simulations. The high-velocity fluctuation region around the lipline is thicker
in the S-1 and S-2 simulations than in the S-3 and S-4 simulations. The thicker 
high-velocity fluctuation region affects the center of the jet flow, with a
shorter region with lower levels of velocity fluctuation from the S-1 and S-2
simulations than the indicated from the velocity contours from S-3 and S-4
simulations.

\begin{figure}[htb!]
\subfloat[S-1 simulation.]{	
	\includegraphics[width=0.48\linewidth]{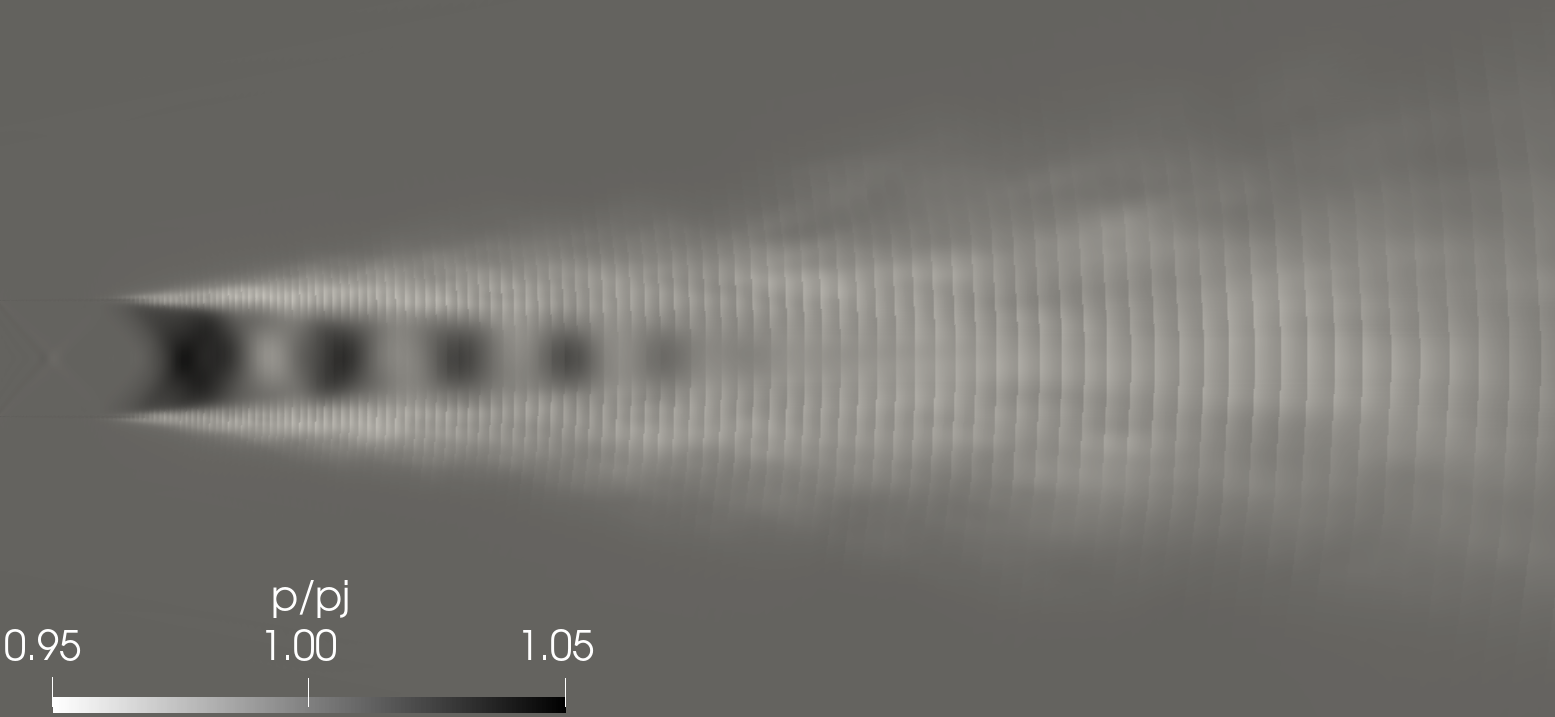}
	\label{fig.res_presa}
	}
\subfloat[S-2 simulation.]{
	\includegraphics[width=0.48\linewidth]{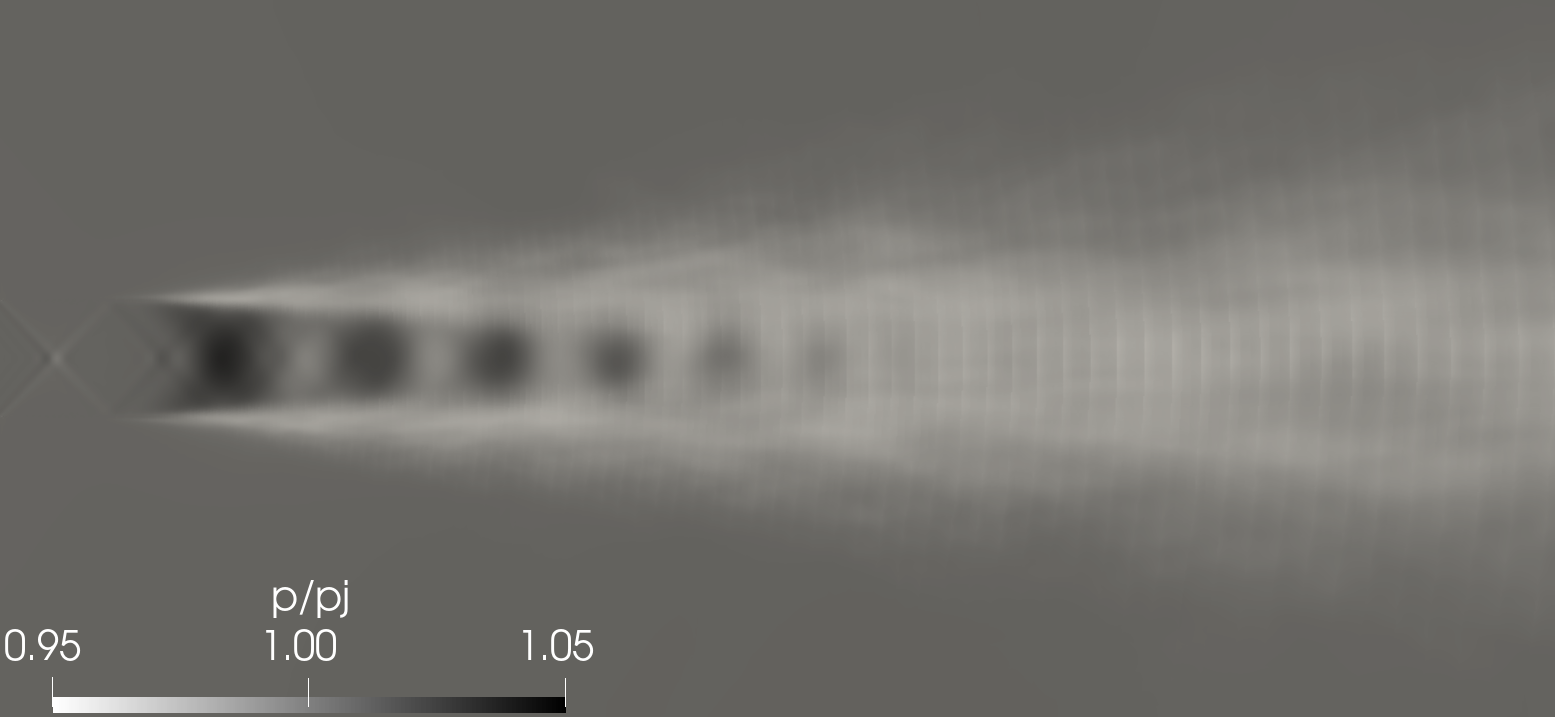}
	\label{fig.res_presb}
	}
\\
\subfloat[S-3 simulation.]{	
	\includegraphics[width=0.48\linewidth]{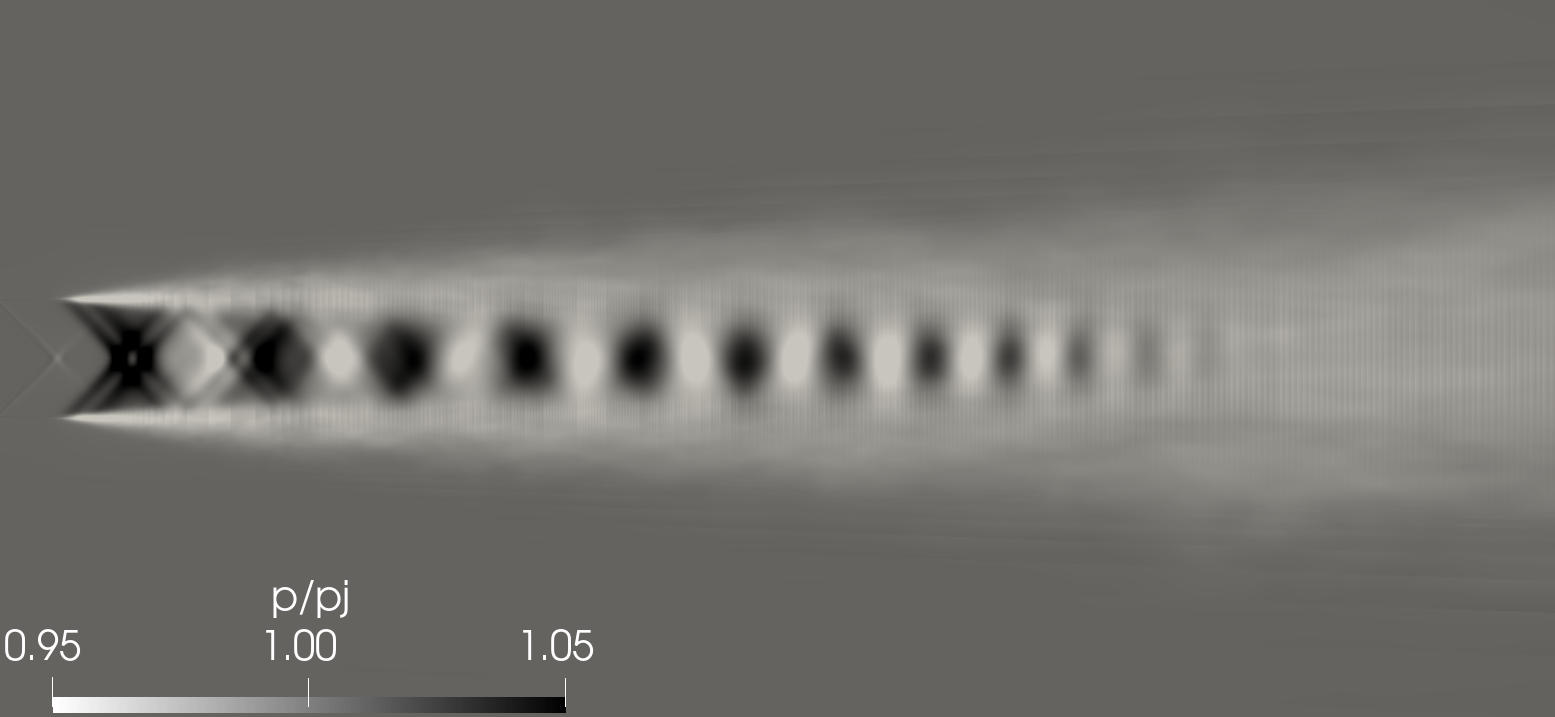}
	\label{fig.res_presc}
	}
\subfloat[S-4 simulation.]{
	\includegraphics[width=0.48\linewidth]{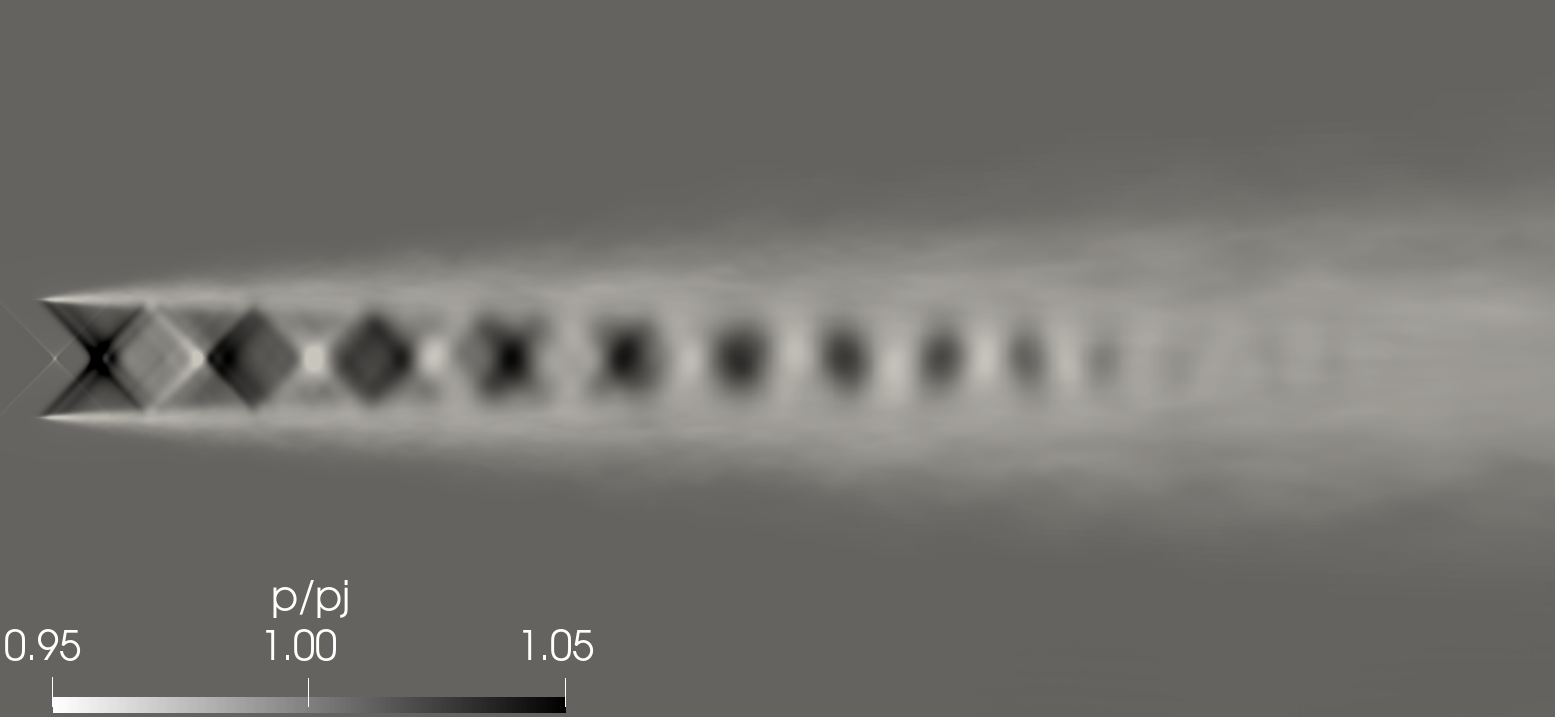}
	\label{fig.res_presd}
	}
\caption{Contours of mean pressure on cutplane in $z/D=0$.}
\label{fig.res_pres}
\end{figure}

The mean velocity and RMS of velocity fluctuation contours provided a 
qualitative analysis of the mean characteristics of the jet flow, such as the
jet potential core and the streamwise velocity spreading. The RMS of the
longitudinal velocity fluctuation contours allowed the inspection of the region
where the velocity fluctuation increases in the jet lipline, with the thickness
of the large velocity fluctuation region that affected the flow in the jet
centerline. The other important aspect of supersonic flows that could be
investigated is the shock waves. The visualization of the shock waves is 
performed with mean pressure contours, Fig.~\ref{fig.res_pres}. The mean 
pressure contours from the S-1 and S-2 simulations, Figs.\ \ref{fig.res_presa} 
and \ref{fig.res_presb}, present a small number of shock waves in the flow
field. The pressure contours from the S-2 simulation present a higher 
resolution for the visualization of the first shock waves when compared to the
pressure contours from the S-1 simulation. The mean pressure contours from the 
S-3 and S-4 simulations, Figs.\ \ref{fig.res_presc} and \ref{fig.res_presd},
present an important increment of shock wave repetitions compared to the S-1
and S-2 simulations. The comparison between the pressure contours from the S-3
and S-4 simulations indicate that the pressure waves from the S-4 simulation 
are better defined than those from the S-3 simulation. However, the S-3
simulation presents stronger shock waves, represented by larger regions with
different pressure levels.

\begin{figure}[htb!]
\centering
\subfloat[Centerline]{
	\includegraphics[width=0.45\linewidth]{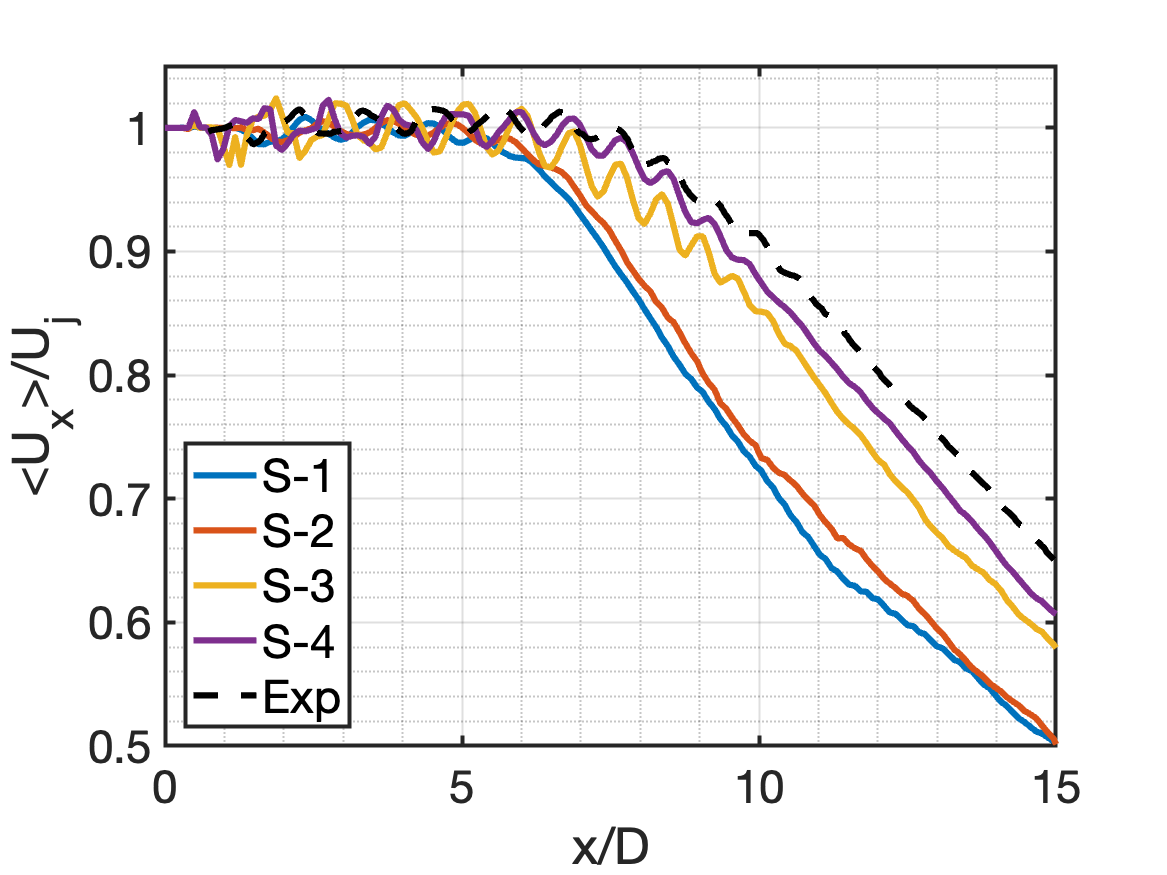}
	\label{fig.res1}	
	}%
\subfloat[Centerline]{
	\includegraphics[width=0.45\linewidth]{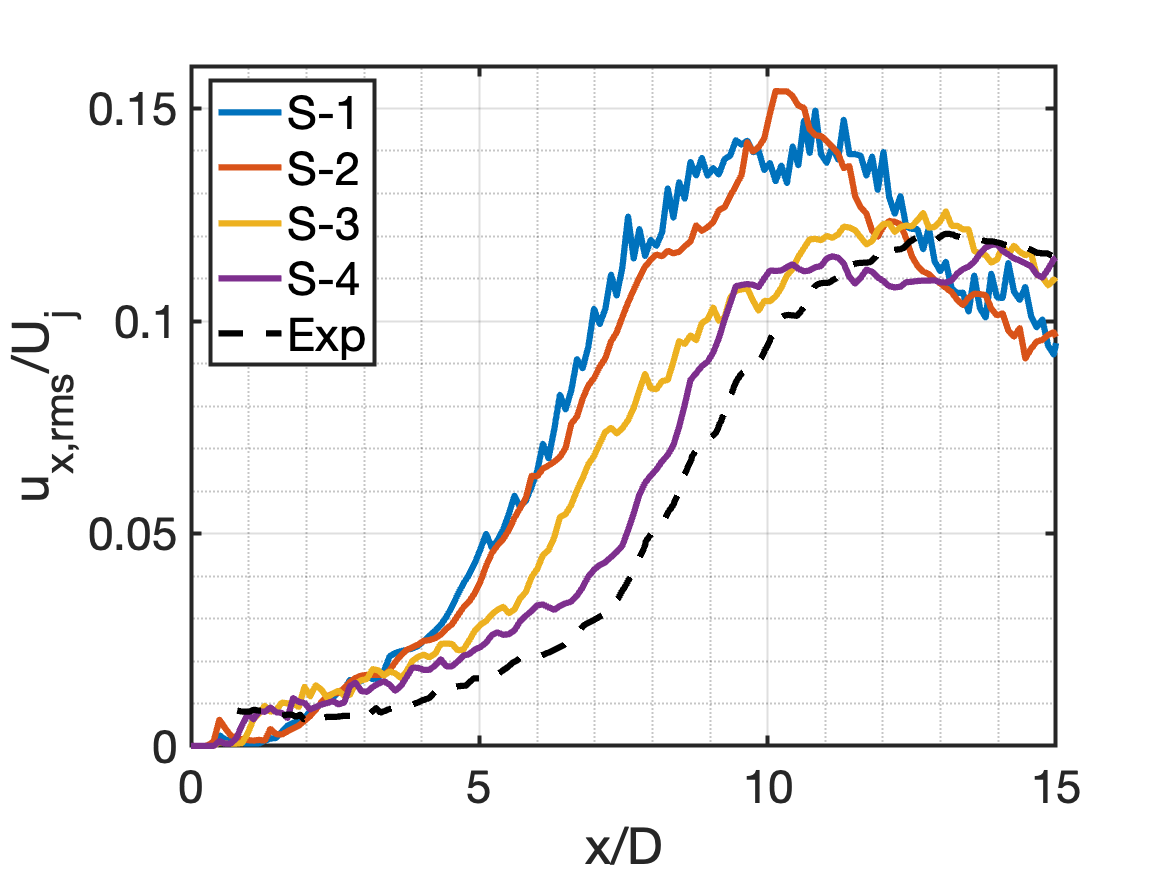}
	\label{fig.res2}	
	}
\newline
\subfloat[Lipline]{
	\includegraphics[width=0.45\linewidth]{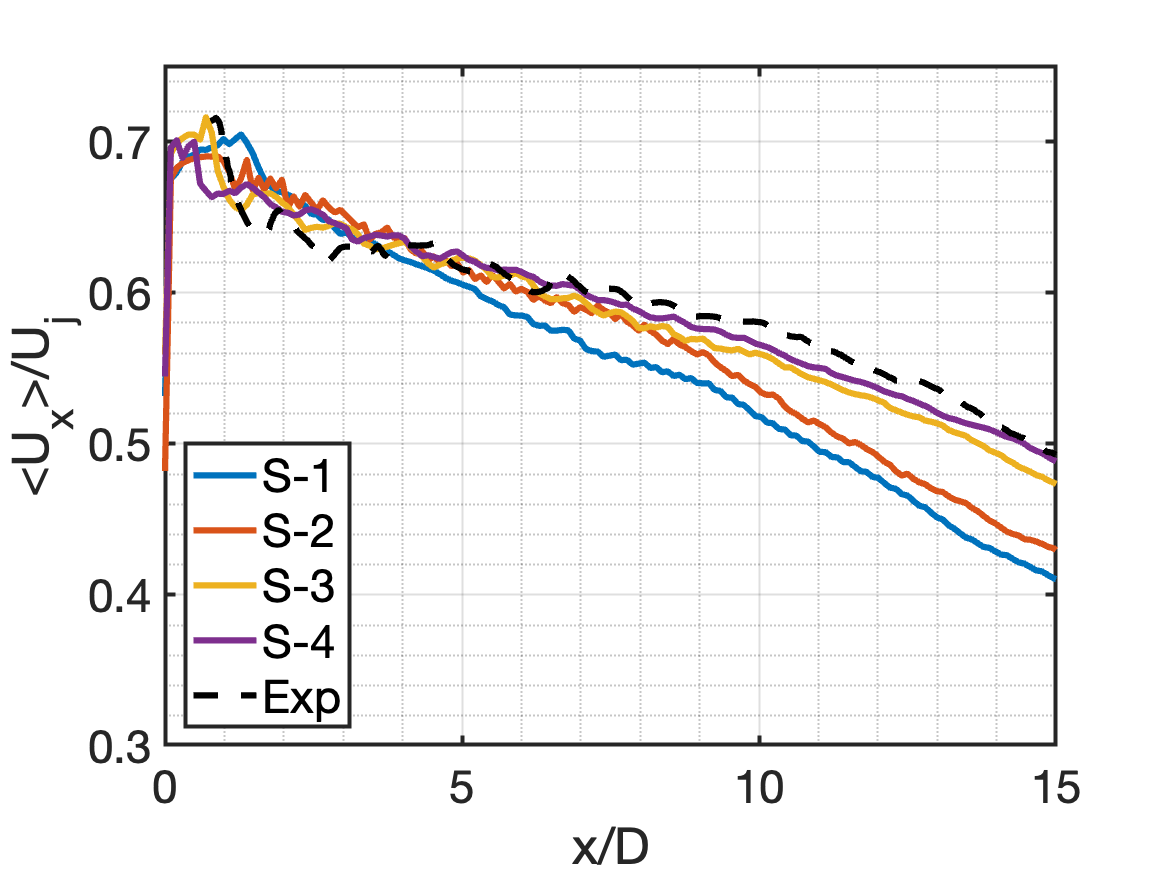}
	\label{fig.res3}
	}%
\subfloat[Lipline]{
	\includegraphics[width=0.45\linewidth]{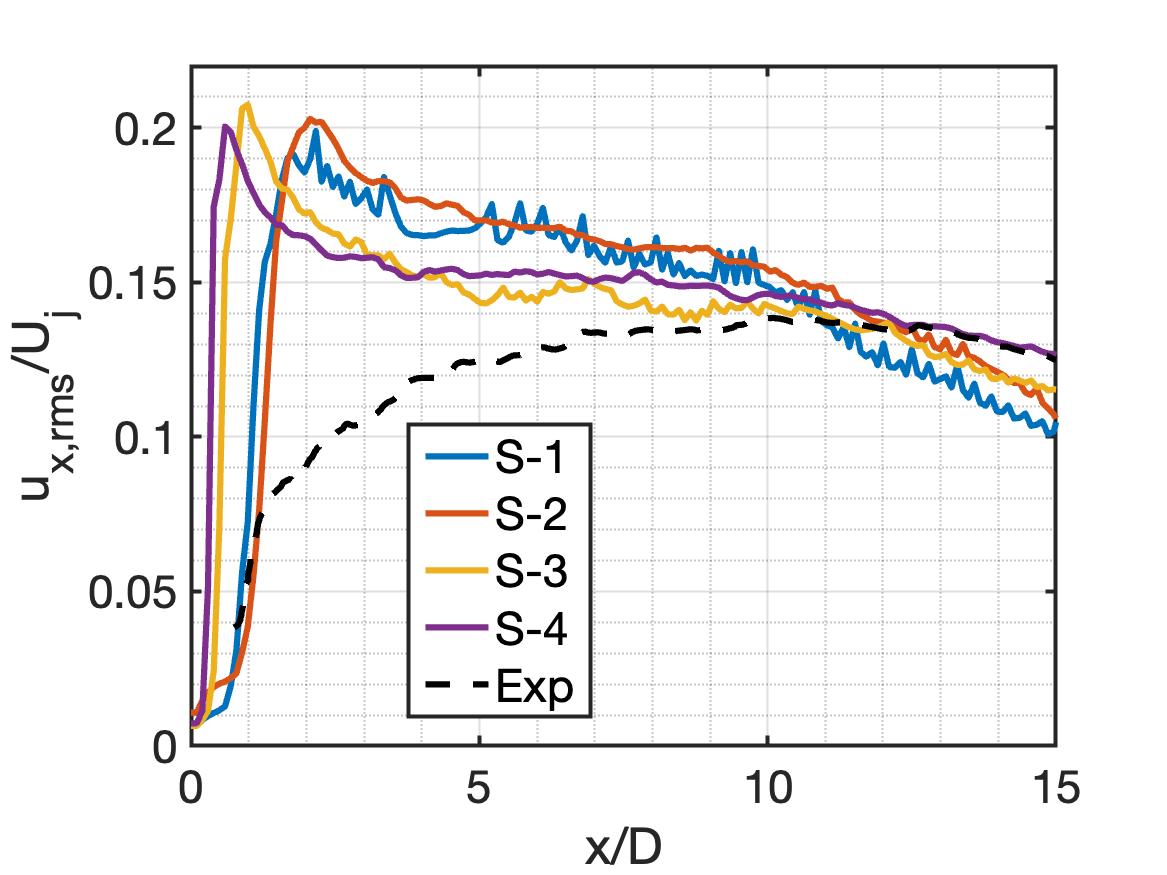}
	\label{fig.res4}	
	}
\caption{Mean longitudinal velocity component distribution (left) and RMS
         of longitudinal velocity fluctuation (right) in the jet centerline
         $y/D=0$ (top) and lipline $y/D=0.5$ (bottom).}
\label{fig.res1_4}
\end{figure}

The mean longitudinal velocity component and RMS of the longitudinal velocity
fluctuation distributions are presented at the centerline and lipline of the
jet in Fig.~\ref{fig.res1_4}. The computed data are compared to the
experimental reference \cite{BridgesWernet2008}. In Fig.~\ref{fig.res1}, 
the distribution of mean longitudinal velocity $<U>/U_j$ is presented along 
the centerline of the jet. In the same way, observed from the mean velocity
contours, Fig.~\ref{fig.res_mvelx}, the mean velocity distributions indicate
a distinction between the numerical results from the S-1 and S-2 simulations
and the S-3 and S-4 simulations. The mean velocity distribution from the 
S-1 and S-2 simulations present a shorter core jet, represented by the velocity 
slope change occurring closer to the inlet section than the data from the S-3
and S-4 simulations. Another difference is that the velocity distribution from
the S-3 and S-4 simulations present a saw-edged behavior due to the shock waves
in the initial development of the jet flows. Such behavior is hardly visualized
on the velocity distribution from the S-1 and S-2 simulations. The RMS of the
longitudinal velocity fluctuation distribution at the jet centerline,
Fig.~\ref{fig.res2} presents some differences between the numerical
simulations. The velocity fluctuation distribution from the S-1 and S-2
simulations present a change in the velocity slope occurring closer to 
the jet inlet section than the S-3 and S-4 simulations, and they also reach
peak velocity fluctuations above the levels of the S-3 and S-4 simulations.

The mean longitudinal velocity component distribution at the jet lipline,
Fig.~\ref{fig.res3} presents a similar shape between the four numerical 
simulations. The difference observed is the larger negative velocity slope
presented by the S-1 and S-2 simulations compared to the S-3 and S-4 
simulations, which present a velocity slope in agreement with the experimental
reference. The RMS of the longitudinal velocity fluctuations distribution at
the jet lipline, Fig.~\ref{fig.res4} is the only data set evaluated where a
significant difference to the experimental reference is observed. Close to 
the jet inlet section, the velocity fluctuation distribution presents a 
high-velocity slope starting at almost zero to the peak values of the velocity
distributions. The experimental reference presents a high-velocity slope that
is reduced after $x/D=1.0$ and only reaches its maximum value close to 
$x/D=10.0$. The increase in the resolution of the numerical simulations leads
to the high-velocity slope occurring closer to the inlet section than the 
simulations with smaller resolution. In the numerical data presented in 
Figs.~\ref{fig.res1}, \ref{fig.res2}, and \ref{fig.res3}, a monotone improvement
towards the experimental data is observed with the increase in the resolution
of the numerical simulations. The opposite behavior is observed in the 
numerical data from Fig.~\ref{fig.res4}, which can be associated with the top
hat profile imposed at the inlet boundary condition.

\begin{figure}[htb!]
\centering
\subfloat[$x/D=2.5$]{
	\includegraphics[trim = 30mm 0mm 20mm 0mm, clip, height=0.26\linewidth]{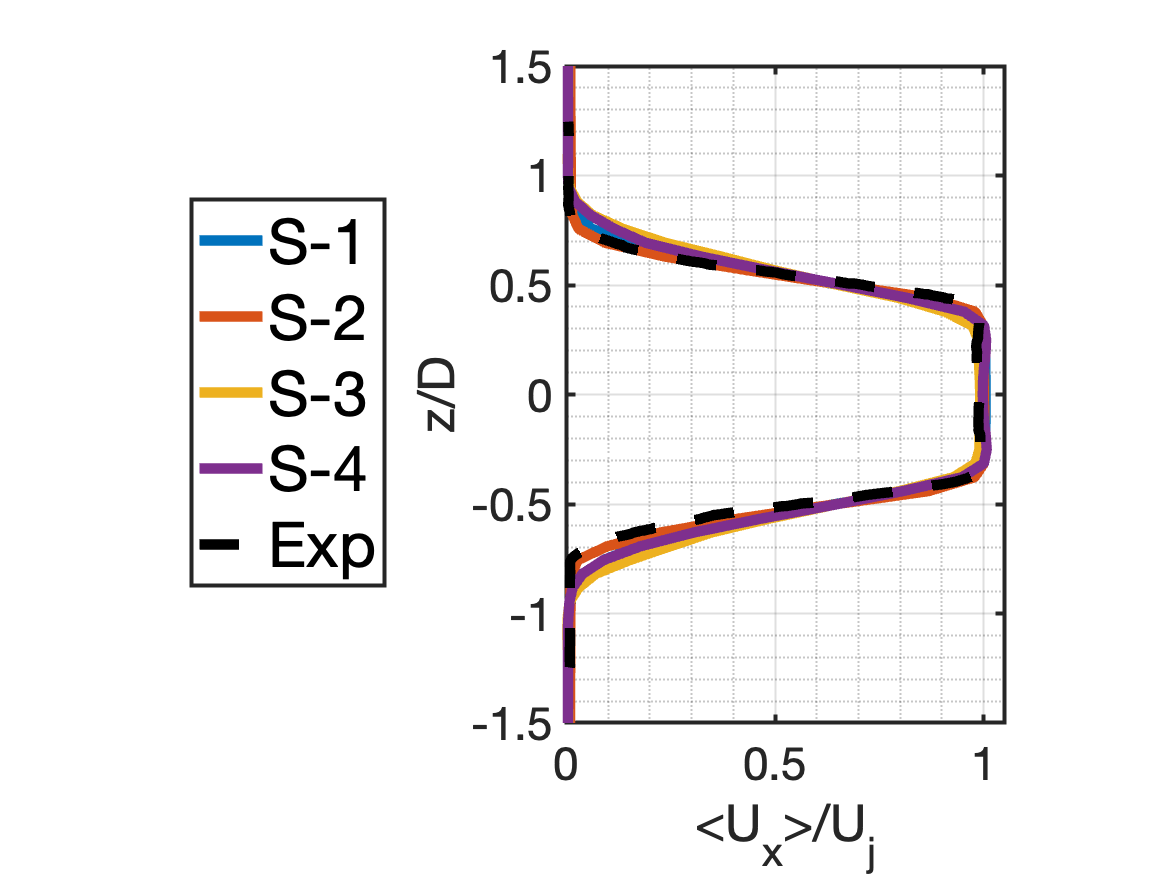}
	\label{fig.res5a}	
	}
\subfloat[$x/D=5$]{
	\includegraphics[trim = 30mm 0mm 55mm 0mm, clip, height=0.26\linewidth]{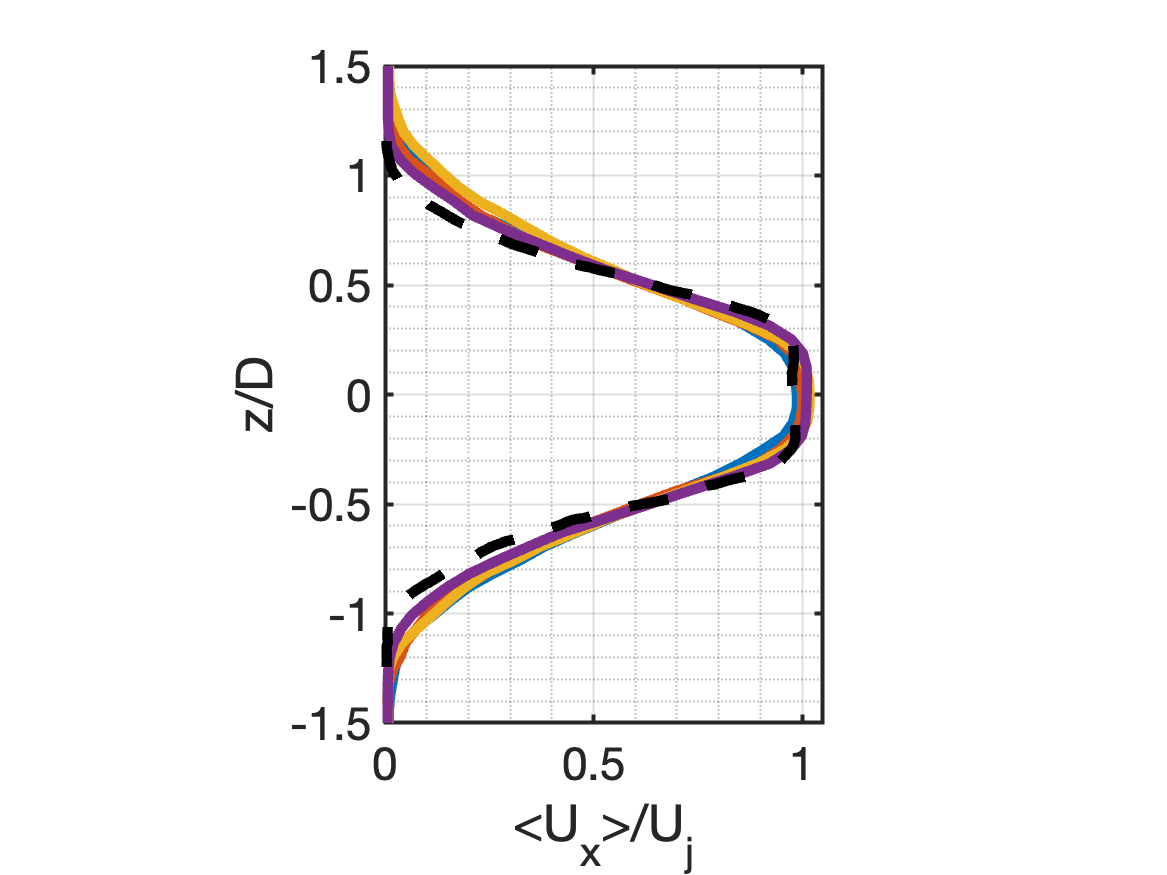}
	\label{fig.res5b}	
	}
\subfloat[$x/D=10$]{
	\includegraphics[trim = 30mm 0mm 55mm 0mm, clip, height=0.26\linewidth]{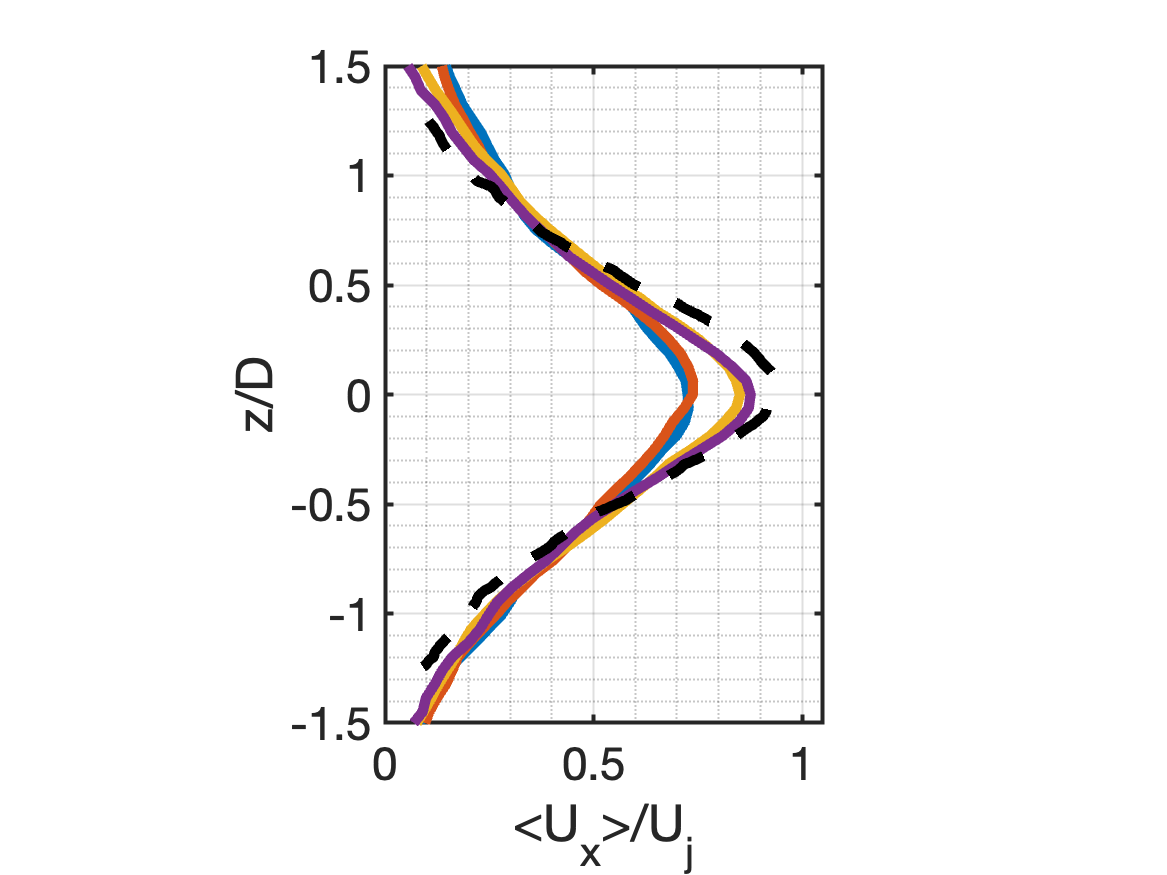}
	\label{fig.res5c}
	}
\subfloat[$x/D=15$]{
	\includegraphics[trim = 30mm 0mm 55mm 0mm, clip, height=0.26\linewidth]{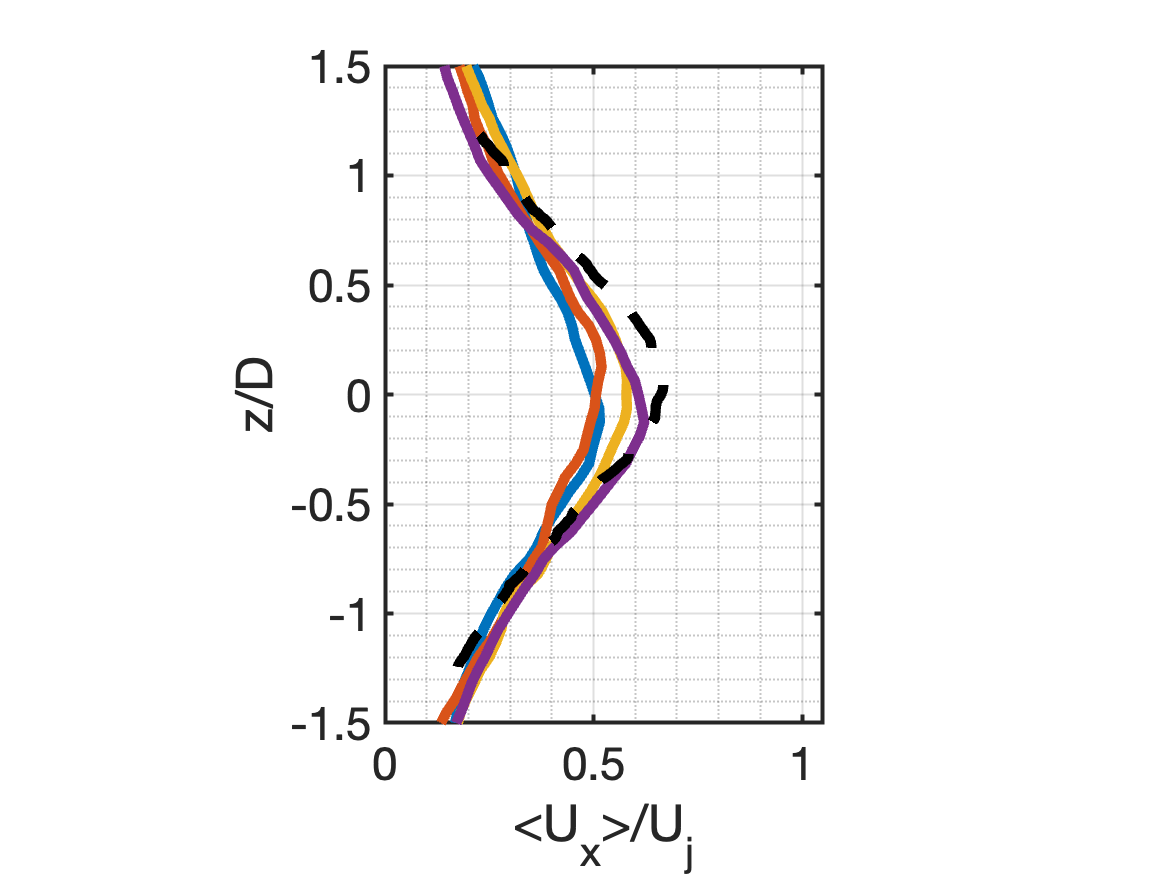}
	\label{fig.res5d}	
	}
\newline
\subfloat[$x/D=2.5$]{
	\includegraphics[trim = 30mm 0mm 20mm 0mm, clip, height=0.26\linewidth]{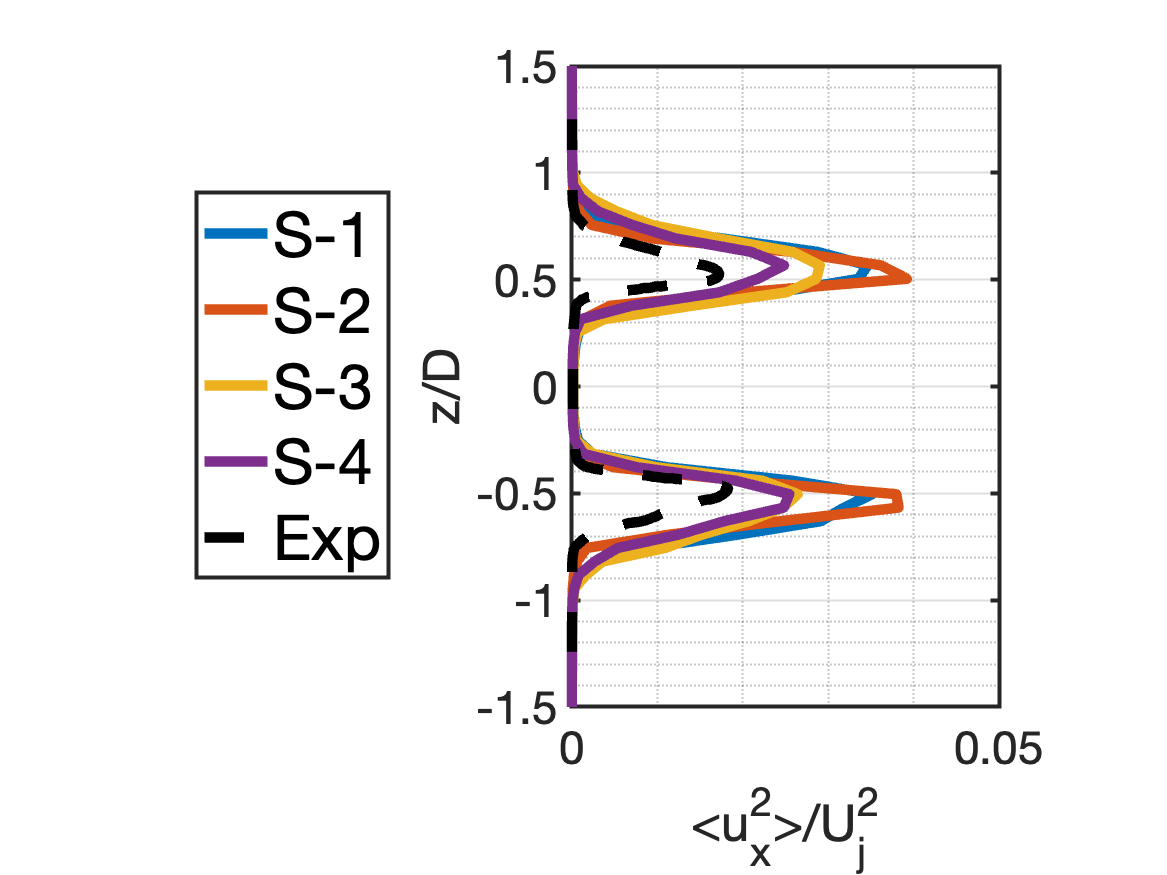}
	\label{fig.res5e}	
	}
\subfloat[$x/D=5$]{
	\includegraphics[trim = 30mm 0mm 55mm 0mm, clip, height=0.26\linewidth]{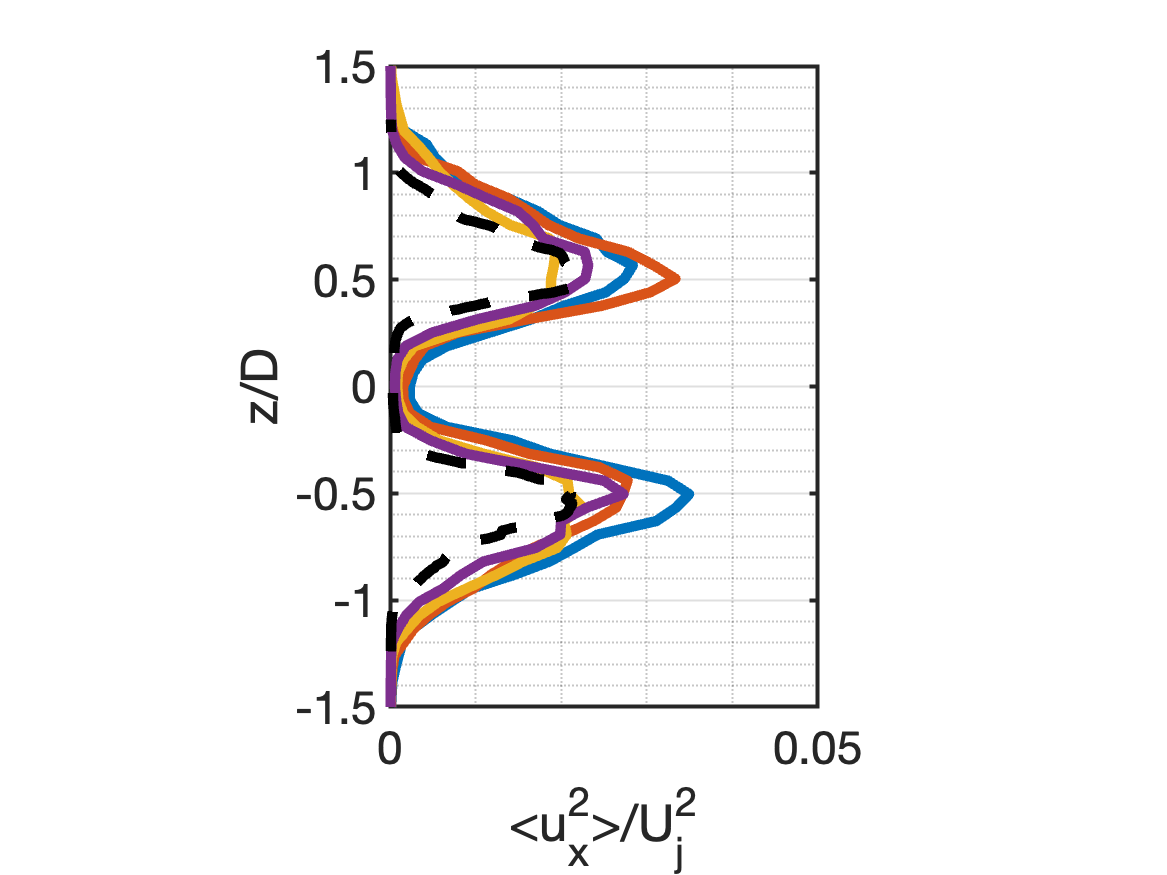}
	\label{fig.res5f}	
	}
\subfloat[$x/D=10$]{
	\includegraphics[trim = 30mm 0mm 55mm 0mm, clip, height=0.26\linewidth]{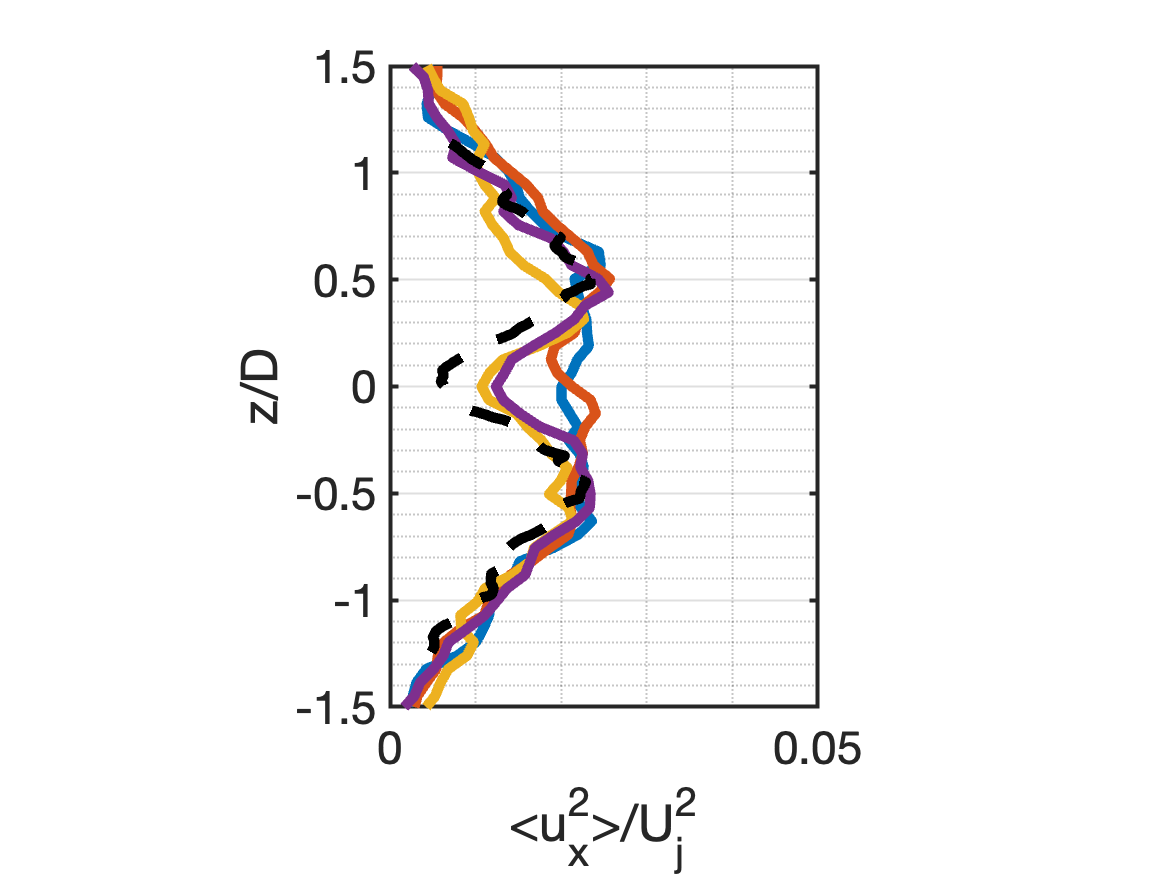}
	\label{fig.res5g}
	}
\subfloat[$x/D=15$]{
	\includegraphics[trim = 30mm 0mm 55mm 0mm, clip, height=0.26\linewidth]{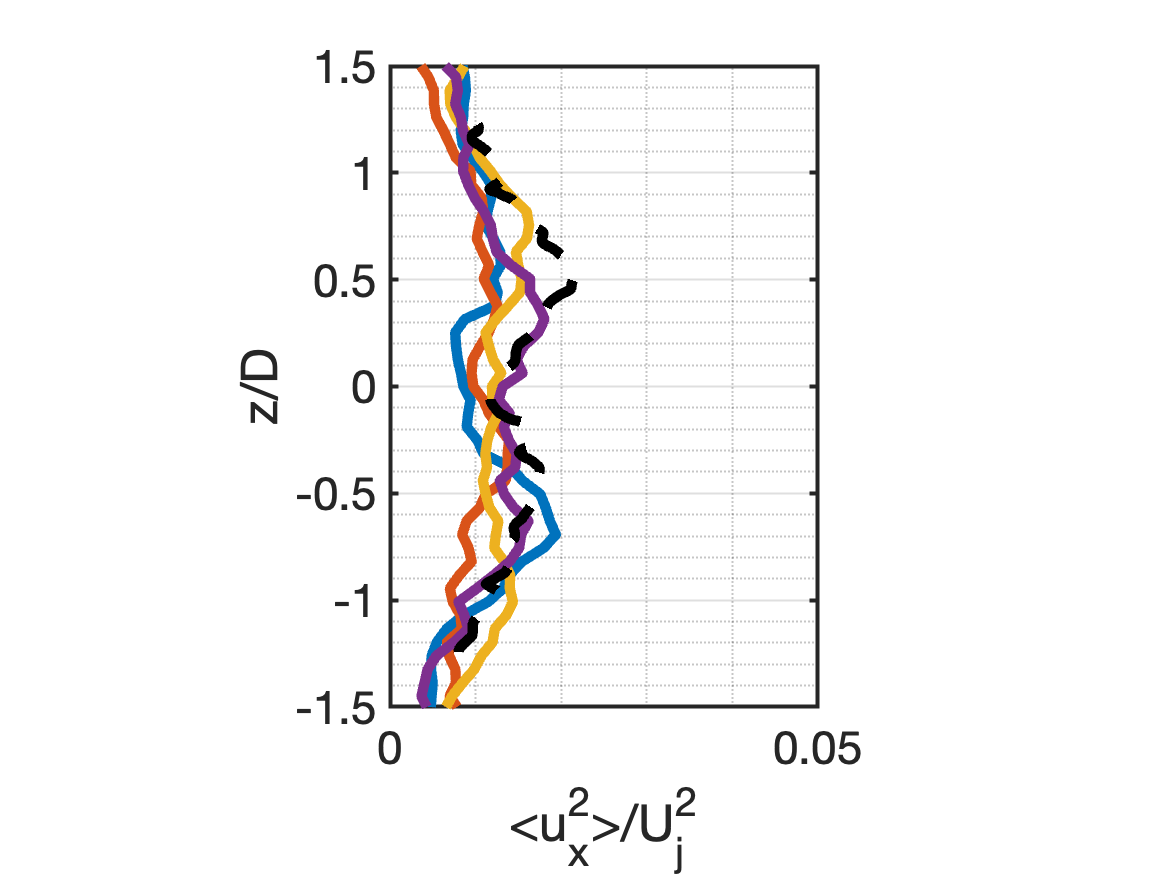}
	\label{fig.res5h}	
	}
\newline
\subfloat[$x/D=2.5$]{
	\includegraphics[trim = 30mm 0mm 20mm 0mm, clip, height=0.26\linewidth]{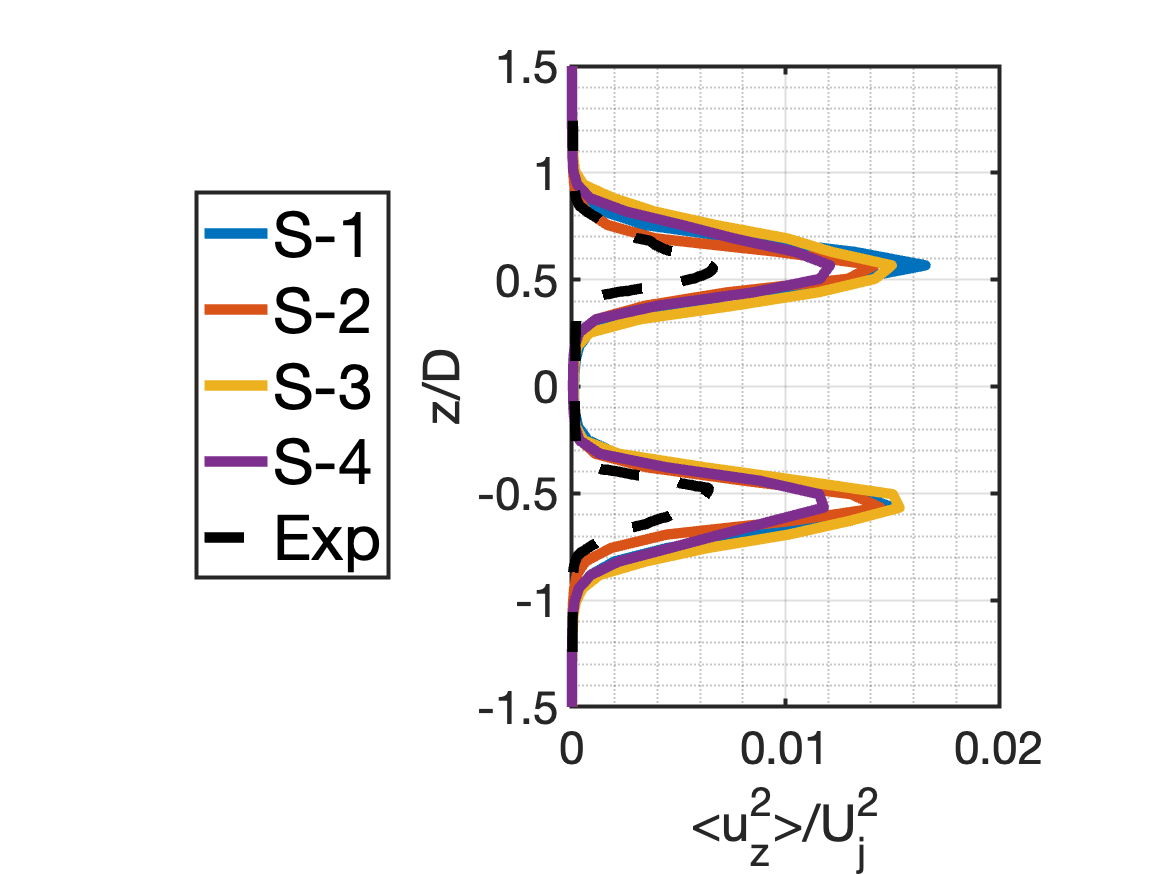}
	\label{fig.res5i}	
	}
\subfloat[$x/D=5$]{
	\includegraphics[trim = 30mm 0mm 55mm 0mm, clip, height=0.26\linewidth]{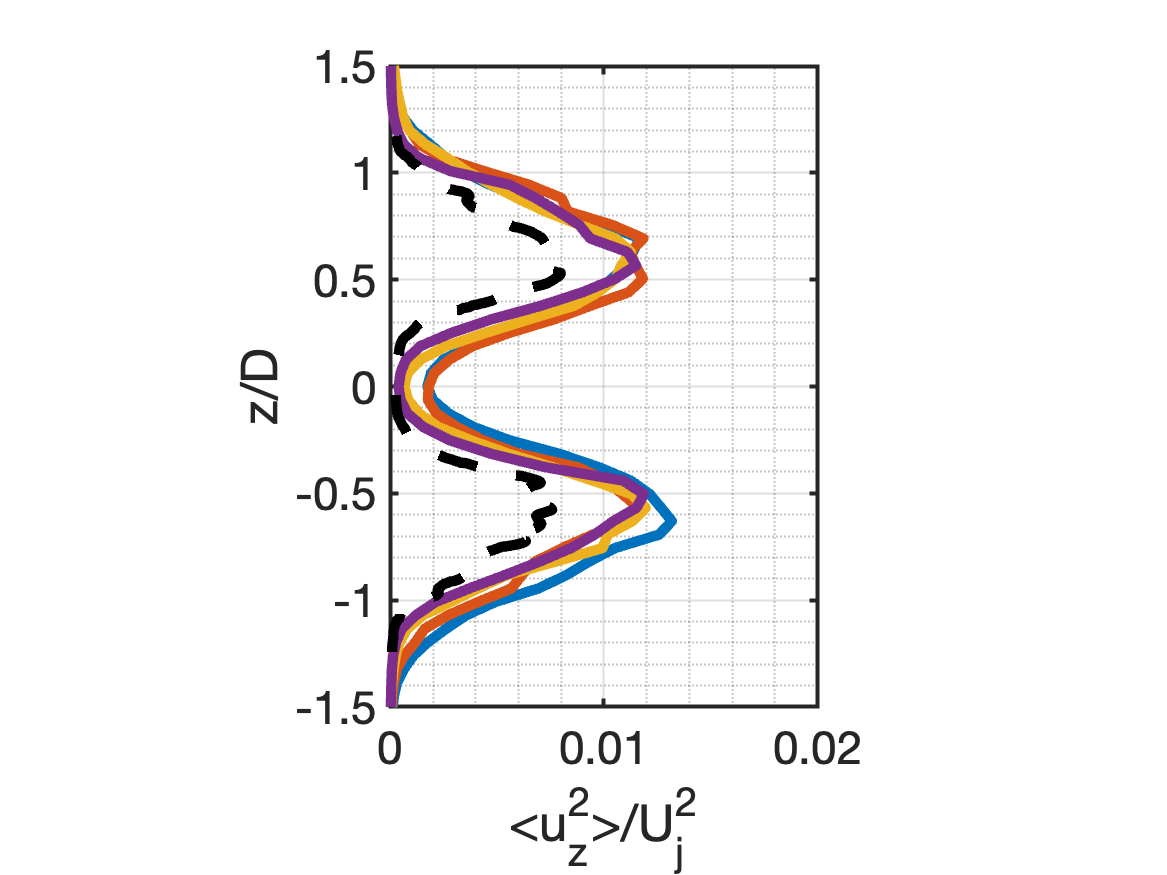}
	\label{fig.res5j}	
	}
\subfloat[$x/D=10$]{
	\includegraphics[trim = 30mm 0mm 55mm 0mm, clip, height=0.26\linewidth]{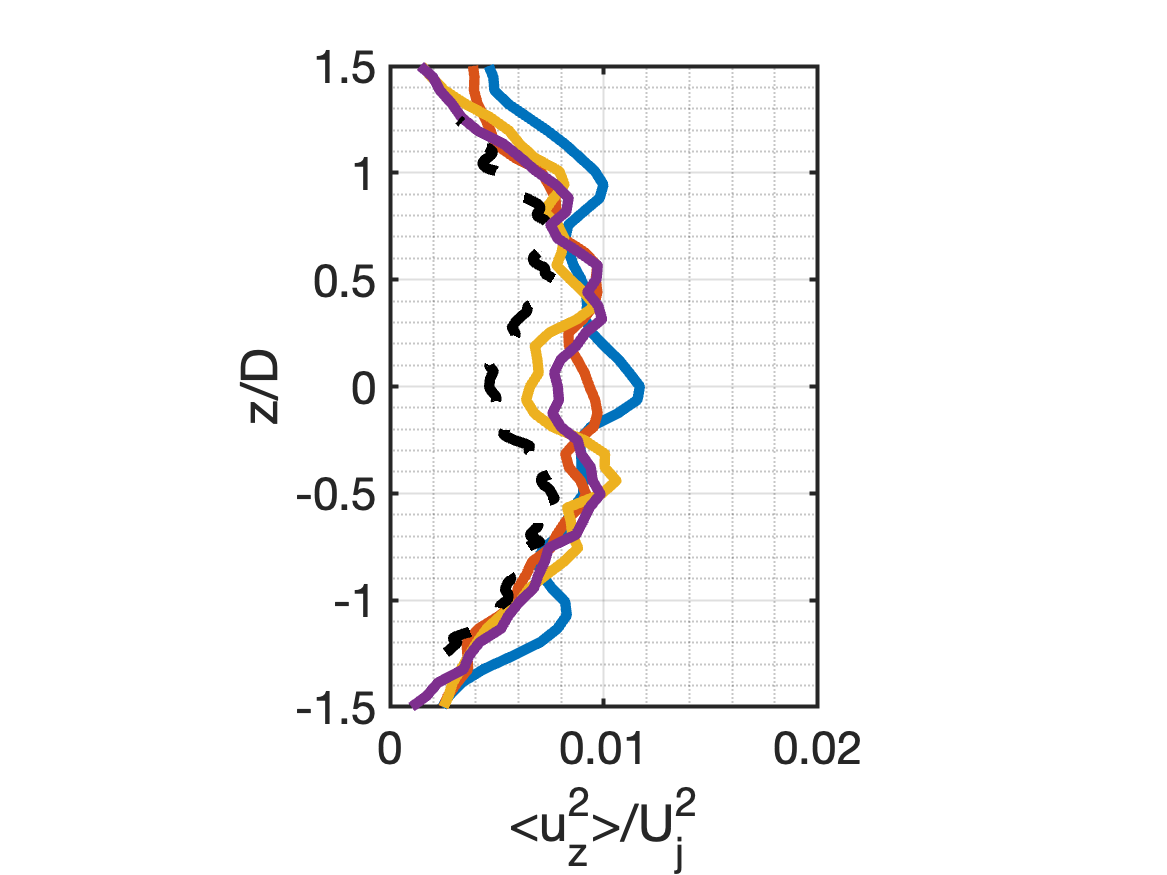}
	\label{fig.res5k}
	}
\subfloat[$x/D=15$]{
	\includegraphics[trim = 30mm 0mm 55mm 0mm, clip, height=0.26\linewidth]{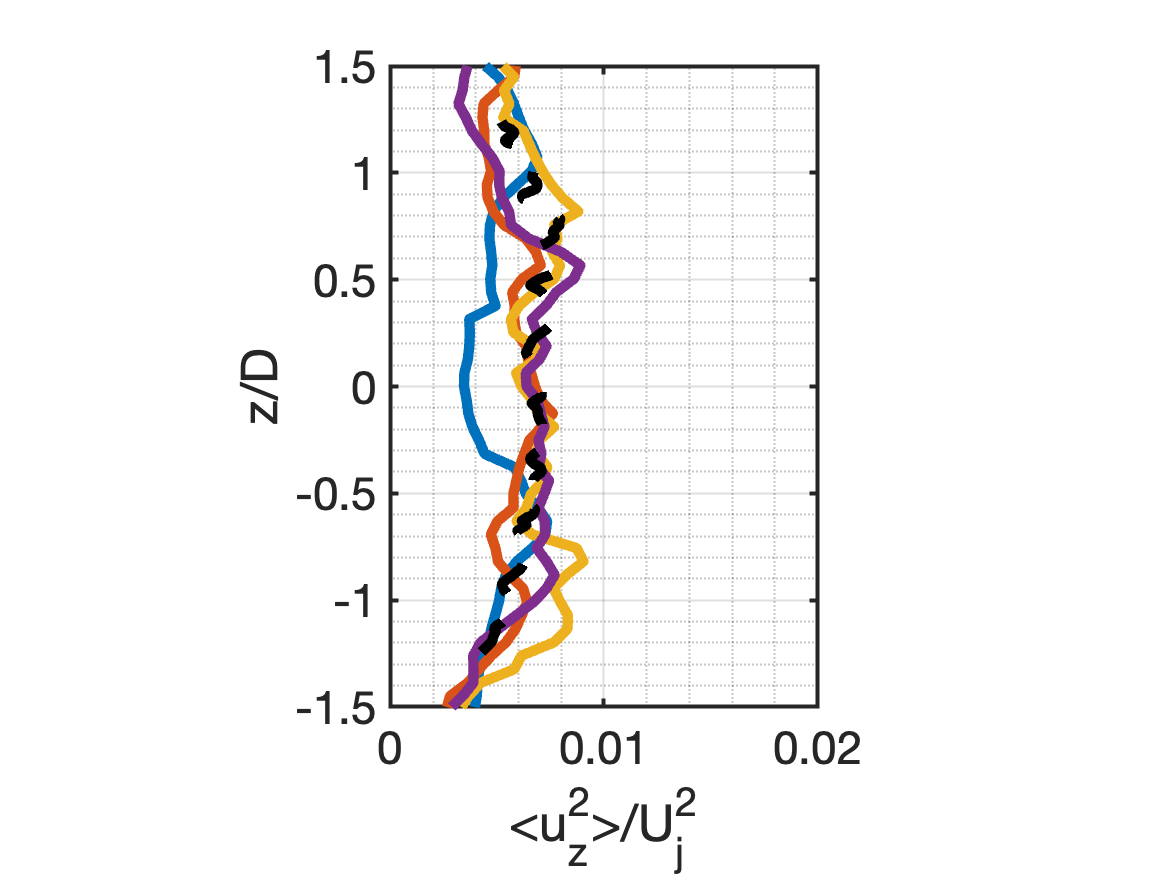}
	\label{fig.res5l}	
	}
\newline
\subfloat[$x/D=2.5$]{
	\includegraphics[trim = 30mm 0mm 20mm 0mm, clip, height=0.26\linewidth]{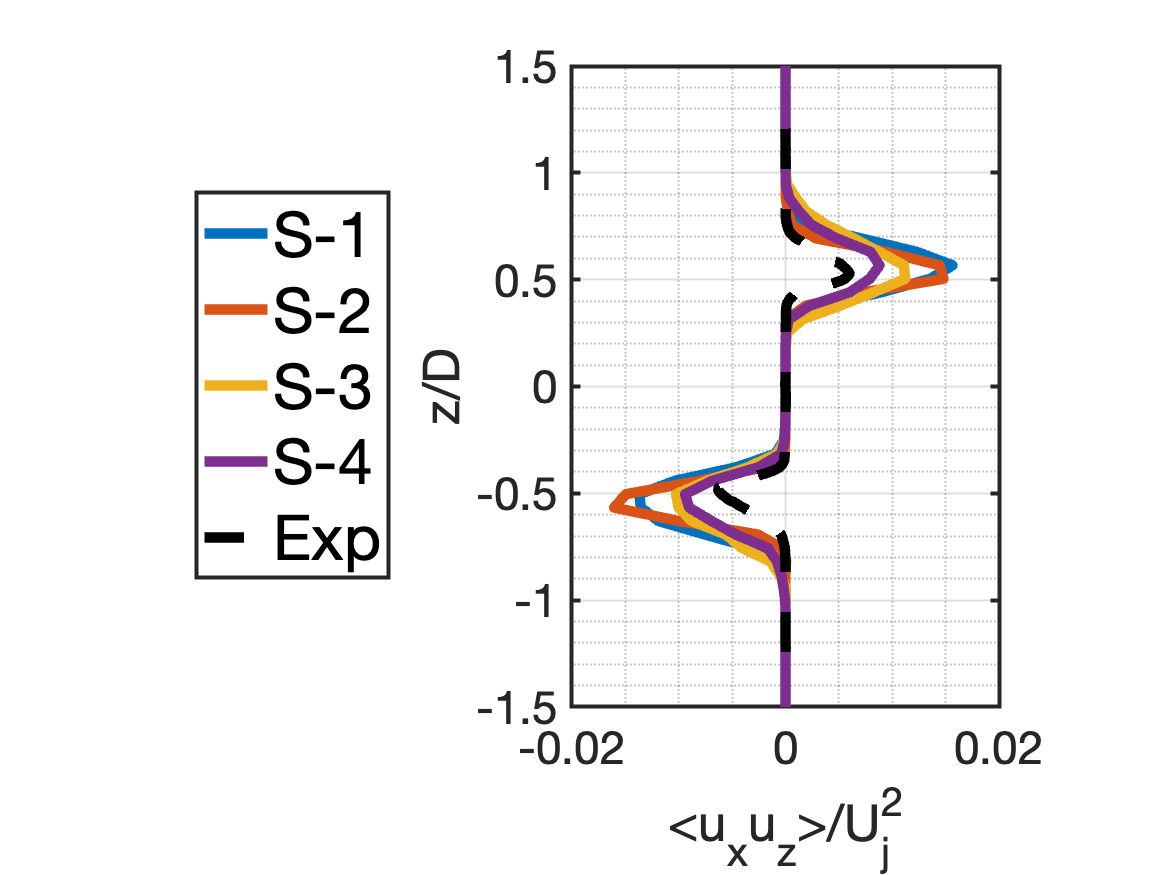}
	\label{fig.res5m}	
	}
\subfloat[$x/D=5$]{
	\includegraphics[trim = 30mm 0mm 55mm 0mm, clip, height=0.26\linewidth]{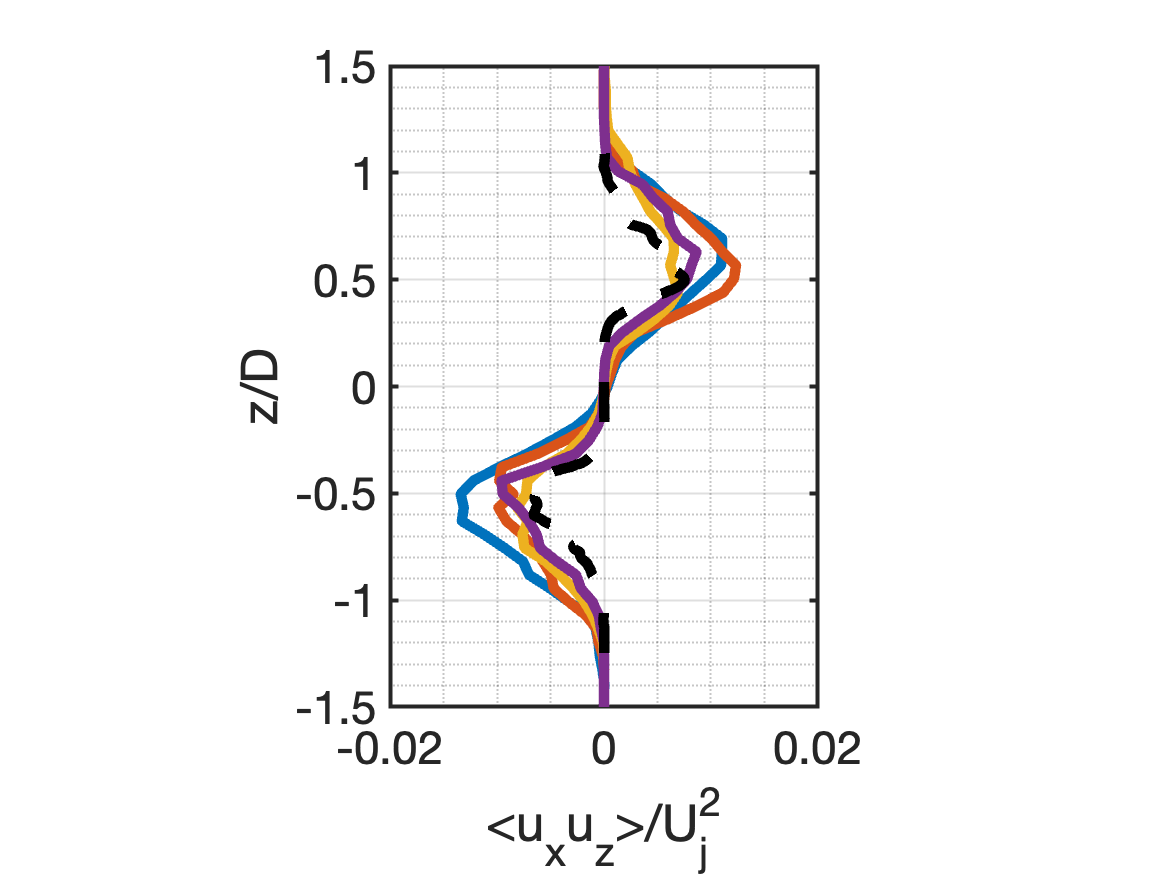}
	\label{fig.res5n}	
	}
\subfloat[$x/D=10$]{
	\includegraphics[trim = 30mm 0mm 55mm 0mm, clip, height=0.26\linewidth]{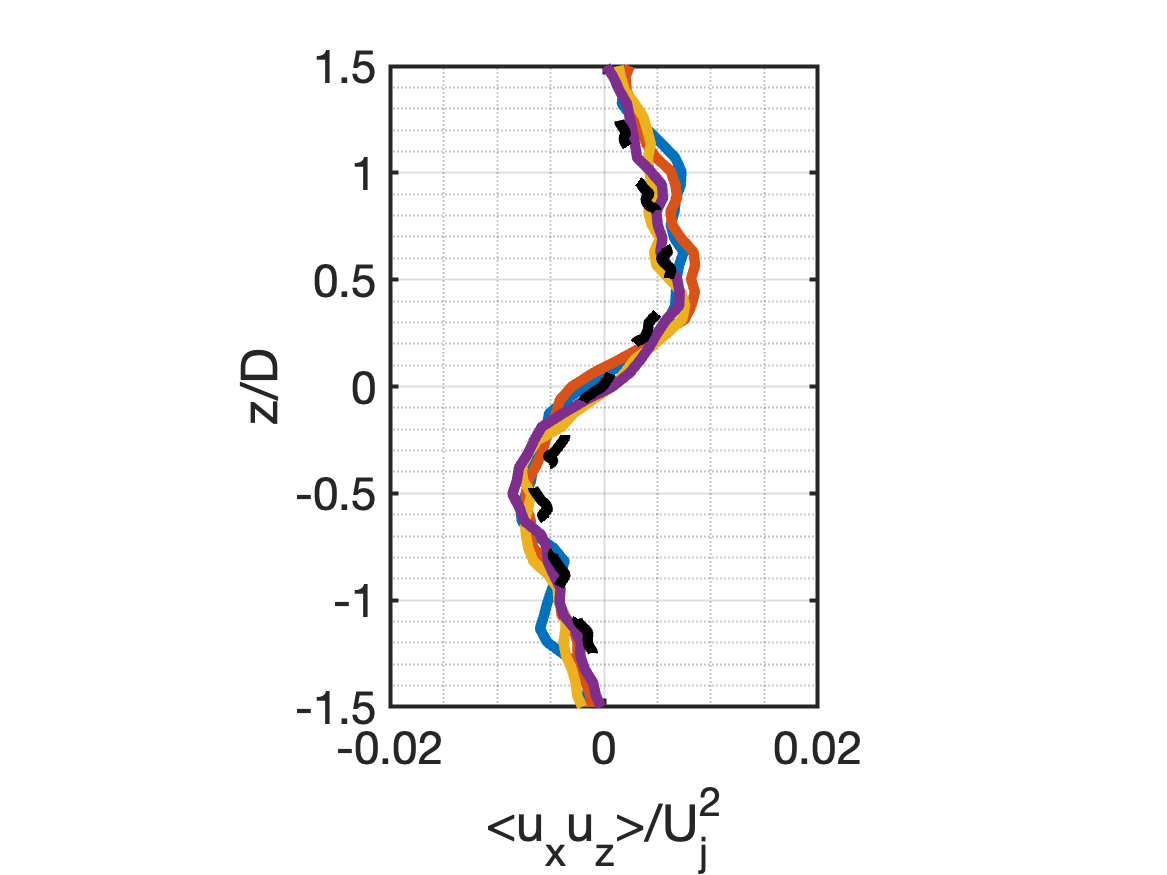}
	\label{fig.res5o}
	}
\subfloat[$x/D=15$]{
	\includegraphics[trim = 30mm 0mm 55mm 0mm, clip, height=0.26\linewidth]{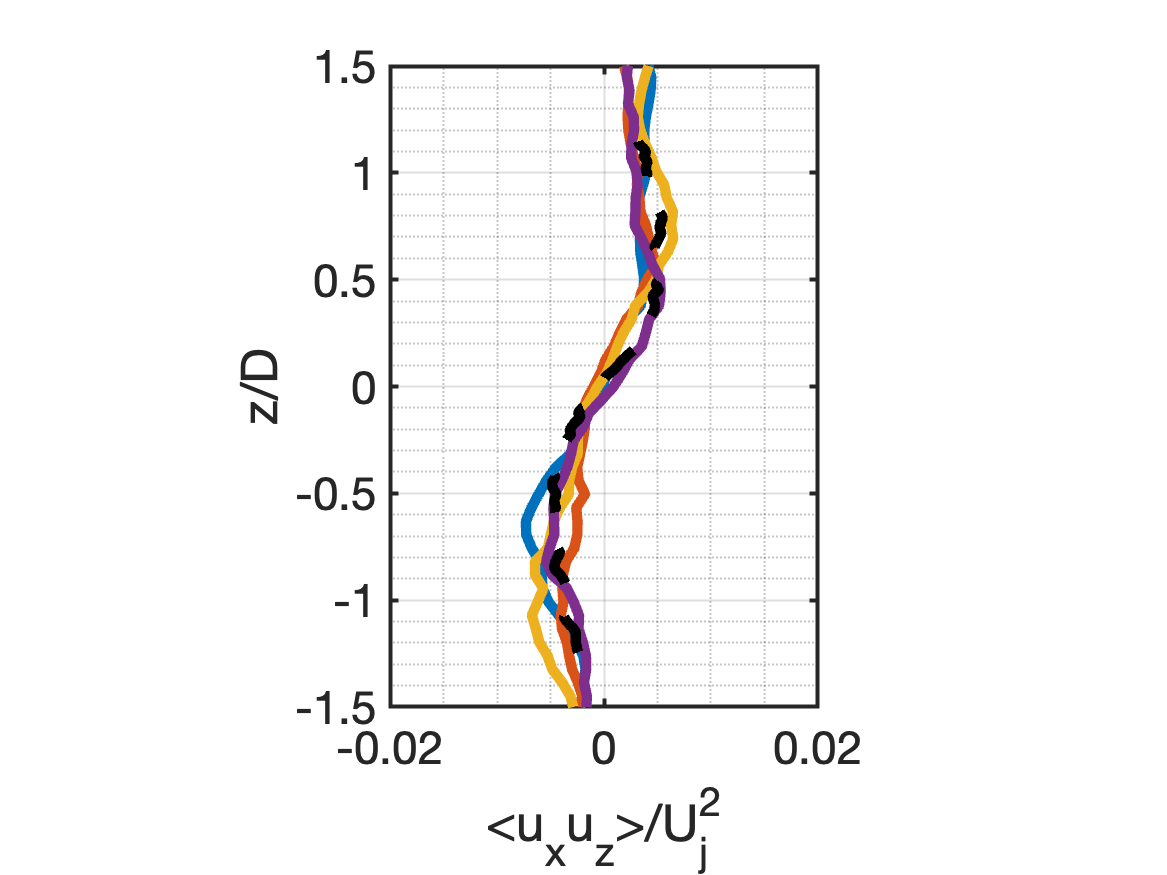}
	\label{fig.res5p}	
	}
\newline
\caption{Profiles of mean longitudinal velocity component, RMS of longitudinal
        velocity fluctuation, RMS of transversal velocity fluctuation, and mean
        shear-stress tensor component (from top to bottom) at four streamwise
        positions $x/D=2.5$, $x/D=5$, $x/D=10$ and $x/D=15$ (from left to
        right).}
\label{fig.res5}
\end{figure}

The development of the jet flow is presented as transversal distributions at
four streamwise stations: $x/D=2.5$, $x/D=5.0$, $x/D=10.$, and $x/D=15.0$,
Fig.~\ref{fig.res5}. The mean longitudinal velocity component distributions 
are presented in Figs.~\ref{fig.res5a} to \ref{fig.res5d}. The velocity 
distributions from the four numerical simulations hold a similar shape.
However, it is possible to verify differences in the velocity slopes from the
numerical data compared to the experimental reference at $z/D\approx 0.5$ and
$z/D\approx -0.5$. In the station $x/D=10.0$, the differences between the
velocity distributions from the numerical simulations are visible. At the jet
centerline, it is possible to observe smaller velocity levels from the S-1 and
S-2 simulations than the S-3 and S-4 simulations. In the last station, the
numerical simulations present lower velocity levels than the experimental
reference close to the jet centerline. The differences are related to the mesh
coarsening the streamwise direction and the fact that an inviscid inlet
profile is imposed, which does present the same behavior indicated by the
experiments at the lipline region, $z/D\approx \pm 0.5$.

The RMS of longitudinal velocity fluctuation profiles are presented in 
Figs.~\ref{fig.res5e} to \ref{fig.res5h}. In the first section, it is possible
to observe the larger velocity fluctuation levels from the numerical simulations
compared to the experimental reference. The differences between the numerical
simulations and the experimental reference are reduced in the second station. 
In the third station, there are differences between the numerical simulations 
and the experimental reference close to the jet centerline, where the velocity
fluctuation values of the S-1 and S-2 simulations are larger than those from 
the S-3 and S-4 simulations, which are larger than the experimental data. The
velocity fluctuation levels from the numerical simulations and the experimental
reference are similar in the last station. The RMS of transversal velocity  
fluctuation profiles are presented in Figs.~\ref{fig.res5i} to \ref{fig.res5l}.
The velocity distributions from the numerical simulations and the experimental
reference present a similar behavior to those from the longitudinal velocity
fluctuation profiles. The shear-stress tensor component distributions are
presented in Figs.~\ref{fig.res5m} to \ref{fig.res5p}. Once more, a similar
behavior from the two previous analyses from the longitudinal and radial
velocity fluctuation profiles is observed. One difference is that after the
second station was evaluated, the numerical data from all the simulations and 
the experimental reference were similar when such behavior was observed only
after the third station in the velocity fluctuation profiles.


The qualitative analysis of the four numerical simulations indicates a 
significant improvement in the representation of physical features from the 
S-3 and S-4 simulations when compared to the S-1 and S-2 cases. The 
comparison between the S-4 and S-3 simulations, indicated that S-4
calculations are capable of producing longer jet potential cores and better
defined shock waves than the ones from the S-3 case. The quantitative 
analysis shows a monotone improvement of the present computational results, that is, the calculated values of the various properties move in the direction of the experimental
reference data with the increased resolution of the simulation. There is one exception to this statement, and it has to do with the
longitudinal velocity fluctuation distribution at the lipline. Due to the
characteristics presented by the inlet condition, the behavior observed in 
this region is different from the one observed in the experimental reference data. In other words, 
the increase in the numerical resolution moves the velocity fluctuation distribution along the lipline away
from the experimental data. Due to significant improvements obtained in
the qualitative and the quantitative analyses with the S-4 simulation, the 
setup with the M-3 mesh, and a third-order accurate spatial discretization 
is chosen for the development of the inflow condition study.

\subsection{Inlet Condition Study}

The inlet condition study is performed with the comparison of the S-4, S-5,
and S-6 simulations. The three simulations are performed with the same numerical
setup. They utilize the M-3 mesh with a third-order accurate spatial 
discretization. The S-4 simulation utilizes the inviscid inflow condition, {\em i.e.}, the
same inflow condition used in the resolution study. The S-5 simulation employs the mean 
profile of the primitive properties obtained from a Reynolds-averaged Navier-Stokes (RANS) simulation of the
flow inside a nozzle that is assumed to be upstream of the present inflow boundary. The S-6 simulation utilizes the same mean RANS profile from the S-5
simulation superimposed by a tripping disturbance of the velocity variables
in the boundary layer.

\FloatBarrier

\begin{figure}[htb!]
\centering
\subfloat[Centerline]{
	\includegraphics[width=0.45\linewidth]{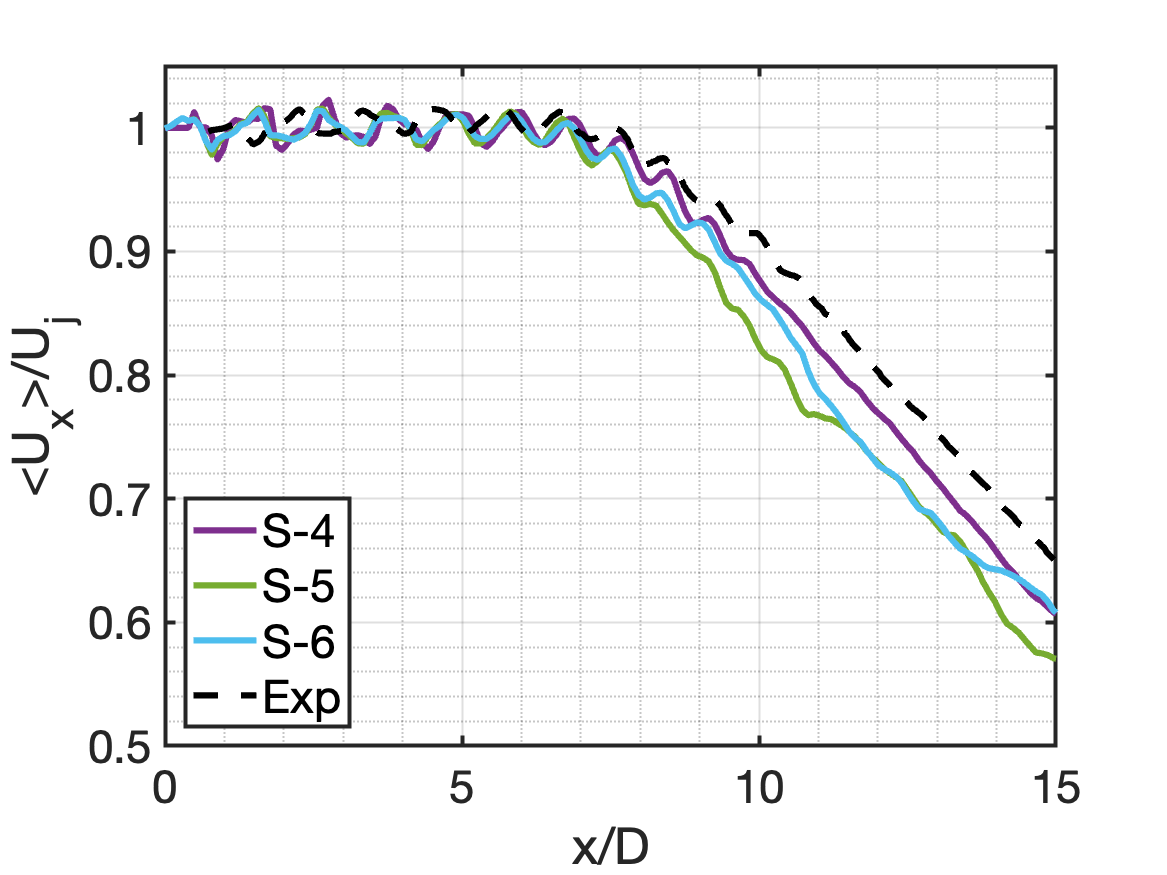}
	\label{fig.res101}	
	}%
\subfloat[Centerline]{
	\includegraphics[width=0.45\linewidth]{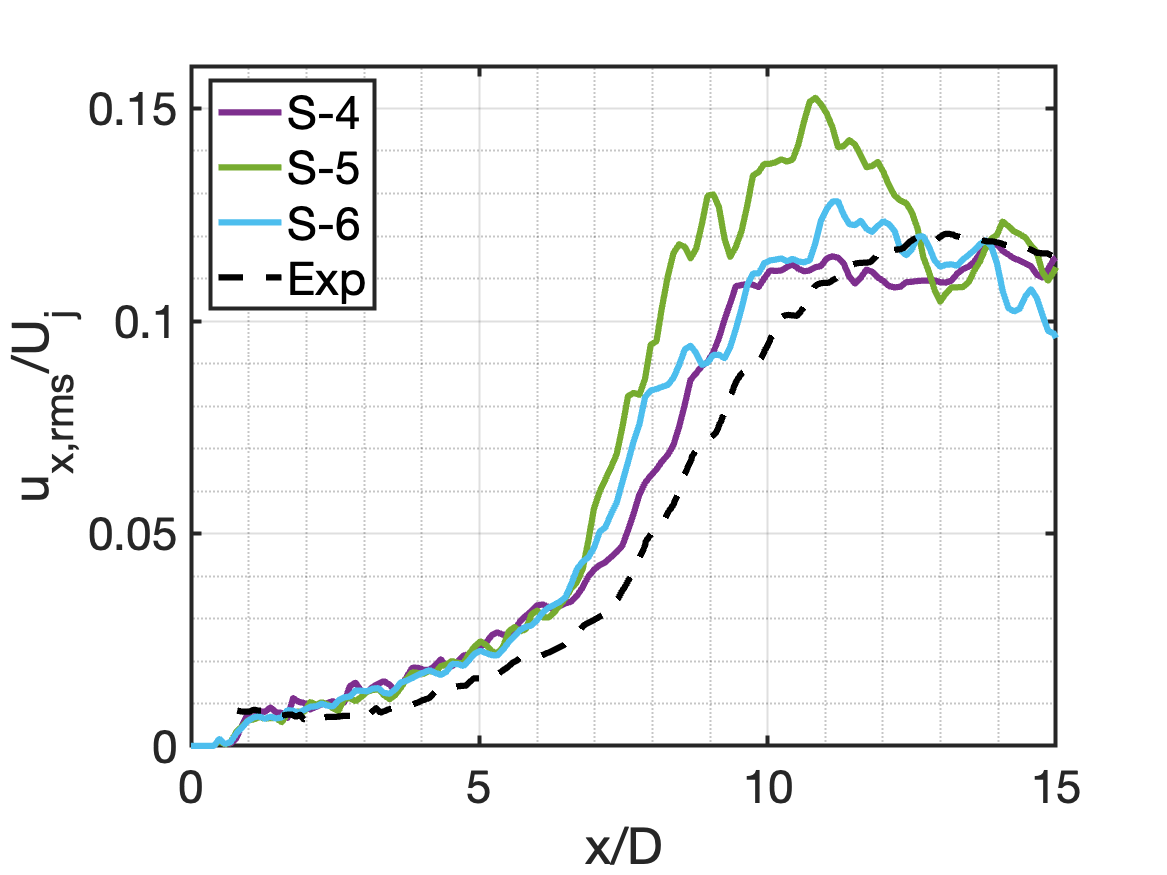}
	\label{fig.res102}	
	}
\newline
\subfloat[Lipline]{
	\includegraphics[width=0.45\linewidth]{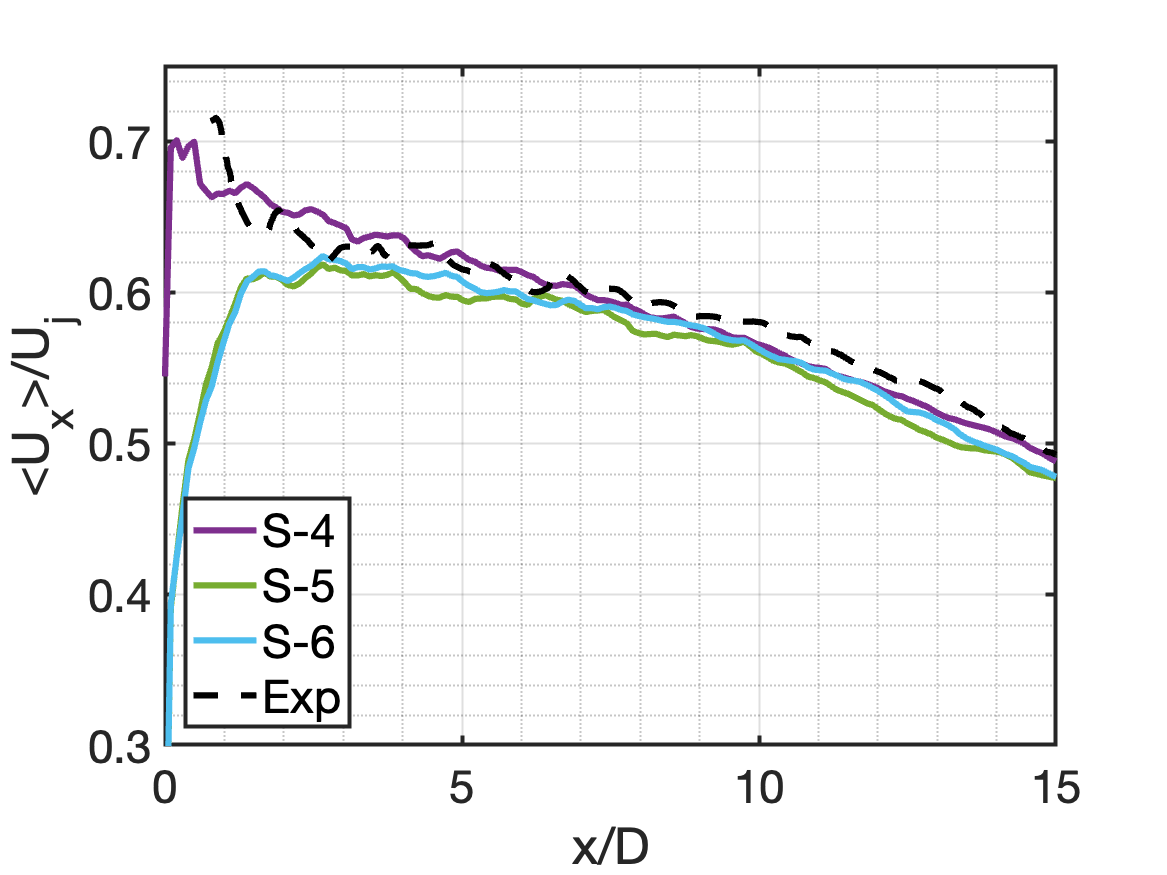}
	\label{fig.res103}
	}%
\subfloat[Lipline]{
	\includegraphics[width=0.45\linewidth]{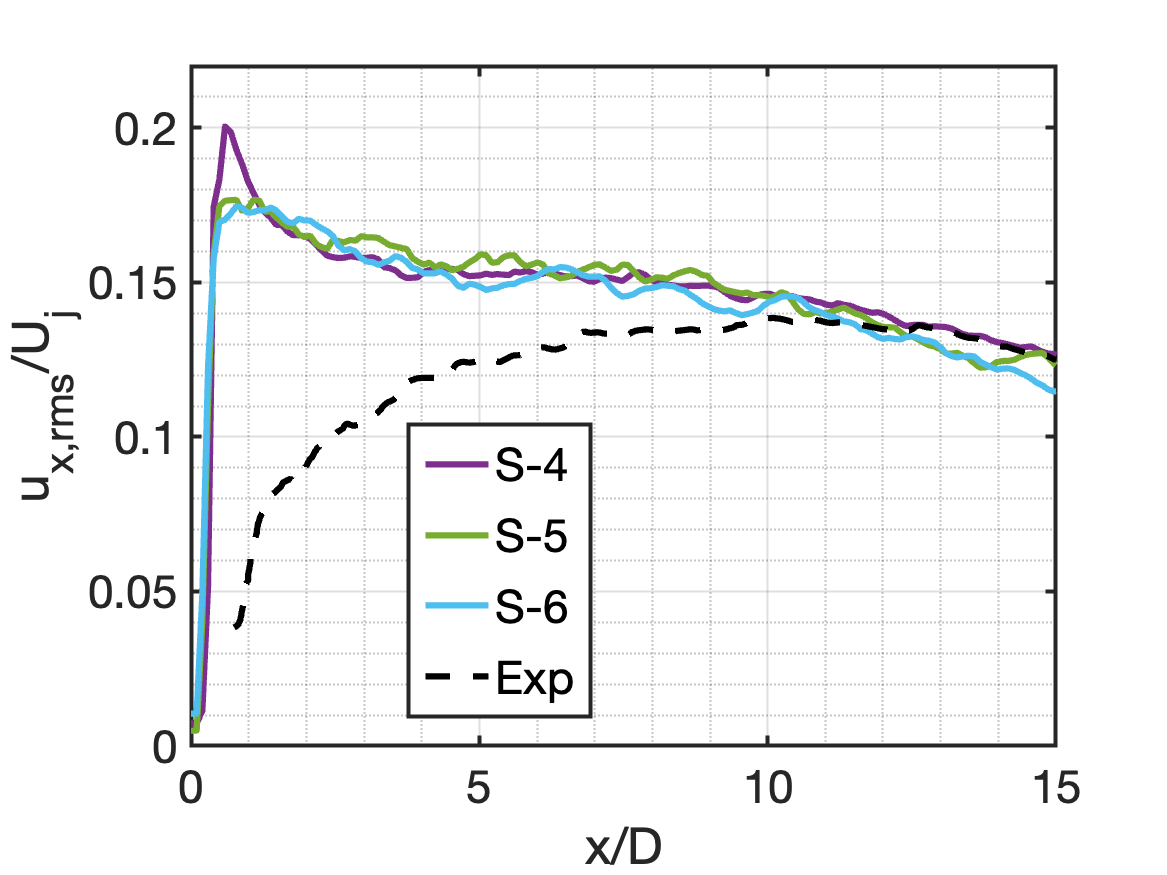}
	\label{fig.res104}	
	}
\caption{Mean longitudinal velocity component distribution (left) and RMS of
         longitudinal velocity fluctuation (right) at the jet centerline
         $y/D=0$ (top) and lipline $y/D=0.5$ (bottom).}
\label{fig.res101_104}
\end{figure}

The longitudinal velocity component distributions at the jet centerline and
lipline of the three simulations with different inflow conditions are presented
in Fig.~\ref{fig.res101_104}. The mean longitudinal velocity component 
distribution at the centerline, Fig.~\ref{fig.res101}, shows similarities
between the three numerical datasets. The S-5 and S-6 simulations present a
change in the velocity slope associated with the interaction of the mixing
layers at the jet centerline starting in a station closer to the jet inlet 
than the S-4 simulation. The RMS data from the longitudinal velocity
fluctuation at the jet centerline, Fig.~\ref{fig.res102}, indicate a similar
behavior to those observed in the mean velocity distribution. The velocity 
fluctuation slope change occurs closer to the jet inlet section. The peak 
values from the S-5 and S-6 simulations are higher than those presented by 
the S-4 simulation.

The velocity profiles at the jet lipline present significant differences when
comparing the three numerical simulations. The mean longitudinal velocity
component distributions at the jet lipline, Fig.~\ref{fig.res103}, are distinct
among calculations. While the S-4 simulation produced velocity 
profile peaks in the initial sections,  
the S-5 and S-6 simulations present a small velocity magnitude in the initial
sections, $x/D<1.0$, then the velocity data reach the values from the S-4
simulation and match the experimental reference. The lipline velocity
distribution indicates that the boundary layer obtained from the nozzle
simulation is thicker than that obtained in the experiments. The RMS of the
longitudinal velocity fluctuations at the lipline, Fig.~\ref{fig.res104},
present small differences from the numerical simulations. When the mean RANS
profile is imposed, the peak of velocity fluctuation observed is reduced when
compared to the simulation with the inviscid profile. However, the 
high-velocity slope, that does not exist in the experiment, remains. The high-velocity slope may be associated with the
transition mechanism due to the imposition of a laminar flow in the boundary
condition. The setup with the tripping method ended up having very small effects in the current calculations.

%
\begin{figure}[htb!]
\centering
\subfloat[$x/D=2.5$]{
	\includegraphics[trim = 30mm 0mm 20mm 0mm, clip, height=0.26\linewidth]{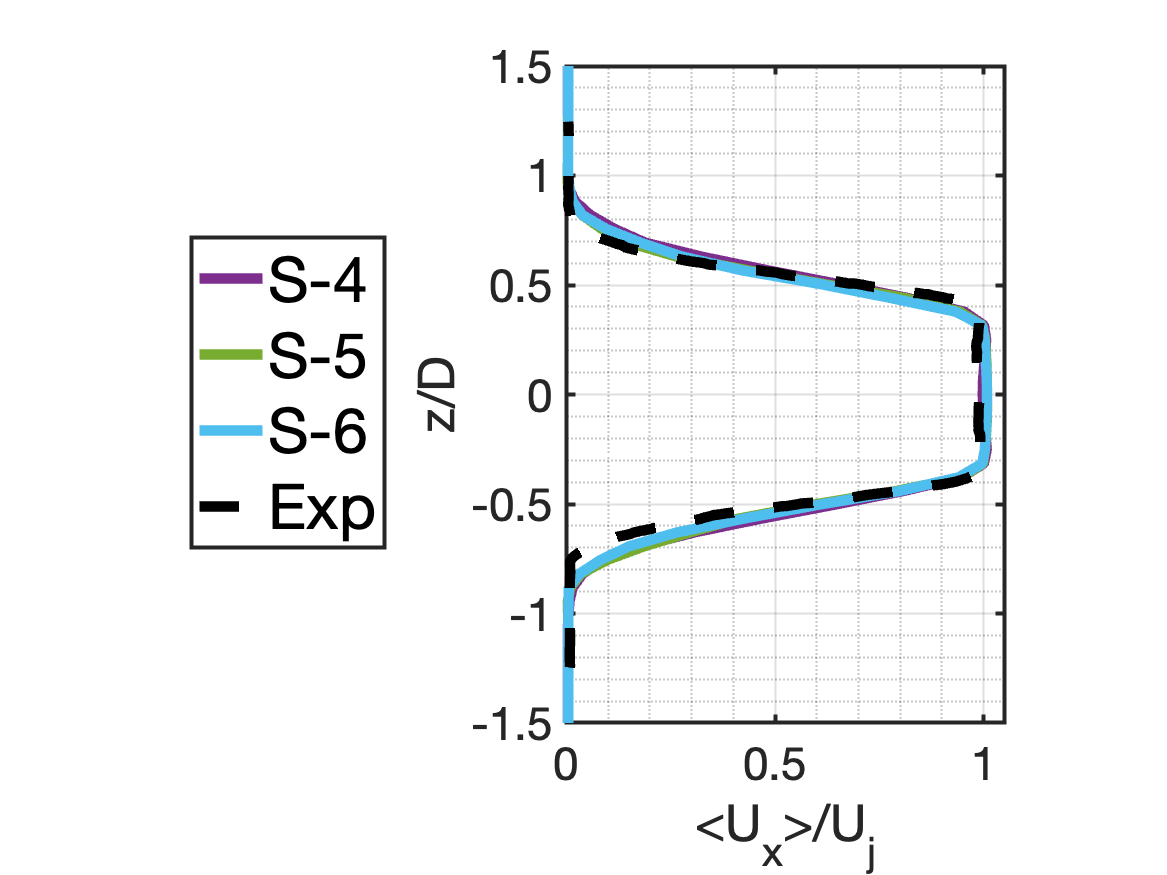}
	\label{fig.res105a}	
	}
\subfloat[$x/D=5$]{
	\includegraphics[trim = 30mm 0mm 55mm 0mm, clip, height=0.26\linewidth]{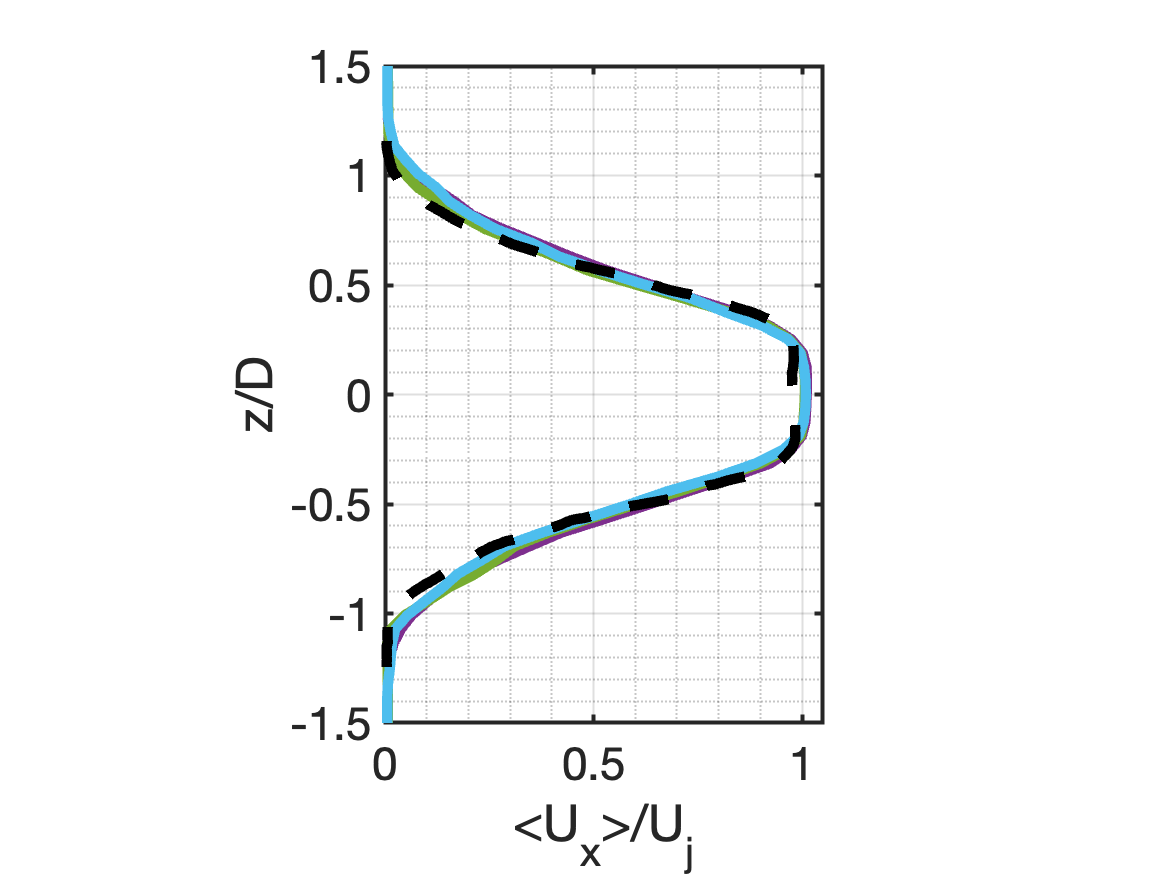}
	\label{fig.res105b}	
	}
\subfloat[$x/D=10$]{
	\includegraphics[trim = 30mm 0mm 55mm 0mm, clip, height=0.26\linewidth]{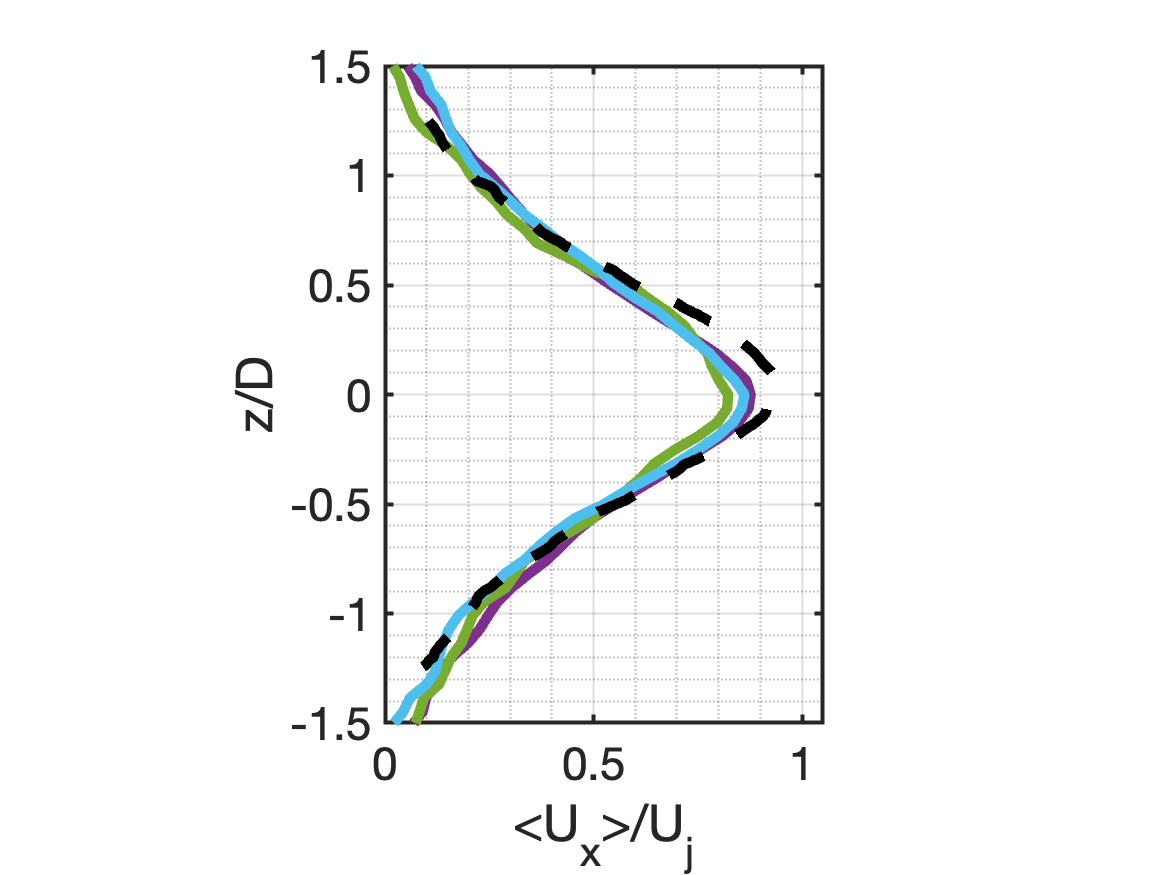}
	\label{fig.res105c}
	}
\subfloat[$x/D=15$]{
	\includegraphics[trim = 30mm 0mm 55mm 0mm, clip, height=0.26\linewidth]{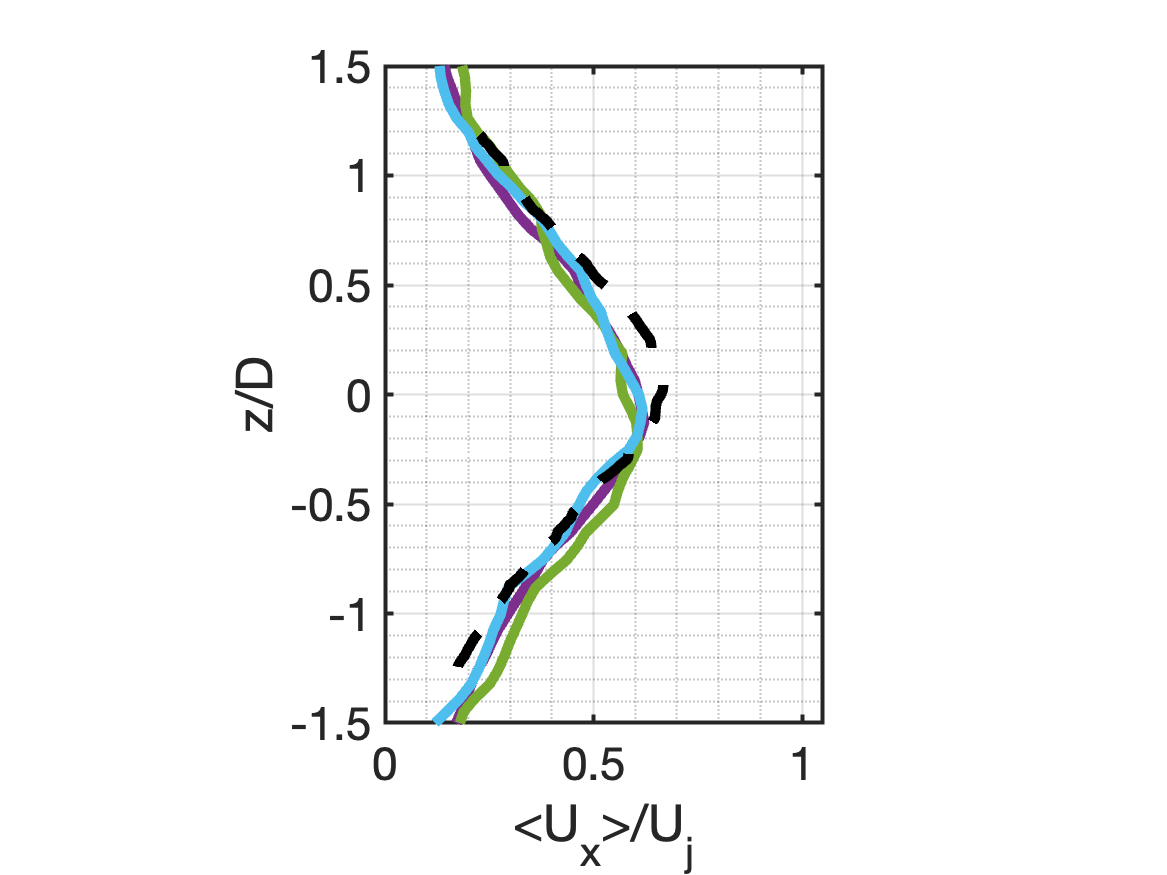}
	\label{fig.res105d}	
	}
\newline
\subfloat[$x/D=2.5$]{
	\includegraphics[trim = 30mm 0mm 20mm 0mm, clip, height=0.26\linewidth]{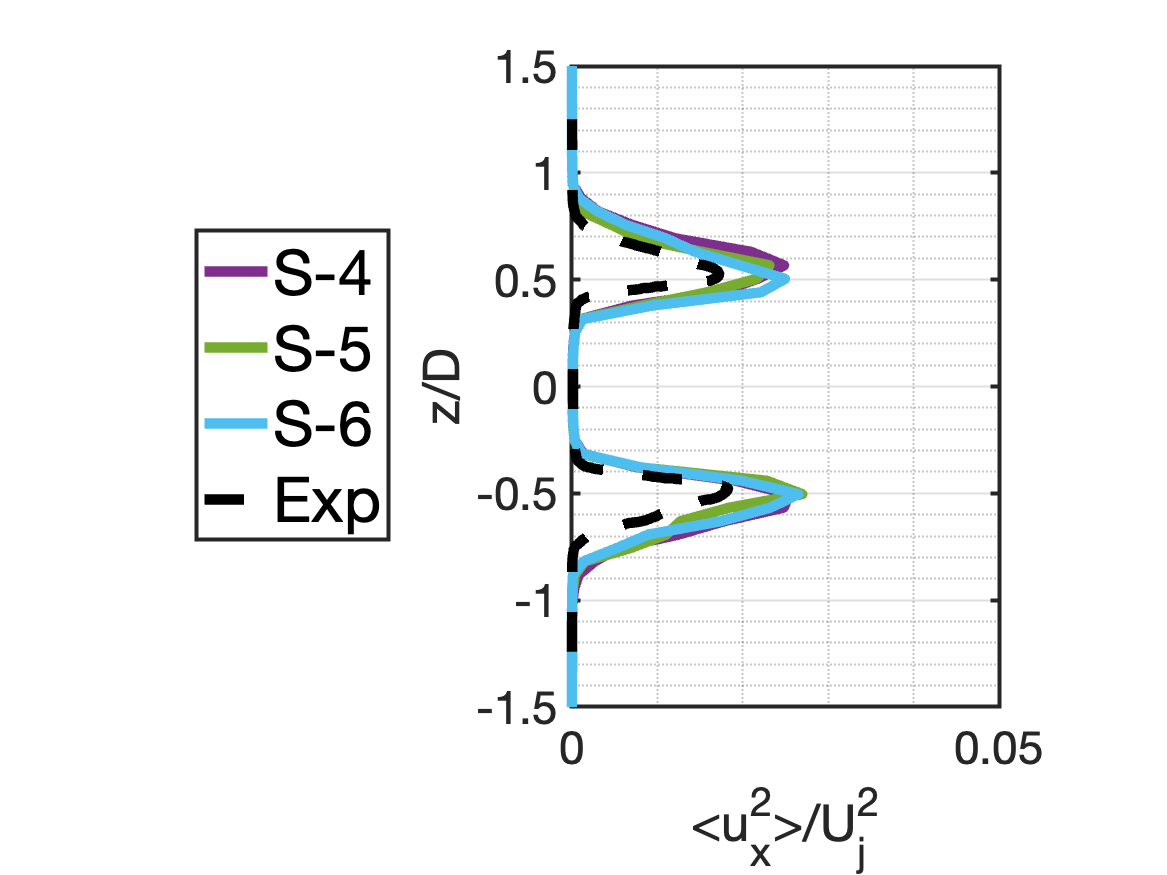}
	\label{fig.res105e}	
	}
\subfloat[$x/D=5$]{
	\includegraphics[trim = 30mm 0mm 55mm 0mm, clip, height=0.26\linewidth]{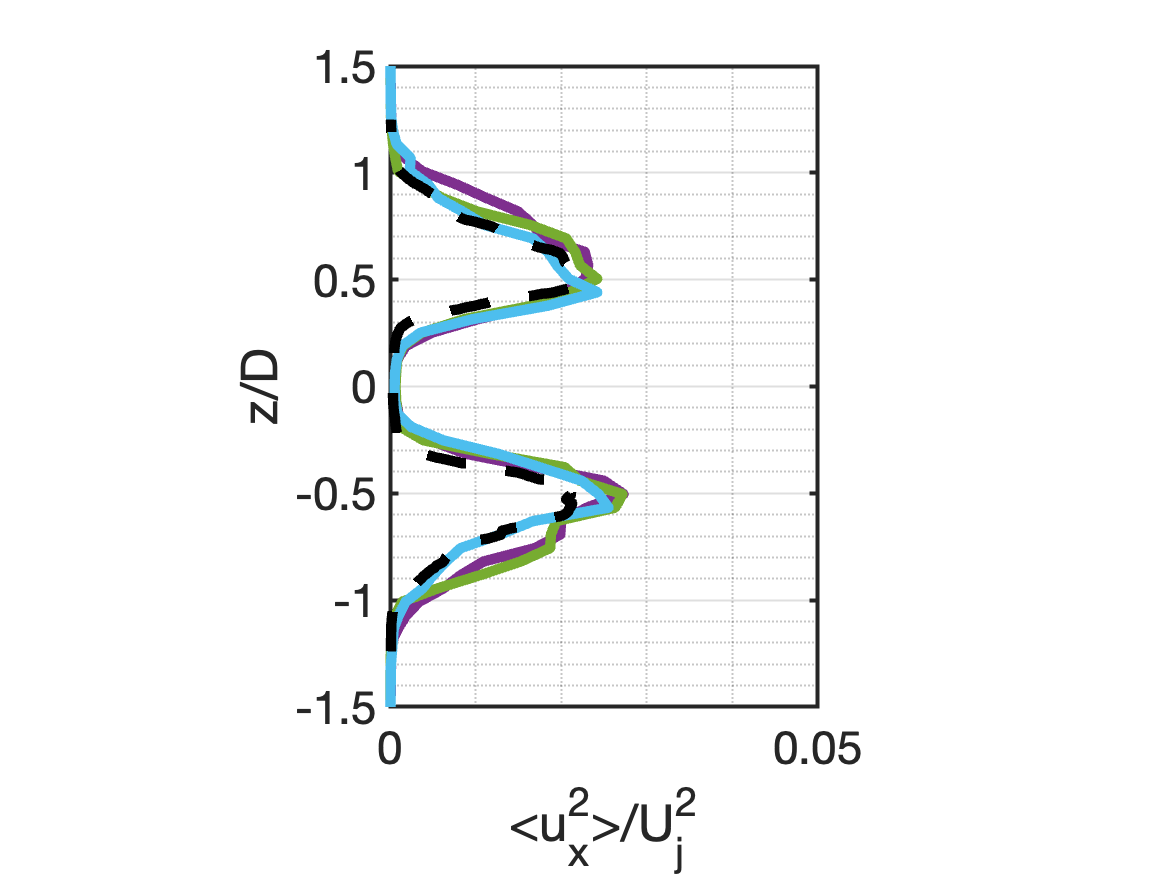}
	\label{fig.res105f}	
	}
\subfloat[$x/D=10$]{
	\includegraphics[trim = 30mm 0mm 55mm 0mm, clip, height=0.26\linewidth]{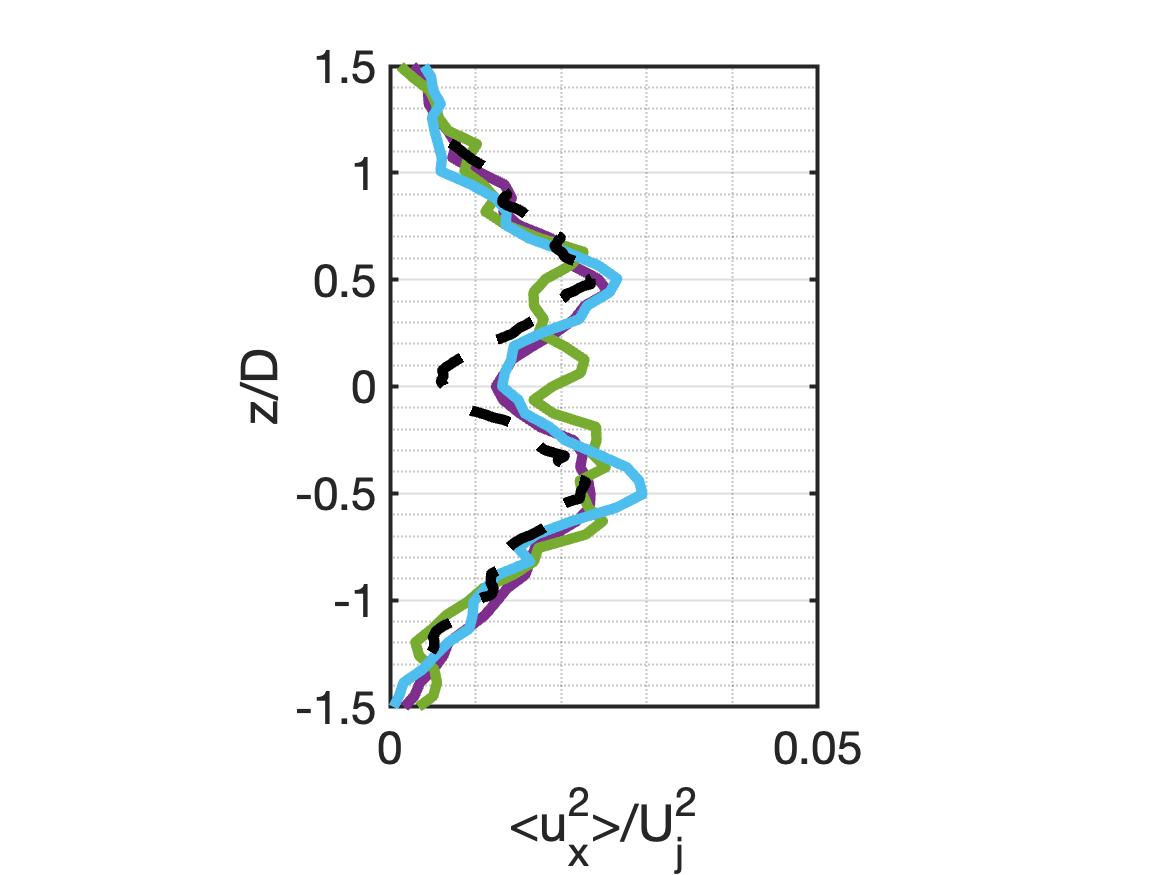}
	\label{fig.res105g}
	}
\subfloat[$x/D=15$]{
	\includegraphics[trim = 30mm 0mm 55mm 0mm, clip, height=0.26\linewidth]{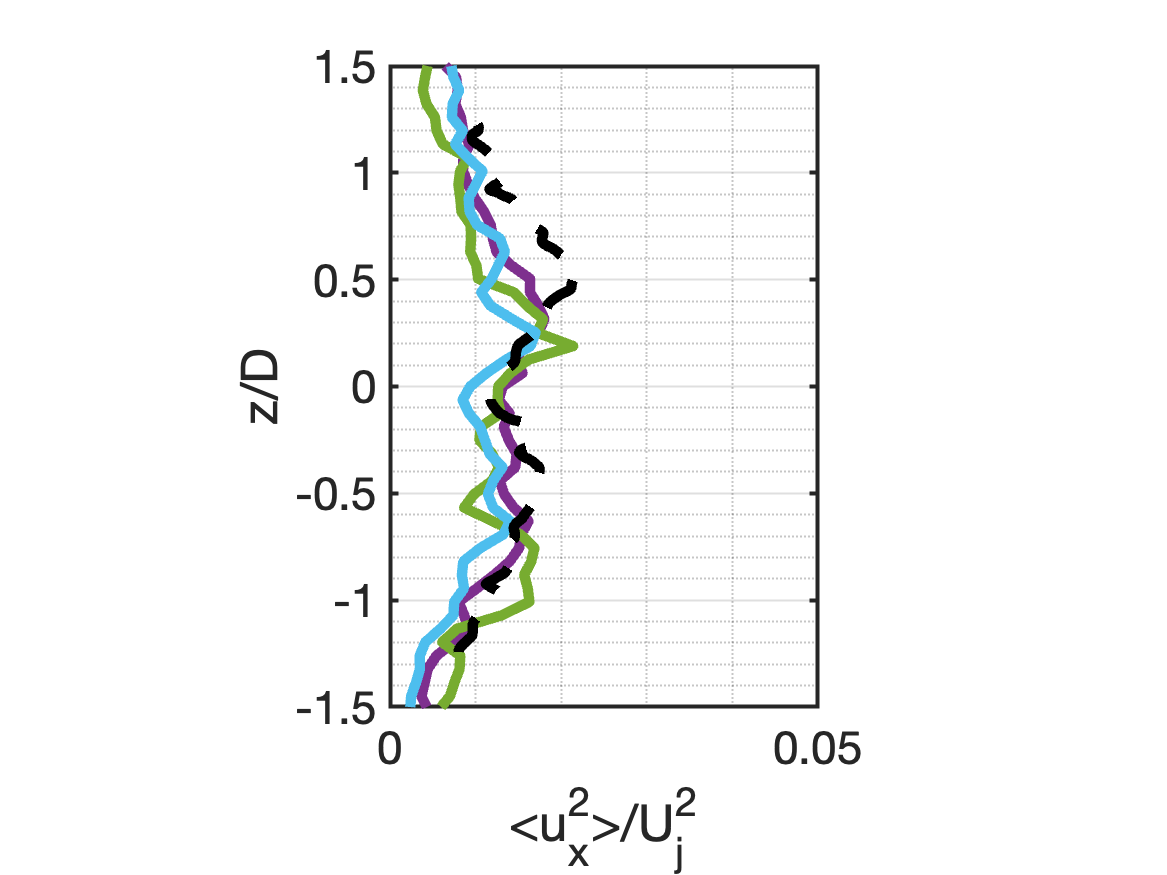}
	\label{fig.res105h}	
	}
\newline
\subfloat[$x/D=2.5$]{
	\includegraphics[trim = 30mm 0mm 20mm 0mm, clip, height=0.26\linewidth]{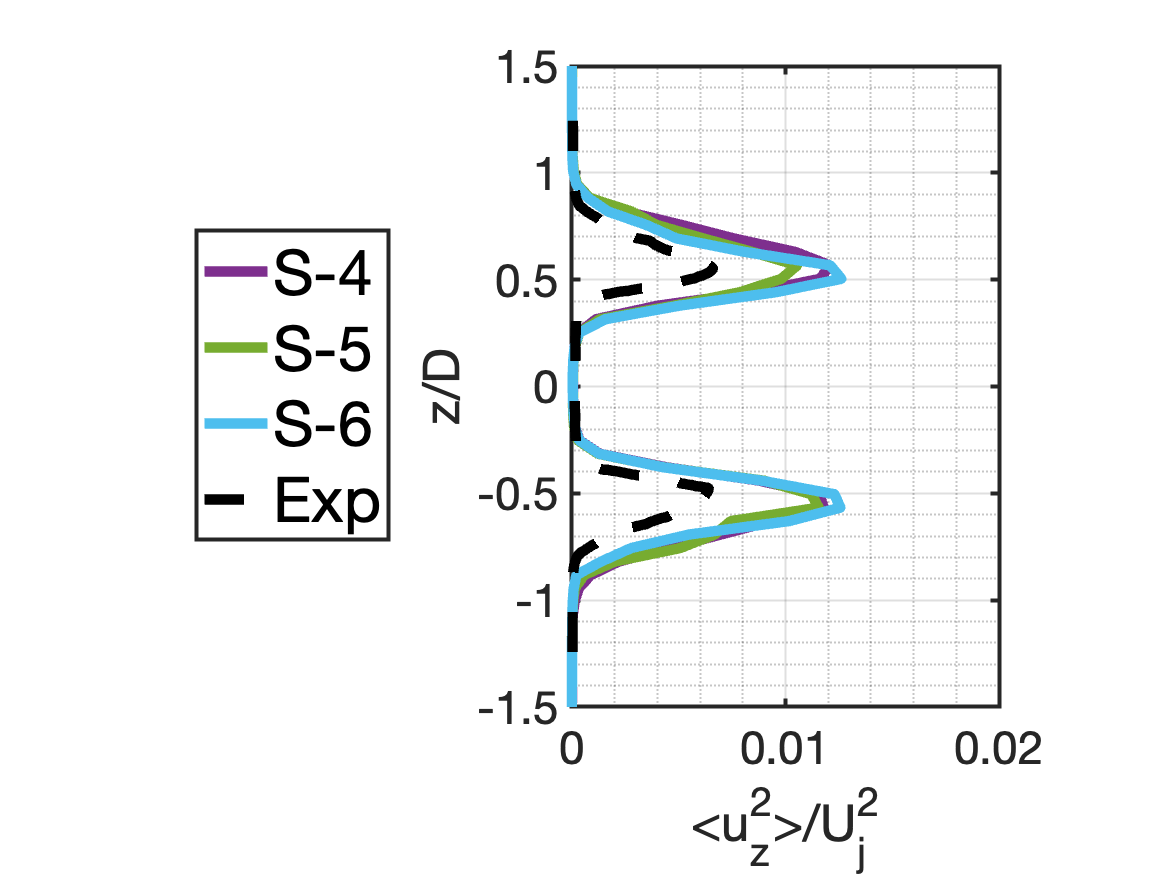}
	\label{fig.res105i}	
	}
\subfloat[$x/D=5$]{
	\includegraphics[trim = 30mm 0mm 55mm 0mm, clip, height=0.26\linewidth]{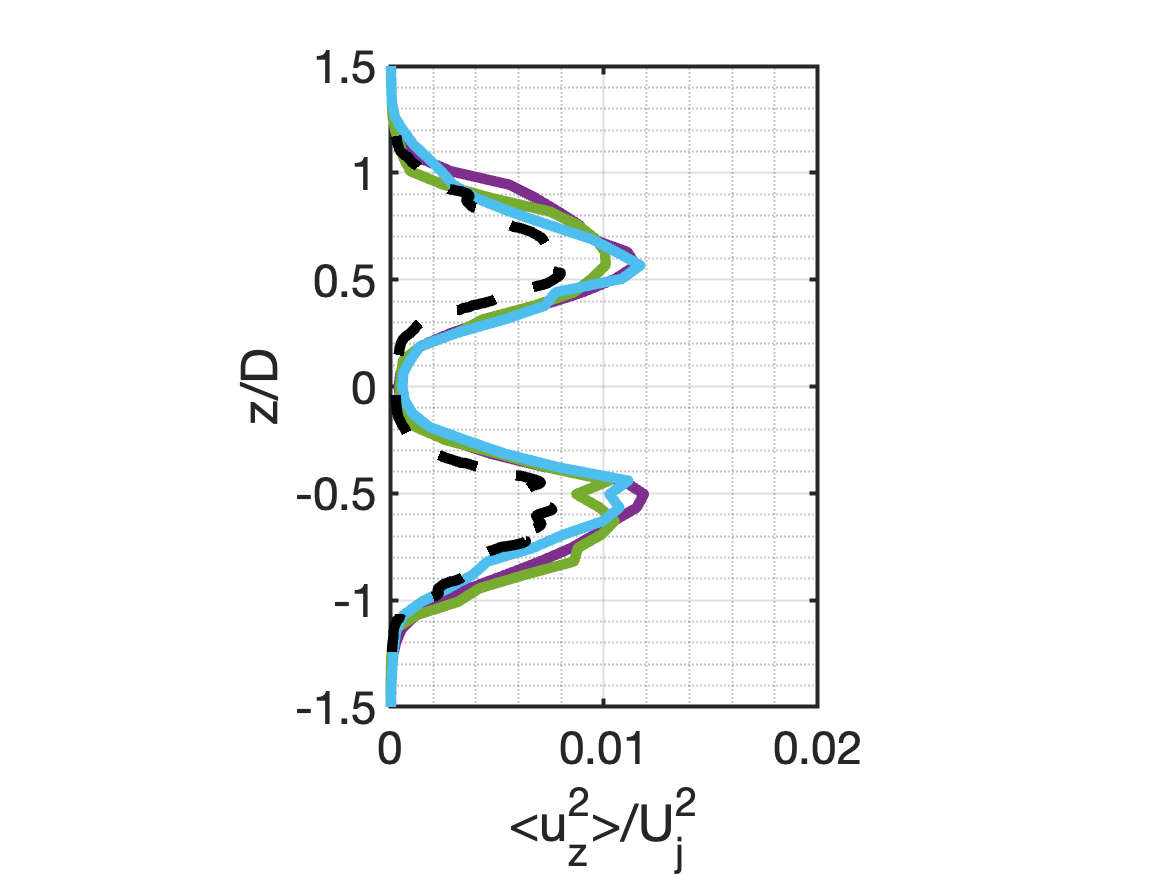}
	\label{fig.res105j}	
	}
\subfloat[$x/D=10$]{
	\includegraphics[trim = 30mm 0mm 55mm 0mm, clip, height=0.26\linewidth]{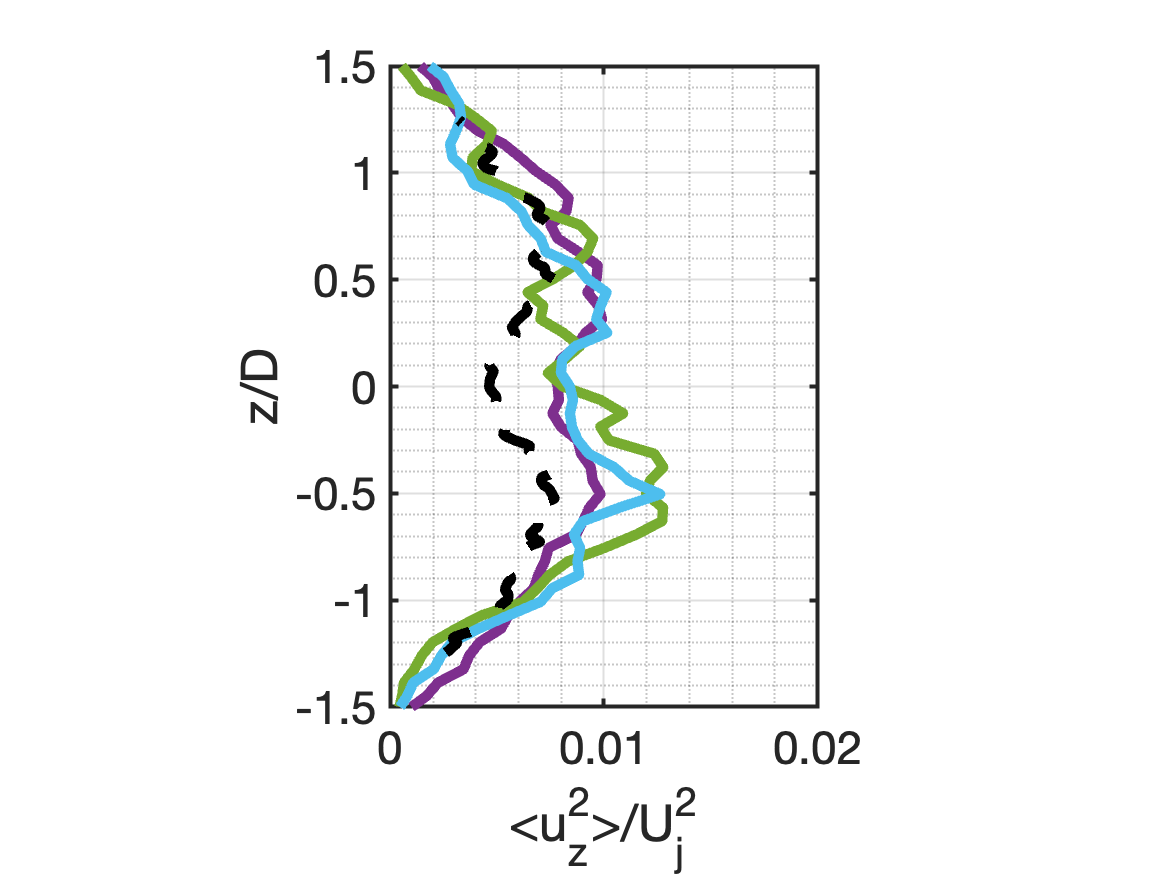}
	\label{fig.res105k}
	}
\subfloat[$x/D=15$]{
	\includegraphics[trim = 30mm 0mm 55mm 0mm, clip, height=0.26\linewidth]{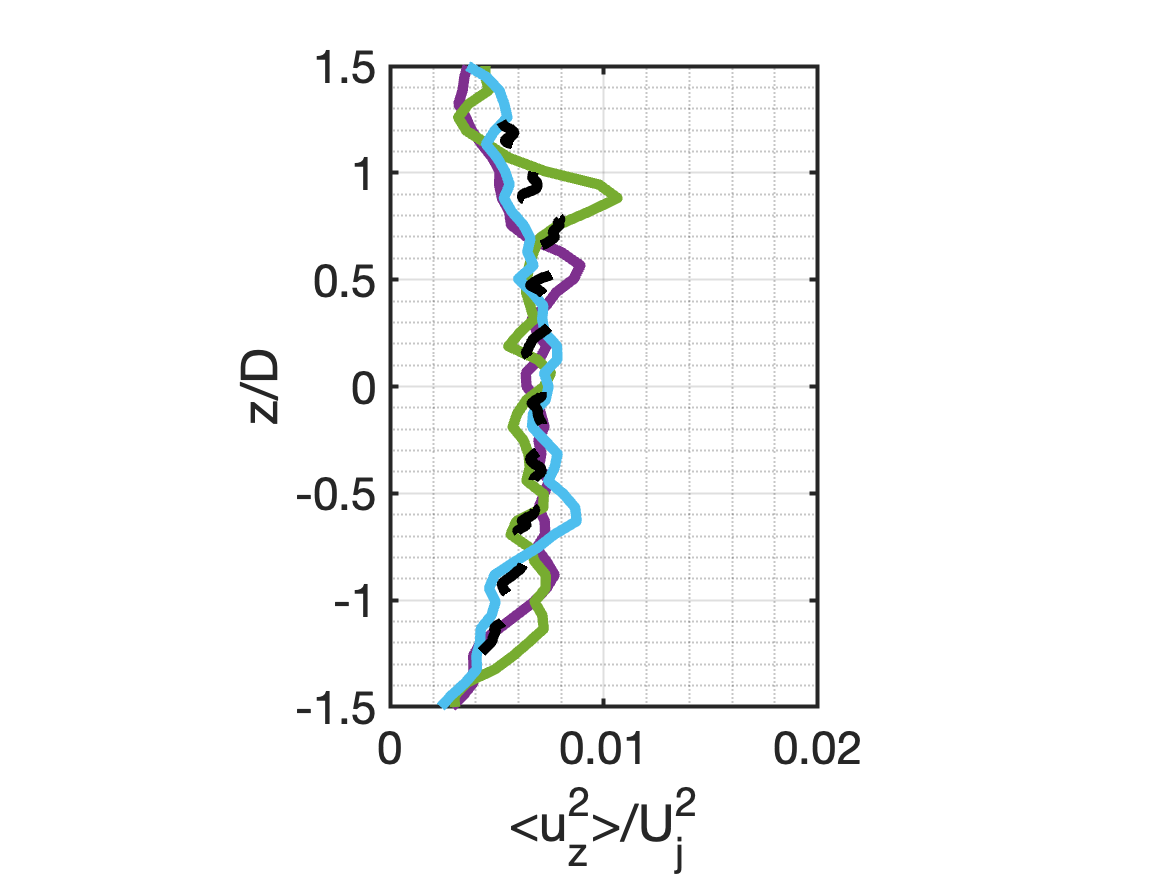}
	\label{fig.res105l}	
	}
\newline
\subfloat[$x/D=2.5$]{
	\includegraphics[trim = 30mm 0mm 20mm 0mm, clip, height=0.26\linewidth]{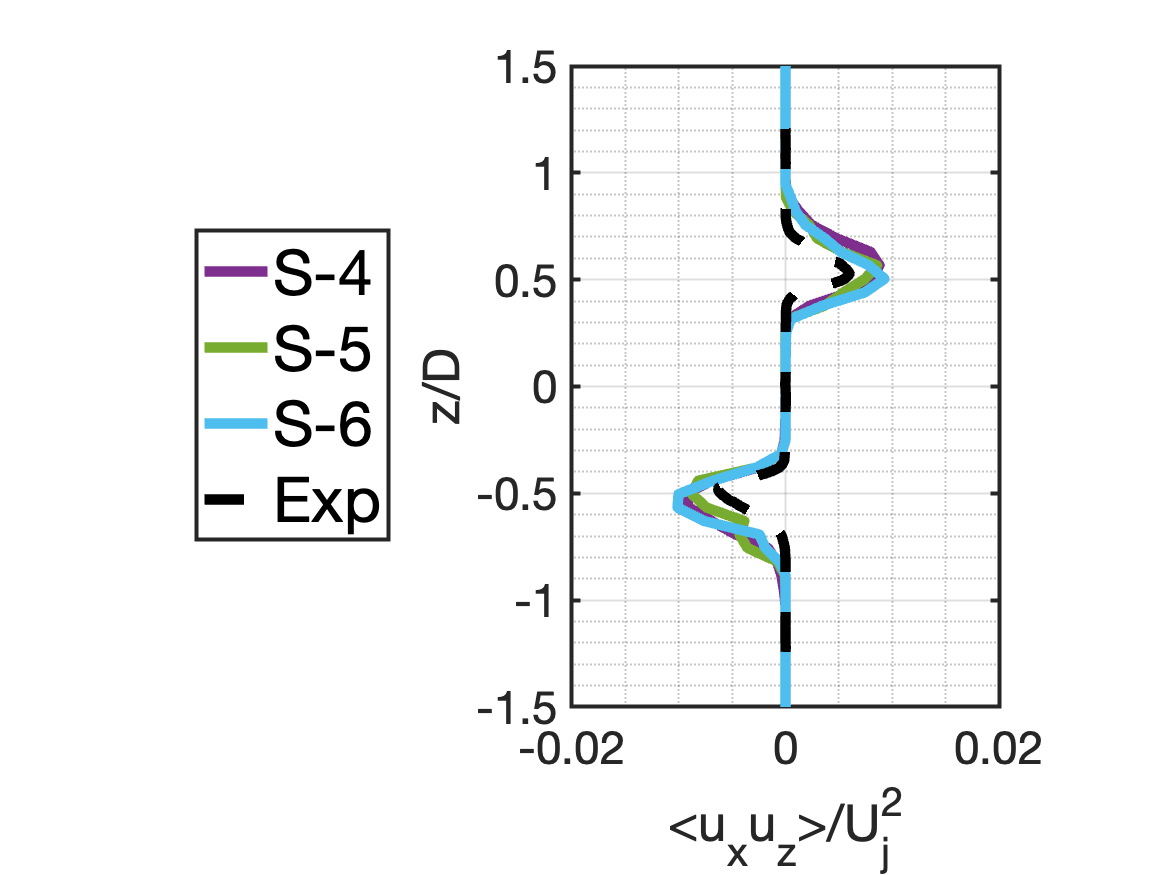}
	\label{fig.res105m}	
	}
\subfloat[$x/D=5$]{
	\includegraphics[trim = 30mm 0mm 55mm 0mm, clip, height=0.26\linewidth]{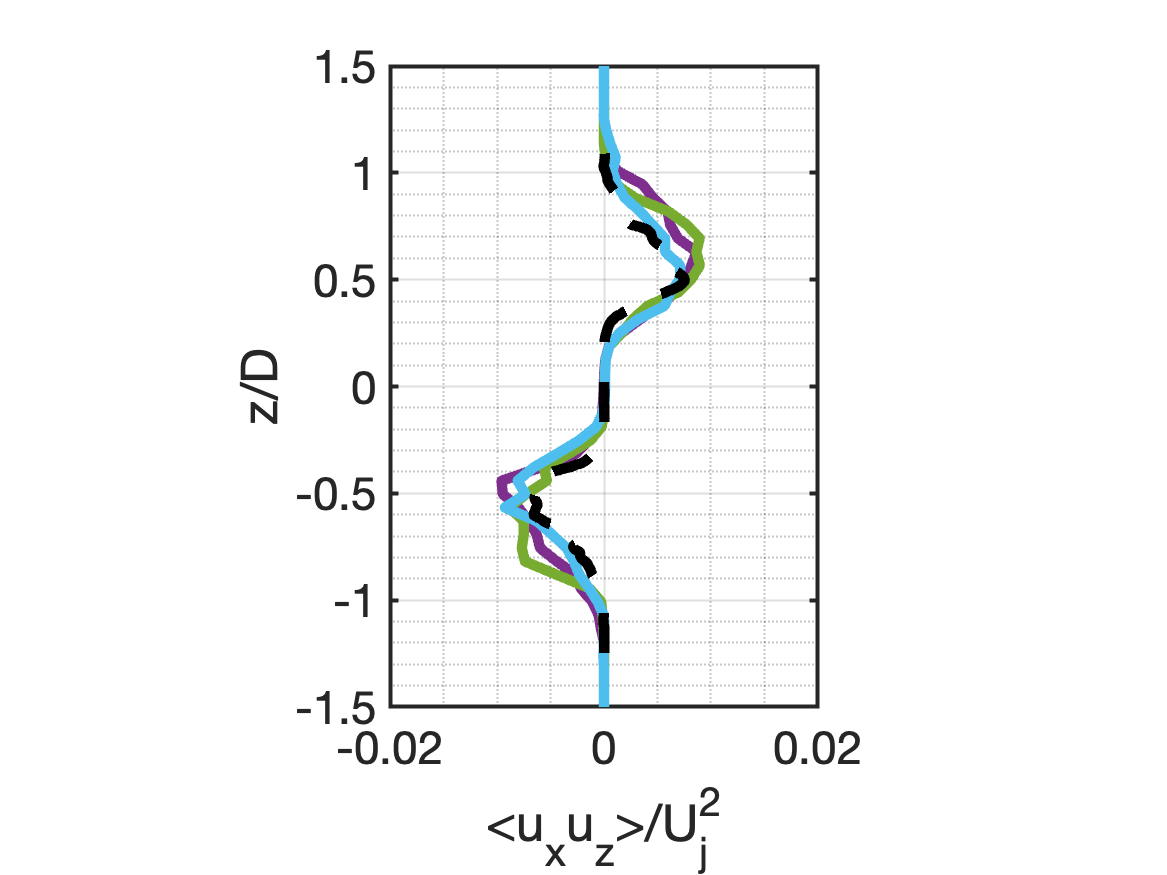}
	\label{fig.res105n}	
	}
\subfloat[$x/D=10$]{
	\includegraphics[trim = 30mm 0mm 55mm 0mm, clip, height=0.26\linewidth]{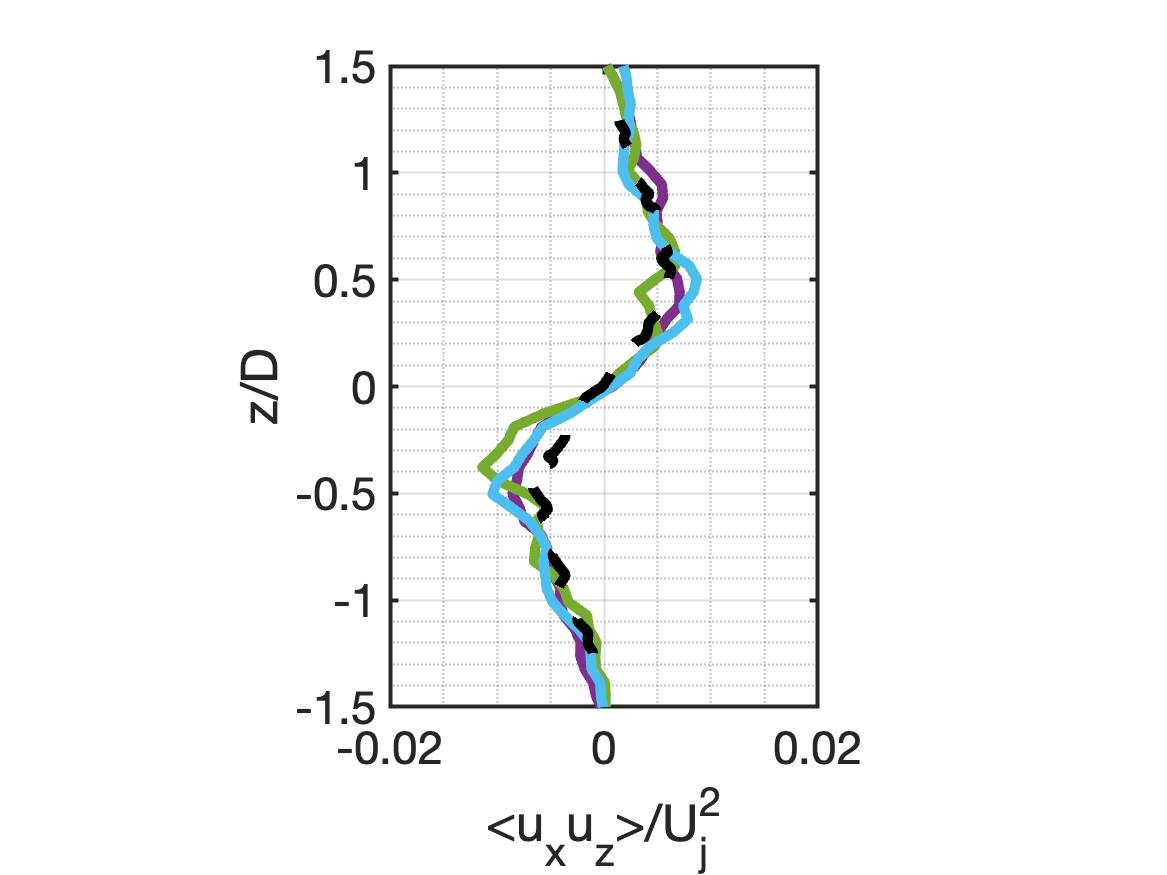}
	\label{fig.res105o}
	}
\subfloat[$x/D=15$]{
	\includegraphics[trim = 30mm 0mm 55mm 0mm, clip, height=0.26\linewidth]{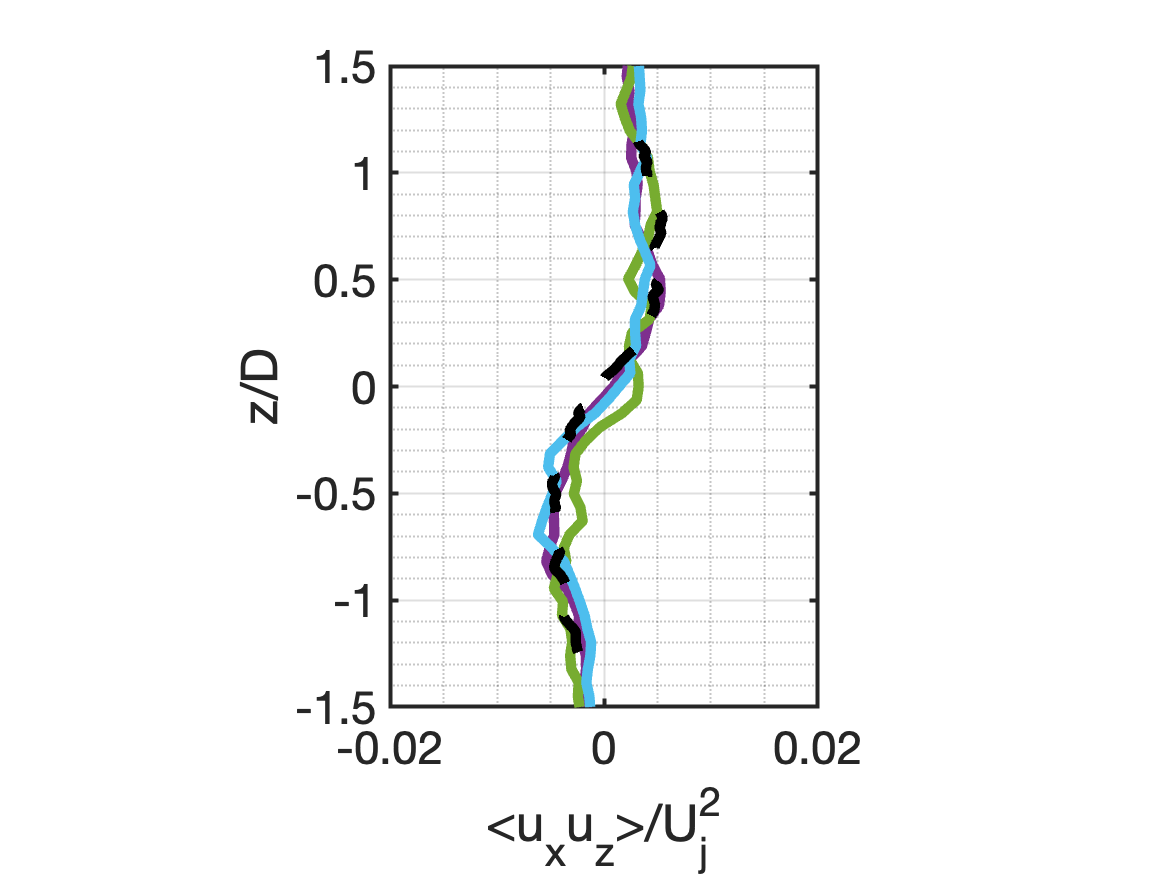}
	\label{fig.res105p}	
	}
\newline
\caption{Profiles of mean longitudinal velocity component, RMS of longitudinal
         velocity fluctuation, RMS of transversal velocity fluctuation, and
         mean shear-stress tensor component (from top to bottom) at four
         streamwise positions $x/D=2.5$, $x/D=5$, $x/D=10$ and 
$x/D=15$ (from left to right).}
\label{fig.res105}
\end{figure}

\FloatBarrier

The development of the jet flows from the three numerical simulations is
investigated in Fig.~\ref{fig.res105}. The mean longitudinal velocity
component's transversal profiles are presented in Figs.~\ref{fig.res105a} to 
\ref{fig.res105d}. In most of the stations, the velocity profiles are nearly
identical. Only in the third station, at $x/D=10.0$, is it possible to observe
smaller velocity levels at the jet centerline from the S-5 and S-6 simulations
compared to the S-4 simulation. The velocity difference agrees with 
what was observed in Fig.~\ref{fig.res101}. The velocity distribution at the
jet lipline indicates differences between the simulations, 
Fig.~\ref{fig.res103}. However, those difference occurs for $x/D<2.5$, so 
they cannot be observed from the transversal profiles.

In general, the RMS of longitudinal velocity fluctuation, 
Figs.~\ref{fig.res105e} to \ref{fig.res105h}, the RMS of the transversal 
velocity fluctuation, Figs.~\ref{fig.res105i} to \ref{fig.res105l}, and the
mean shear-stres tensor component, Figs.~\ref{fig.res105m} to \ref{fig.res105p},
present a small difference between the three numerical simulations. The 
velocity distribution from the numerical simulations with the three inflow
conditions indicate that the change in the inflow condition from the inviscid
profile to the mean RANS profile produce differences to the mean velocity
and RMS of the velocity fluctuation in the region of the lipline close to the
jet inlet section. The mean RANS profile boundary condition does not affect the
flow characteristics far from the jet inlet section, where the profile with
the inviscid profile already presented an agreement with experimental data.
The superimposition of the noise-tripping to the mean RANS profile was not 
capable of producing important changes to the velocity distribution when
compared to the mean RANS profile only.

\section{Concluding Remarks}

The study of the effects of inlet conditions on the simulation of supersonic
jet flows is developed in this work. Resolution analysis, using an inviscid
inlet profile, indicates that the calculations using M-3 mesh with a 
third-order accurate spatial discretization can reproduce aspects of the 
jet flow physics with a high level of agreement with experimental data. The 
resolution improvement positively affects the representation of the potential
core and the core length of the jet. The mixing layer development starts 
closer to the jet inlet section, and the spreading rate decreases when 
refining the study. The shock waves have a higher sharpness with increased
sets. One can conclude that, in general, the properties investigated got 
close to experimental data with improved \textit{hp} refinement.

The increased resolution of the simulation led to velocity profiles that got
far from experimental data for both mean longitudinal velocity and fluctuation
of longitudinal velocity component close to the inlet section. Due to the
proximity of the jet inlet boundary condition and simplified physics imposed 
by the inviscid profile, which does not represent the condition obtained in 
the experiments, the authors are convinced that the mismatch with experimental
data is related to the inlet condition imposed.


The solution to this issue is addressed by improving the physical 
representation of the jet inlet condition. Two boundary conditions are 
investigated. In the first inflow method, the mean profiles of the primitive
properties obtained from a RANS simulation of the nozzle flow are utilized as
the inflow condition in the LES calculation. The mean profile is steady. In the
second inflow method, the mean RANS profile is superimposed by a noise-tripping
procedure in the velocity properties inside the boundary layer. The two inflow
conditions are simulated with the numerical setup that produced the highly
resolved simulation and are compared with the simulation with the inviscid
profile.

The comparison of the numerical data from the three numerical simulations 
shows that the imposition of a mean profile from a RANS simulation could
produce different mean velocity and velocity fluctuation distributions close 
to the inlet section. A mean velocity reduction is observed in the lipline
of the jet flow, and the peak RMS values in the lipline are also reduced with
the employment of the mean RANS profile. While the velocity distributions were
affected by the inflow method close to the inlet section, far from the inlet
section, the influence of the inflow condition is negligible. The lack of 
influence far from the inlet section is important since, in this region, the
numerical data with the inviscid profile already presented agreement with the
experimental reference. The noise-tripping procedure could not produce
additional improvements to the velocity distributions. Its influence on the
results is negligible, suggesting that it needs to be better adjusted for the
supersonic jet flow.

Due to the low computational cost of the mean RANS profile and the improvement
to the numerical solution that it brought to the numerical simulations, it is
the choice for the development of other jet flow simulations. The sequence of
the study should focus on how to superimpose unsteady information into the 
RANS profile that could put the numerical results even closer to the 
experimental data. One direct step could be to adjust the parameters of the 
noise-tripping procedure and observe if it produces any change to the numerical 
results.

\section*{Acknowledgments}

The authors acknowledge the support for the present research provided by Conselho Nacional 
de Desenvolvimento Cient\'{\i}fico e Tecnol\'{o}gico, CNPq, under the Research Grant No.\ 
309985/2013-7\@. The work is also supported by the computational resources from the Center 
for Mathematical Sciences Applied to Industry, CeMEAI, funded by Funda\c{c}\~{a}o de Amparo 
\`{a} Pesquisa do Estado de S\~{a}o Paulo, FAPESP, under the Research Grant No.\ 2013/07375-0\@. 
The authors further acknowledge the National Laboratory for Scientific Computing (LNCC/MCTI, 
Brazil) for providing HPC resources of the SDumont supercomputer. This work was also granted 
access to the HPC resources of IDRIS under the allocation A0152A12067 made by GENCI. The first 
author acknowledges authorization by his employer, Embraer S.A., which has allowed his 
participation in the present research effort. The doctoral scholarship provide by FAPESP to the 
third author, under the Research Grant No.\ 2018/05524-1, is thankfully acknowledged. Additional 
support to the fourth author under the FAPESP Research Grant No.\ 2013/07375-0 is also gratefully 
acknowledged.

\bibliography{bibfile-scitech-2024}

\end{document}